\documentclass[english]{article}
\usepackage{times}
\usepackage[T1]{fontenc}
\usepackage[latin1]{inputenc}
\usepackage{geometry}
\usepackage{verbatim}
\usepackage{rotating}
\usepackage{amsmath}
\usepackage{graphicx}
\usepackage{amssymb}

\makeatletter

\providecommand{\tabularnewline}{\\}

\newcommand{\lyxaddress}[1]{
\par {\raggedright #1
\vspace{1.4em}
\noindent\par}
}

\usepackage{babel}
\makeatother
\begin{document}

\newcommand{\sgrad}{\boldsymbol{\nabla}_{S}}

\newcommand{\degree}{^{\circ}}

\newcommand{\dif}[1]{\vec{#1}-\vec{#1}'}

\newcommand{\dd}[2]{\delta^{#2}\left(#1-#1'\right)}

\newcommand{\vd}[2]{\delta^{#2}\left(\vec{#1}-\vec{#1'}\right)}

\newcommand{\ds}[1]{\delta\left(#1-#1'\right)}

\newcommand{\grad}{\boldsymbol{\nabla}}

\newcommand{\phik}[3]{e^{#3i#1\cdot#2}}

\newcommand{\intn}[4]{\int_{#1}^{#2}d^{n}\vec{#3}\left(#4\right)}

\newcommand{\height}{\mathcal{H}}

\newcommand{\h}{\height}

\newcommand{\f}{\mathcal{F}}

\newcommand{\dx}{\boldsymbol{\Delta}\mathbf{x}}

\newcommand{\av}{\boldsymbol{\alpha}}

\newcommand{\ez}{\mathcal{E}_{0\degree}}

\renewcommand{\vec}{\mathbf}

\providecommand{\cs}[1][SOMETHING]{[CITE:#1]\marginpar{C} }

\providecommand{\linkable}{}


\title{Order of Epitaxial Self-Assembled Quantum Dots: Linear Analysis}

\author{Lawrence H. Friedman}

\maketitle

\lyxaddress{\begin{center}Dept. of Engineering Science and Mechanics, Pennsylvania
State University, 212 Earth and Engineering Science Building, University
Park, Pennsylvania 16802\\
\linkable{lfriedman@psu.edu}\par\end{center}}

\noindent \textbf{keywords:} quantum dots, strained films, epitaxial
growth, semiconductors

\begin{abstract}
Epitaxial self-assembled quantum dots (SAQDs) are of interest for
nanostructured optoelectronic and electronic devices such as lasers,
photodetectors and nanoscale logic. Spatial order and size order of
SAQDs are important to the development of usable devices. It is likely
that these two types of order are strongly linked; thus, a study of
spatial order will also have strong implications for size order. Here
a study of spatial order is undertaken using a linear analysis of
a commonly used model of SAQD formation based on surface diffusion.
Analytic formulas for film-height correlation functions are found
that characterize quantum dot spatial order and corresponding correlation
lengths that quantify order. Initial atomic-scale random fluctuations
result in relatively small correlation lengths (about two dots) when
the effect of a wetting potential is negligible; however, the correlation
lengths diverge when SAQDs are allowed to form at a near-critical
film height. The present work reinforces previous findings about anisotropy
and SAQD order and  presents as explicit and transparent mechanism
for ordering with corresponding analytic equations. In addition, SAQD
formation is by its nature a stochastic process, and various mathematical
aspects regarding statistical analysis of SAQD formation and order
are presented.
\end{abstract}

\section{Introduction}

\label{sec:Introduction}Epitaxial self-assembled quantum dots (SAQDs)
represent an important step in the advancement of semiconductor fabrication
at the nanoscale that will allow breakthroughs in optoelectronics
and electronics.~\cite{Bimberg99,Pchelyakov2000,Grundmann2000,Petroff2001,Liu2001,Heinrichsdorff1997,Bimberg2002,Ledentsov2002,Friesen2003,Cheng2003,Krebs2003,Sakaki2003}
Most frequent optoelectronic applications are high efficiency lasers
with exotic wavelengths or photodetectors.~\cite{Bimberg99,Grundmann2000,Petroff2001,Liu2001,Heinrichsdorff1997,Bimberg2002,Ledentsov2002,Cheng2003,Krebs2003,Sakaki2003}
SAQDs are the result of a transition from 2D growth to 3D growth in
strained epitaxial films such as $\mbox{Si}_{x}$$\mbox{Ge}_{1-x}$/Si
and $\mbox{In}_{x}\mbox{Ga}_{1-x}\mbox{As}$/GaAs. This process is
known as Stranski-Krastanow growth or Volmer-Webber growth.~\cite{Spencer:1991we,Bimberg99,Brunner:2002gf,Freund:2003ih}. 

In applications, order is a key factor. There are two types of order,
spatial and size. Spatial order refers to the regularity of SAQD dot
placement, and it is necessary for nano-circuitry applications. Size
order refers to the uniformity of SAQD size which determines the voltage
and/or energy level quantization of SAQDs. It is reasonable to expect
that these type of order are linked, and it is important to understand
the factors that determine SAQD order. Further understanding should
help in the design and simulation of both spontaneous {}``bottom
up'' self-assembly and directed or guided self-assembly to enhance
SAQD order.~\cite{Shiryaev:1997iz,Kumar:fk,Friedman:2006bc,Krishna:1999ri,Hull:2003yi,Guise:patterning,Niu:2006fk,Zhao:2006qy}Here,
an elaboration of and further application of a linear analysis of
SAQD order~\cite{Friedman:fk} is presented. The work reported here
forms the basis of a non-linear theory and modeling of SAQD order
that will be reported in future work.

In~\cite{Friedman:fk} it was reported that one could calculate a
correlation function using a linearized model of SAQD formation. This
correlation function included two correlation lengths that could be
used to describe SAQD order. It was also found that one effect of
a hypothesized wetting potential was to enhance SAQD order when growth
occurs near the critical film height for 3D growth. Here, these results
are expanded to create a more rigorous linearized theory of SAQD order
that will inform non-linear theories. In particular, the model is
generalized to any model that combines local energy effects such as
surface energy density and non-local elastic destabilization, and
the procedure for predicting order based on any linear theory with
peak wavelengths is presented. The hypothesized effect of elastic
anisotropy in~\cite{Friedman:fk} is verified with calculations using
linear anisotropic elasticity theory.~\cite{Obayashi:1998fk,Ozkan:1999gf}
Details such as statistical fluctuation and convergence are also addressed
along with a discussion of the possible forms of linear anisotropic
terms in SAQD growth kinetics, and the effect of an atomic-scale cutoff
in the continuum theory is addressed. Finally, the order enhancing
effect of growing near the critical threshold is explored in more
detail using calculations appropriate to Ge/Si SAQDs.

In the literature, two modes of SAQD formation are generally discussed,
the thermal nucleation mode and the nucleationless mode.~\cite{Tersoff:1994fk,Spencer:1993vt,Baribeau:2006fj}
In the thermal nucleation mode, a 2D film surface is metastable, and
the formation of individual quantum dots is thermally activated.~\cite{Tersoff:1994fk}.
This growth mode leads to the formation of individual quantum dots
as uncorrelated or loosely correlated discrete events at essentially
random locations. In the nucleationless mode, the 2D film surface
transitions from stable (or metastable) to unstable. In this mode,
dots form everywhere at once appearing at first as a cross-hatched
ripple-like disturbance on the 2D film surface and then maturing into
recognizable individual dots.\cite{Tersoff:1994fk,Srolovitz:1989fu,Spencer:1993vt,Gao:1999ve,Sutter:2000kx}
These two modes are probably connected via an encompassing conceptual
and mathematical model%
\footnote{It is likely that there a transition from stable, to metastable and
finally to unstable. The analysis presented in~\cite{Golovin:2003ms}
would appear to support such a view where the film height acts as
the control parameter driving the transition. There is also some controversy
regarding whether all dot growth is nucleationless or not.~\cite{Ramasubramaniam:2005oq,Sutter:2000kx}%
}, and perhaps some of what is observed experimentally is in fact a
hybrid mechanism. In agreement with intuition, it appears that the
nucleationless mode leads to a more ordered dot pattern than the thermal
nucleation mode that is dominated by randomness.~%
\footnote{Compare various figures in~\cite{Baribeau:2006fj,Berbezier:2003mw,Brunner:2002gf,Gao:1999ve,Bortoleto:2003zh}.%
} Thus, the presented analysis applies to the nucleationless mode. 

There are various implementations of nucleationless growth models~\cite{Spencer:1993vt,Liu:2003kx,Zhang:2003tg,Wang:2004dd,Tekalign:2004jh,Friedman:2006bc,Ramasubramaniam:2005oq},
although, there is also a great deal of commonality among these models.
In particular, they all include a non-local elastic effect and local
surface energies and/or local wetting energies. Here, a linear analysis
of quantum dot order resulting from this class of model is presented.
Particular note is taken of the effects of stochastic initial conditions
crystal anisotropy in general, elastic anisotropy in particular, and
the effect of varying film height as a control parameter as first
introduced in~\cite{Golovin:2003ms}. A simple model similar to \cite{Spencer:1993vt,Liu:2003kx,Zhang:2003tg,Tekalign:2004jh,Friedman:2006bc}
is presented to produce numerical examples and explore the effects
of the average film height. Concurrently, a more abstract and general
model is presented and analyzed that includes non-local elastic strain
effects, and a local combined surface and wetting energy. The linear
model with stochastic initial conditions and deterministic film height
evolution will pave the way for more sophisticated analysis involving
a non-linear model of stochastic film height evolution. 

As previously stated, one of the goals in the present work is to further
explore the role of the wetting potential during growth near the stability
threshold in film height. A wetting potential has been included in
the analysis and simulations in~\cite{Zhang:2003tg,Golovin:2003ms,Liu:2003kx,Spencer:1993vt}.
Although somewhat controversial, the wetting potential plays an important
phenomenological role. It ensures that growth takes place in the Stranski-Krastanow
mode: that a 3D unstable growth occurs only after a critical layer
thickness is achieved, and that a residual wetting layer persists.
The physical origins and consequences of the wetting potential are
discussed in~\cite{Beck:2004yq,Spencer:1993vt}. The analysis presented
here is usable in models that neglect the wetting potential by simply
setting it to zero. Another possibility is simply that the wetting
potential is simply an approximation to the stabilizing effect of
intermixing.~\cite{Tu:2004tg} That said, if the wetting potential
is real, the present analysis shows that it is beneficial to SAQD
order to grow near the critical layer thickness.

The presented analytic formulas and linear analysis are intended to
complement existing numerical models of SAQD order.~\cite{Holy:1999th,Liu:2003kx,Liu:2003ig,Springholz:2001nx}
and to form a basis for future non-linear analytic analysis of SAQD
order. The current findings agree with previous work on the beneficial
effects of elastic anisotropy to enhance in-plane order. 

The linear analysis, of course, represents a simplification of the
film evolution, and it applies only to the initial stages of SAQD
formation when the nominally flat film surface becomes unstable and
transitions to three-dimensional growth. However, the small surface
fluctuation stage of SAQD growth determines the initial seeds of order
or disorder in an SAQD array; thus, the small fluctuation stage should
have an important influence on the final outcome.  At later stages
when surface fluctuations are large, there is a natural tendency of
SAQDS to either order or ripen~\cite{Golovin:2003ms,Liu:2003kx,Liu:2003qi,Wang:2004dd,Ross:1998fk}
Ordering systems tend to evolve slowly due to critical slowing down~\cite{Wang:2004dd},
while ripening tends to diminish order further.~\cite{Liu:2003kx}
Thus, it is possible that the linear model could, in fact, yield good
predictions of SAQD order. The simplification and linearizion facilitates
the development of analytic solutions that are most transparent, easily
portable to multiple material systems and have no effective limit
on system size. Finally, it is virtually impossible to have a thorough
understanding of the full non-linear model without first having a
thorough understanding of the linear behavior.

The remainder of the paper is organized as follows. Section~\ref{sec:Modeling}
presents the physical assumptions and mathematical approximations
used to model film growth. Section~\ref{sec:Correlation-Functions}
discusses the stochastic initial conditions and the resulting correlation
functions and correlation lengths. Section~\ref{sec:Application-to-Materials},
presents a procedure for estimating SAQD order with an application
to Ge dots on a Si substrate. Section~\ref{sec:Conclusions} presents
conclusions, while Appendices~\ref{sec:Diffusion-Potential}-\ref{sec:Atomic-Scale-Cutoff}
present additional calculational details.

\section{Modeling}

\label{sec:Modeling}The formation of SAQDs is modeled as a deterministic
surface diffusion process with stochastic initial conditions. The
resulting equations and ultimately the sought after correlation functions
are different depending on whether the film surface is treated as
one-dimensional isotropic, two-dimensional isotropic or two-dimensional
anisotropic. The 1D and 2D isotropic cases are discussed first, and
then the essential differences of the 2D anisotropic model are presented.
The stochastic initial conditions need to be expressed in terms of
the correlation functions that are also use to analyze order; consequently,
the discussion of the initial conditions is deferred to Sec.~\ref{sub:Stochastic-Initial-Conditions}.

It should be noted that the results presented here are fairly general.
There has been a good deal of recent work refining the modeling of
nucleationless growth processes to incorporate various phenomenological
aspects of SAQD growth. For example, the inclusion of orientation-dependent
surface energy~\cite{Zhang:2003tg}, strain-dependent surface energy~\cite{Ramasubramaniam:2005oq}
and explicit modeling of atomic species segregation and film-substrate
inter diffusion.~\cite{Tersoff:2003fk} Two models are presented
here. One is a simple concrete example. It is the simplest model one
can use including elastic effects surface energy and wetting energy.
The second model is more abstract and describes the general case of
a local potential energy that depends on both the film height and
film height gradient. One effect that is not examined here is that
of mixed 4-fold and two-fold symmetry. Such a mixing can occur due
to diffusional anisotropy or surface energy anisotropy. (Sec.~\ref{par:Surface-and-Wetting}
and Appendix~\ref{sec:Diffusion-Anisotropy}). However, a similar
analysis procedure should work for these cases as well. The general
procedure for possible application to other models is discussed in
Sec.~\ref{sub:Generalizability}. 

The following discussion will use abstract vector notation, \emph{e.g.}
$\vec{k}$ instead of $k_{i}$, etc. Also, because it is sometimes
computational expedient to perform one-dimensional modeling~\cite{Friedman:fk,Wang:2004dd,Kumar:fk,Tu:2004tg},
the case of a one dimensional surface with two dimensional volume
is discussed along with the case of an isotropic 2D surface. To facilitate
this combined discussion, the dimensionality of the surface will be
denoted as $d$. In Secs.~\ref{sub:Reciprocal-Space-Corelation}
and~\ref{sub:Real-Space-Correlation}, $d=1,2$ will be substituted
as appropriate. Finally, much of the calculation involves reciprocal
space. The convention used for the Fourier transforms is\[
f(\vec{x})=\int d^{d}\vec{k}\, e^{i\vec{k}\cdot\vec{x}}f_{\vec{k}}\text{, and }f_{\vec{k}}=(2\pi)^{-d}\int d^{d}\vec{x}\, e^{-i\vec{k}\cdot\vec{x}}f(\vec{x})\]
following the example of~\cite{Spencer:1993vt}.

\subsection{1D and 2D Isotropic model}

\label{sub:Isotropic-model}This discussion pertains to both the 1D
model and the 2D isotropic model. The formation of SAQDs is modeled
as a surface diffusion process where the film height is a function
of the lateral position and time. The system is treated as deterministic
with stochastic initial conditions. First, the general non-linear
governing equations are presented. Then, the linearized form is presented.
Finally, the key behavior is reviewed.

The mathematical model uses film height, $\h(\vec{x},t)$ as the dependent
variable and the horizontal position $\vec{x}$ and time $t$ as the
independent variables. The film height evolves over time due to surface
diffusion driven by a diffusion potential $\mu(\vec{x},t)$ and a
flux of new material $Q$. The surface velocity is thus\begin{equation}
v_{n}=n_{z}\partial_{t}\h=-\sgrad\cdot\mathcal{D}\sgrad\mu(\vec{x},t)+Q\label{eq:governing1}\end{equation}
where $n_{z}$ is the vertical component of the surface normal $n_{z}=[1+(\grad\h)^{2}]^{-1/2}$,
$\sgrad$ is the surface gradient, $\mathcal{D}$ is the diffusivity,
and $\sgrad\cdot$ is the surface divergence.

\subsubsection{Energetics}

\label{sub:Energetics}The diffusion potential $\mu(\vec{x},t)$ must
produce Stranski-Krastanow growth. Thus, it must contain an elastic
term that destabilizes film growth, a surface energy term that stabilizes
planar growth and a wetting energy that ensures a wetting layer. The
diffusion potential can be derived from a total free energy.\begin{eqnarray*}
\mathcal{F} & = & \mathcal{F}_{\text{elast}}+\mathcal{F}_{\text{surf.}}+\mathcal{F}_{\text{wet}}\\
 & = & \int_{\text{volume}}dV\,\omega+\int_{\text{surface}}dA_{\text{surf.}}\,\gamma+\int dA\, W(\h)\end{eqnarray*}
where $\omega$ is the elastic energy density, $\gamma$ is the surface
energy density, $W(\h)$ is the wetting energy density. The last integral
corresponds to $\mathcal{F}_{\text{wet}}$, and whether the integral
should be taken over the film surface or the substrate is ambiguous.
The {}``simple'' model (Sec.~\ref{par:simple-form}) assumes that
the integral is over the substrate, while the {}``general'' model
(Sec.~\ref{par:general-form}) can accommodate both cases.

\paragraph{simple form}

\label{par:simple-form}The simplest possible model results if the
integral corresponding to $\mathcal{F}_{wet}$ is taken over the lateral
positions $\vec{x}$ rather than over the actual free-surface. In
concrete terms, one can use $dV=d^{2}\vec{x}dz$ and $dA_{\text{surf.}}=d^{2}\vec{x}\left[1+\left(\grad\h(\vec{x})\right)^{2}\right]^{1/2}$
to obtain the expression,\begin{equation}
\mathcal{F}=\int_{\text{volume}}d^{2}\vec{x}dz\,\omega[\h](\vec{x},z)+\int_{\vec{x}\text{-\text{plane}}}d^{2}\vec{x}\,\left\{ \left[1+\left(\grad\h(\vec{x})\right)^{2}\right]^{1/2}\gamma+W\left(\h(\vec{x})\right)\right\} ,\label{eq:simplest}\end{equation}
where the {}``$\omega[\h]$'' indicates that the elastic energy
density is a non-local functional of the film height, $\h$. The diffusion
potential $\mu$ can be found, similar to~\cite{Freund:2003ih},
by differentiating $\mathcal{F}$ with respect to the surface motion
(Appendix~\ref{sub:General-form-of}), $\mu(\vec{x})=\Omega\delta\mathcal{F}/\delta\h(\vec{x})$.
Doing so for Eq.~\eqref{eq:simplest} (Appendix~\ref{sub:Simple-Model}),
\begin{equation}
\mu(\vec{x})=\Omega\left[\omega(\vec{x})-\gamma\kappa(\vec{x})+W'\left(\h(\vec{x})\right)\right].\label{eq:mu1}\end{equation}
where $\Omega$ is the atomic volume, $\omega(\vec{x})$ is the elastic
energy density at the film surface (implicitly $\omega[\h]\left(\vec{x},\h(\vec{x})\right)$),
$\kappa=\grad\cdot\left\{ \grad\h(\vec{x})\left[1+\left(\grad\h(\vec{x})\right)^{2}\right]^{-1/2}\right\} $
is the total surface curvature, and $W'(\h)=\partial_{\h(\vec{x})}W\left(\h(\vec{x})\right)$
is the derivative of $W\left(\h(\vec{x})\right)$ evaluated at $\vec{x}$.

\paragraph{general form}

\label{par:general-form}It should be noted that Eq.~\eqref{eq:mu1}
is not the same diffusion potential used in~\cite{Zhang:2003tg}.
The wetting potential used there can be derived by taking $W(\h)$
as an energy density of the free surface, not a density in the $\vec{x}$-plane.
Expressions like Eq.~\eqref{eq:mu1} and Eq.~(1) in~\cite{Zhang:2003tg}
are part of a larger class of surface evolution models with more or
less the same linear behavior. 

The surface and wetting energy can be combined and incorporated into
a more general form, with a total free energy $\mathcal{F}_{sw}$
and a free energy density $F_{sw}(\h,\grad\h)$ that depends on the
film height $\h(\vec{x})$ and the film height slope or orientation
$\grad\h(\vec{x})$. The total free energy is thus\begin{eqnarray}
\mathcal{F} & = & \mathcal{F}_{\text{elast.}}+\mathcal{F}_{sw}\label{eq:swcombo}\\
 & = & \int_{\text{volume}}d^{2}\vec{x}dz\,\omega[\h](\vec{x},z)+\int_{\vec{x}-\text{plane}}d^{2}\vec{x}\, F_{sw}\left(\h(\vec{x}),\grad\h(\vec{x})\right).\nonumber \end{eqnarray}
$F_{sw}$ may not necessarily be the sum of separate surface energy
and wetting energy contributions. It need only be a local function
of $\h$ and $\grad\h$. The corresponding diffusion potential is\begin{equation}
\mu(\vec{x})=\Omega\left[\omega(\vec{x})+F_{sw}^{\left(10\right)}(\vec{x})-\grad\cdot\vec{F}_{sw}^{\left(01\right)}(\vec{x})\right],\label{eq:mu2}\end{equation}
where $F_{sw}^{(mn)}$ indicates the $m^{\text{th}}$ derivative with
respect to $\h$ and the $n^{\text{th}}$ derivative with respect
to $\grad\h$. $F_{sw}^{\left(10\right)}(\vec{x})=\partial_{\h(\vec{x})}F_{sw}\left(\h(\vec{x}),\grad\h(\vec{x})\right)$
and each vector component of $\vec{F}_{sw}^{\left(01\right)}(\vec{x})$
is $\left[\vec{F}_{sw}^{\left(01\right)}(\vec{x})\right]_{i}=\partial_{\left[\grad\h(\vec{x})\right]_{i}}F_{sw}\left(\h(\vec{x}),\grad\h(\vec{x})\right)$.
One can obtain the results of the simple model (Eqs.~\eqref{eq:simplest}
and~\eqref{eq:mu1}) by setting \begin{equation}
F_{sw}=\left[1+\left(\grad\h(\vec{x})\right)^{2}\right]^{1/2}\gamma+W\left(\h(\vec{x})\right).\label{eq:fsw-simp}\end{equation}
A diffusion potential like Eq.~(1) in~\cite{Zhang:2003tg} can be
obtained by setting \[
F_{sw}=\left[1+\left(\grad\h(\vec{x})\right)^{2}\right]^{1/2}\left[\gamma\left(\grad\h(\vec{x})\right)+W\left(\h(\vec{x})\right)\right].\]
 This is different from Eq.~\eqref{eq:fsw-simp} in two ways. First,
the surface energy density depends on the surface orientation. Second,
the Jacobian, $J=\left[1+\left(\grad\h(\vec{x})\right)^{2}\right]^{1/2}$
multiplies both the surface energy density and the wetting potential.
Despite these differences, the common form of the diffusion potential
(Eq.~\eqref{eq:mu2}) among different models suggests that they might
all lead to similar linearized forms and behavior.

\paragraph{Linearization}

\label{par:Linearization}The diffusion potential is now linearized
about the average film height $\bar{\h}$. In general, one can control
the amount of deposited material, and thus the average film height
$\bar{\h}$. It is therefore useful to decompose $\h(\vec{x})$ into
the spatially averaged mean value and fluctuations about the average.
Similar to~\cite{Spencer:1993vt}, \begin{equation}
\h=\bar{\h}+h(\vec{x},t).\label{eq:heights}\end{equation}
In the present calculation, $\bar{\h}$ is specified as constant in
time. This assumption corresponds physically to a fast deposition
and then an anneal. It is not too difficult to generalize to a time
dependent $\bar{\h}$, but that is beyond the scope of this manuscript.
In~\cite{Zhang:2003tg,Ramasubramaniam:2005vn}, deposition and evaporation
is explicitly modeled.

All terms in $\mu(\vec{x},t)$ are now kept to only linear order in
$h(\vec{x},t)$. The elastic energy density $\omega$ is a non-local
functional of $h(\vec{x},t)$~\cite{Tekalign:2004jh}; however, the
equations generating $\omega(\vec{x})$ are translationally invariant.
Thus, it is convenient to use reciprocal space for the linearization.
The curvature is trivially linearized as $\kappa(\vec{x})\rightarrow\nabla^{2}h(\vec{x})$
in real space or $\kappa_{\vec{k}}\rightarrow-k^{2}h_{\vec{k}}$ in
reciprocal space. The linearized elastic strain energy $\omega$ can
be found in reciprocal space as in~\cite{Freund:2003ih} to be $\omega_{\vec{k}}=-2M(1+\nu)\epsilon_{m}^{2}h_{\vec{k}},$
where $M=E/(1-\nu)$ is the biaxial modulus, $E$ is the Young modulus,
$\nu$ is the Poisson ratio, and $\epsilon_{m}$ is the film-substrate
mismatch strain. This formula neglects possible differences in elastic
moduli between the film and substrate as in~\cite{Spencer:1993vt},
but a similar method of analysis should apply to that case as well.
Linearizing Eqs.~\eqref{eq:mu1} and~\eqref{eq:mu2} in reciprocal
space, $\mu_{\vec{k}}$ is proportional to $h_{\vec{k}}$ with a proportionality
coefficient that depends on $\vec{k}$ and $\bar{\h}$.

\begin{equation}
\mu_{\text{lin},\vec{k}}=f(\vec{k},\bar{\h})h_{\vec{k}}\label{eq:mulin}\end{equation}
where $f(\vec{k},\bar{\h})$ for three different isotropic cases,
corresponding to Eqs.~\eqref{eq:mu1} and~\eqref{eq:mu2}, and an
abstracted general form, is given by

\begin{equation}
f(\vec{k},\bar{\h})=\begin{cases}
\Omega\left[-2M(1+\nu)\epsilon_{m}^{2}k+\gamma k^{2}+W''(\bar{\h})\right] & ;\text{ case a (Eq.\,\eqref{eq:mu1}})\\
\Omega\left[-2M(1+\nu)\epsilon_{m}^{2}k+F^{02}k^{2}+F^{20}\right] & ;\text{ case b (Eq.\,\eqref{eq:mu2}})\\
-ak+bk^{2}+c & ;\text{ case c (general)}\end{cases}.\label{eq:fiso}\end{equation}
Due to isotropy, $f(\vec{k},\bar{\h})$ is independent of the direction
of $\vec{k}$, and only the wave number, $k=\left\Vert \vec{k}\right\Vert $,
appears in the right hand side. $F_{sw}^{(20)}$ is the second derivative
of $F_{sw}$ with respect to $\h$, and $F_{sw}^{(02)}$ the second
derivative of $F_{sw}$ with respect to $\grad\h$. $F_{sw}^{(20)}$
and $F_{sw}^{(02)}$ depend on $\bar{\h}$ only; thus they are constants
in the present analysis. See Appendix~\ref{sub:Linearizing-the-general}
for more precise definitions and the derivation of $f(\vec{k},\bar{\h})$.
Using Eq.~\eqref{eq:fsw-simp}, produces $F_{sw}^{(02)}=\gamma$
and $F_{sw}^{(20)}=W''(\bar{\h})$ which is identical to the \emph{simple}
case of Eq.~\eqref{eq:fiso}, a. Case c, labeled as {}``general''
where $a$, $b$, and $c$ depend implicitly on $\bar{\h}$ shows
that $f(\vec{k},\bar{\h})$ for cases~a and~b have the same relatively
simple form. It also emphasizes the dynamic effects as opposed to
the physical causes. There is a destabilizing term, $-ak$, a short
wavelength cutoff term, $bk^{2}$, and a term that stabilizes the
entire spectrum, $c$. 

Despite the label {}``\emph{general,''} there are of course limitations
to the application of Eqs.\emph{~\eqref{eq:mulin}} and\emph{~\eqref{eq:fiso}}.
For example, there has been recent work on the effects of strain-dependent
surface energies.~\cite{Ramasubramaniam:2005oq} The second form
can not represent such an effect because the derivation assumes that
the surface energy only depends on local quantities, ($\h$ and $\grad\h$)
whereas the strain effect is non-local. However, it is reasonable
to conjecture that a more detailed analysis of the effects of a strain
dependent surface energy term would produce a coefficient function
$f(\vec{k},\bar{\h})$ not very different from the case c {}``general''
form of Eq.~\eqref{eq:fiso}. Thus, the following analysis may very
well apply to this more exotic model, but more study is needed to
be certain.

\subsubsection{Dynamics\label{sub:Dynamics}}

As discussed in Sec.~\ref{sub:Energetics}, the dynamics are derived
assuming no flux of new material $\left(Q=0\right)$ and keeping only
terms to linear order in the height fluctuation, $h(\vec{x},t)$.
Under these assumptions, Eq.~\eqref{eq:governing1} can be decomposed
into a trivial equation for $\bar{\h}$ and an equation for the film
height fluctuation by inserting Eq.~\eqref{eq:heights}. \begin{eqnarray}
d\bar{\h}/dt & = & 0\label{eq:flux}\\
\partial_{t}h(\vec{x}) & = & -\grad\cdot\mathcal{D}\grad\mu_{\text{lin}}(\vec{x})\label{eq:surfd}\end{eqnarray}
where $\mu_{\text{lin}}(\vec{x})$ is the inverse Fourier transform
of Eqs.~\eqref{eq:mulin} and~\eqref{eq:fiso}, and it depends implicitly
on the average film height $\bar{\h}$. Note that the time dependence
is implicitly while the coordinate dependence is explicit. The explicit
coordinate dependence serves to distinguish Assuming that the diffusivity
$\mathcal{D}$ is constant, the Fourier transform of Eq.~\eqref{eq:surfd}
gives the linearized differential equation for the evolution of each
Fourier component.\begin{eqnarray}
\partial_{t}h_{\vec{k}} & = & -\mathcal{D}k^{2}\mu_{\vec{k}}=-\mathcal{D}k^{2}f(\vec{k},\bar{\h})h_{\vec{k}}.\label{eq:gov_lin}\end{eqnarray}
Solving Eq.~\eqref{eq:gov_lin},\begin{eqnarray}
h_{\vec{k}}(t) & = & h_{\vec{k}}(0)e^{\sigma_{\vec{k}}t};\label{eq:td}\\
\sigma_{\vec{k}} & = & -\mathcal{D}k^{2}f(\vec{k},\bar{\h}).\label{eq:gcgen}\end{eqnarray}
The surface evolves in reciprocal space as an initial condition, $h_{\vec{k}}(0)$
multiplied by an envelope function, $e^{\sigma_{\vec{k}}t}$. For
most values of $\bar{\h}$, this envelope function has a peak. As
time passes, this peak narrows and can be approximated by a gaussian.
To analyze this behavior, appropriate dimensionless variables are
defined. Then, the stability of the film is discussed. Finally, $\sigma_{\vec{k}}$
is expanded about its peak to aid analytic calculations. 

\begin{figure}
\begin{centering}\includegraphics[width=3.375in,keepaspectratio]{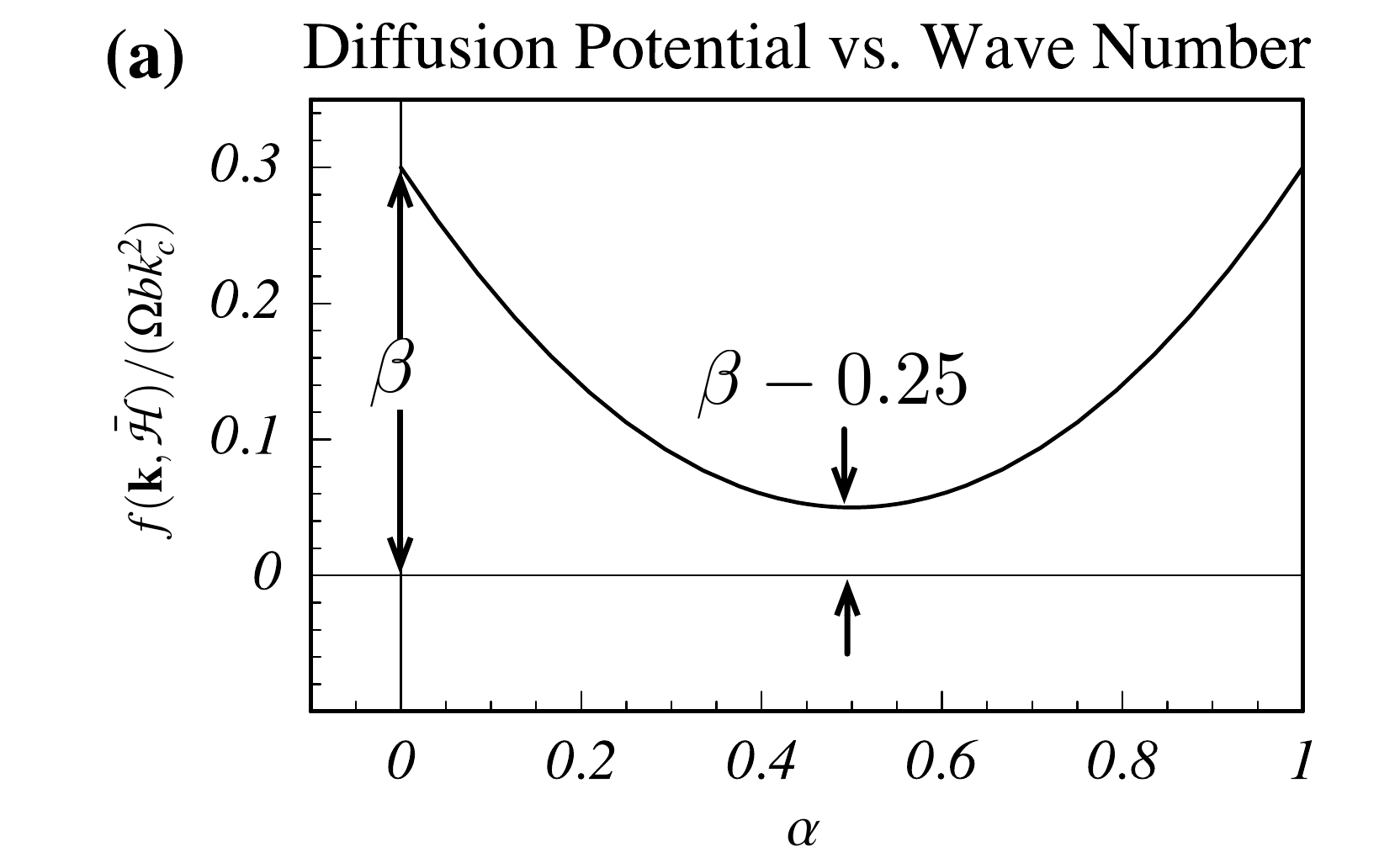}\par\end{centering}

\begin{centering}\includegraphics[width=3.375in,keepaspectratio]{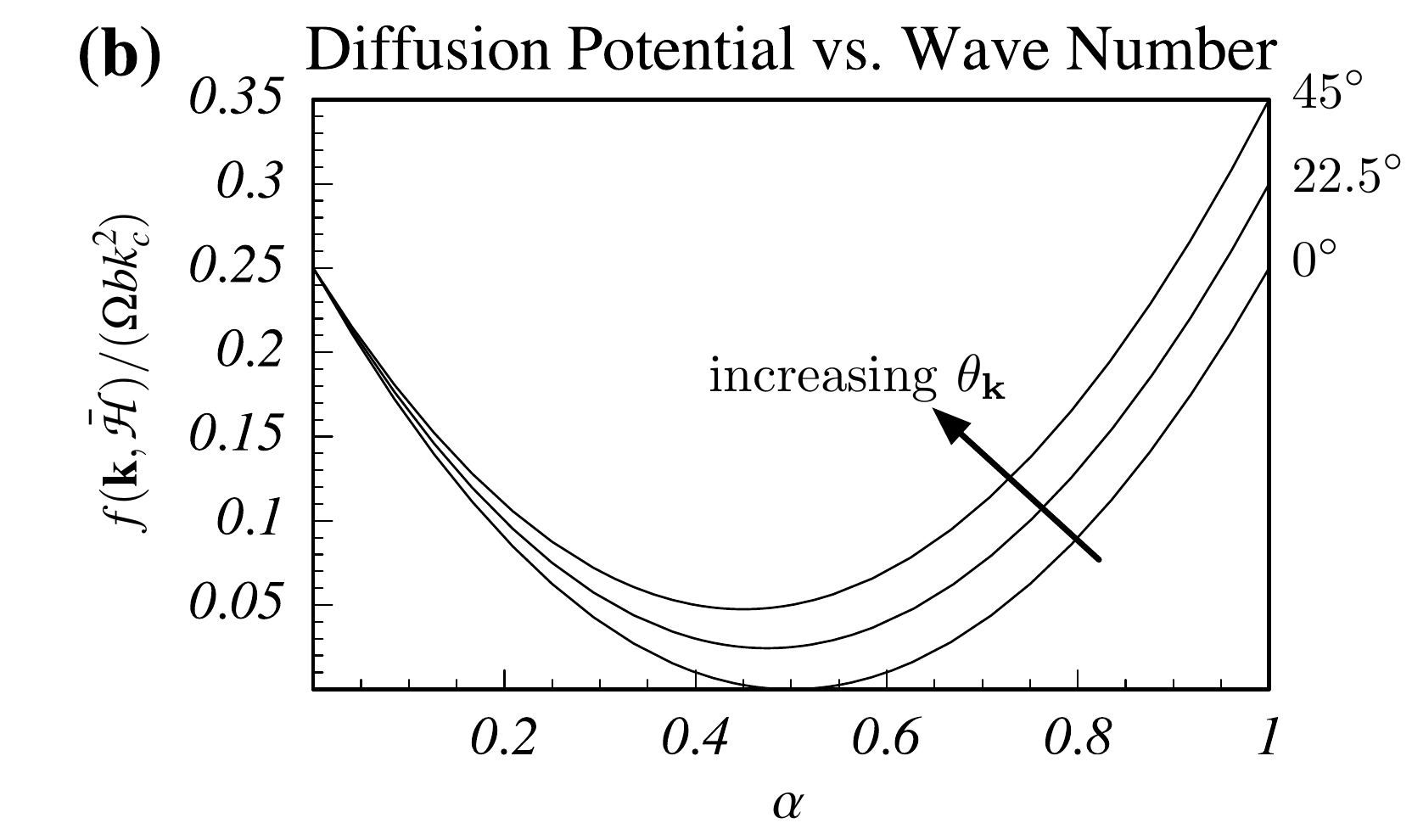}\par\end{centering}

\caption{\emph{\label{fig:Dimensionless-Diffusions-Potential}Dimensionless
diffusion potential prefactors vs. dimensionless wave number.} (a)
The one dimensional or isotropic case with $\beta=0.3$. (b) The elastically
isotropic case with anisotropy $\epsilon_{A}=0.1$ (see Eq.~\eqref{eq:einterp}). }
\end{figure}

\begin{figure}
\begin{centering}\includegraphics[width=3.375in,keepaspectratio]{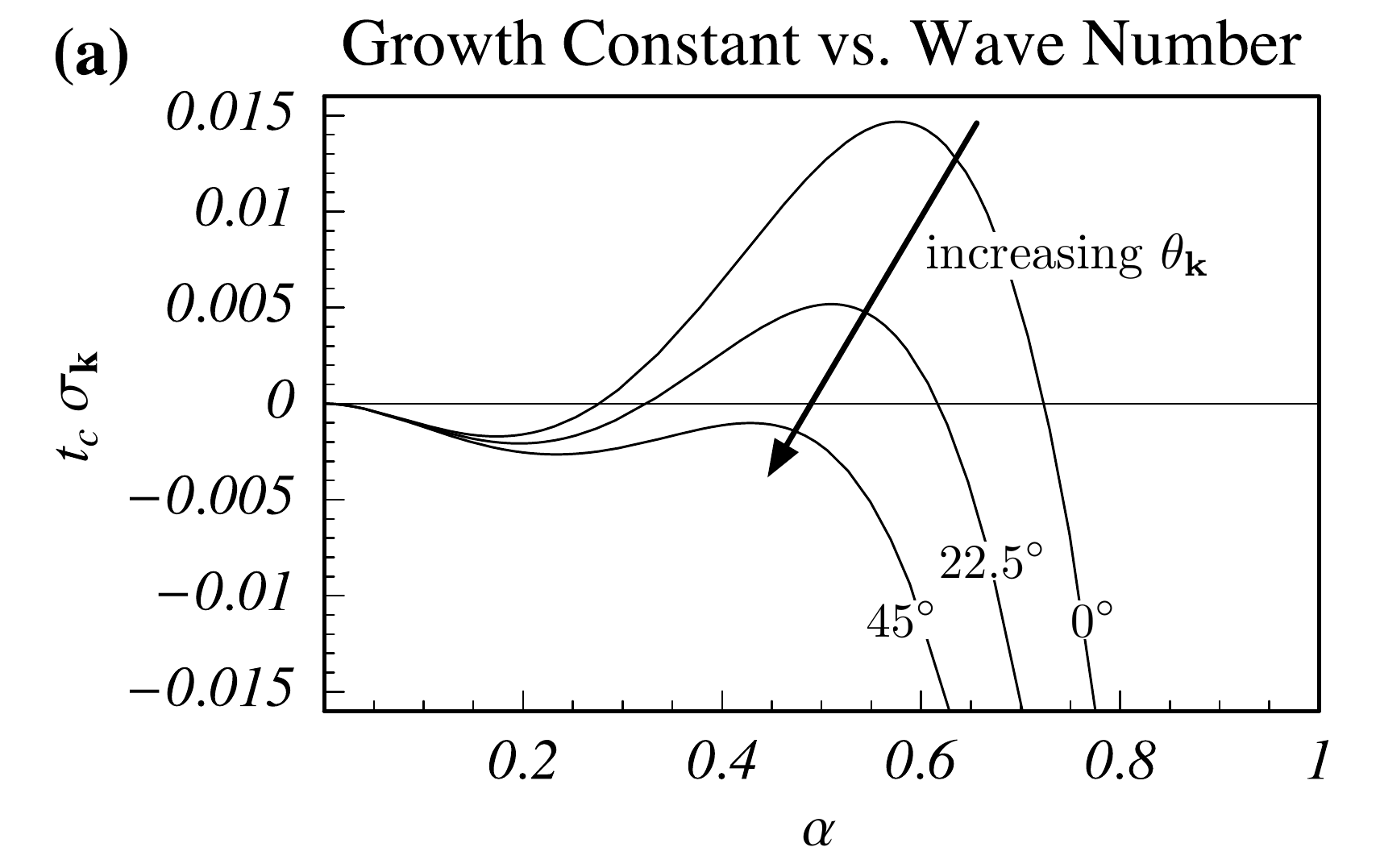}\par\end{centering}

\begin{centering}\includegraphics[width=3.375in,keepaspectratio]{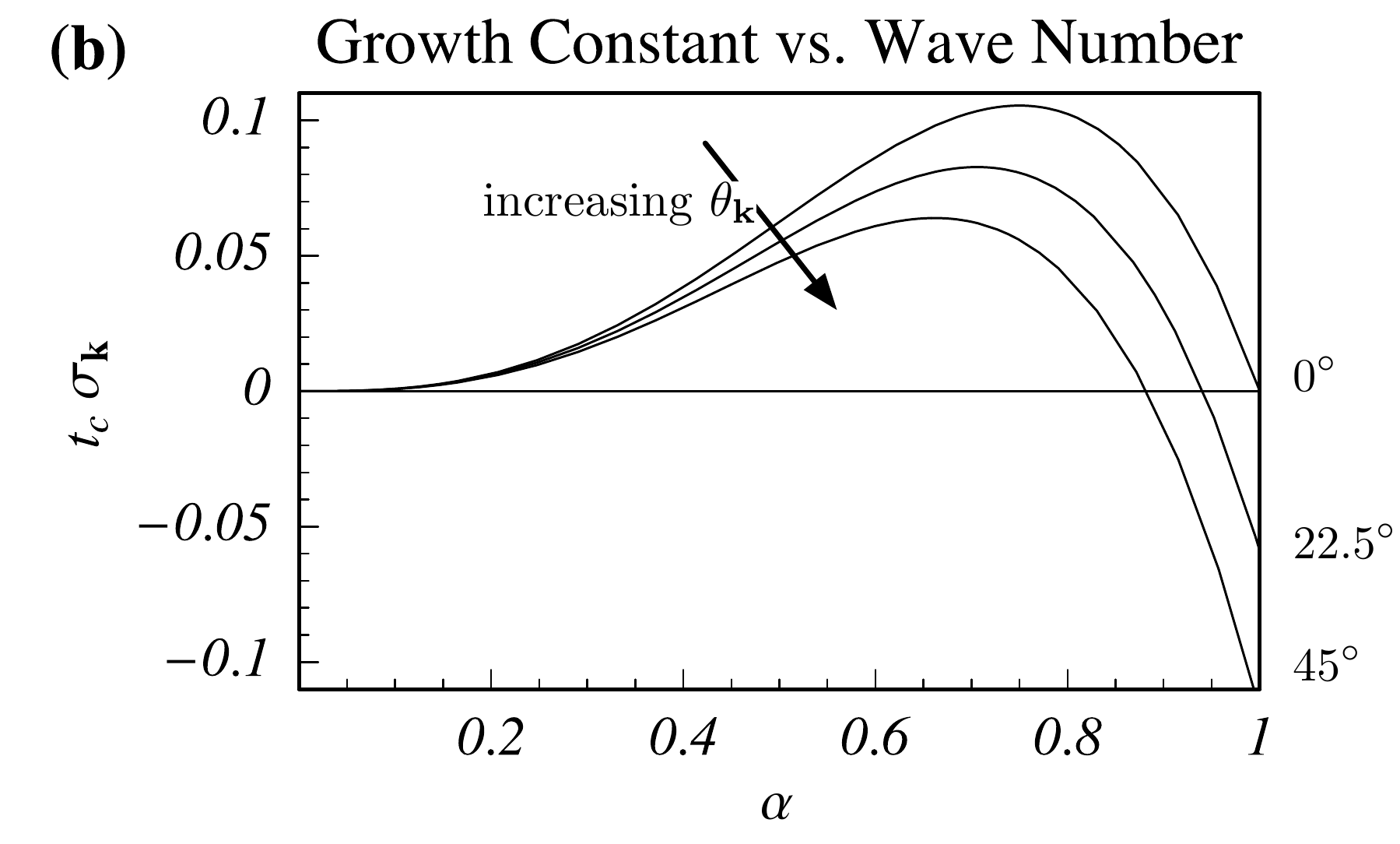}\par\end{centering}

\caption{\emph{\label{fig:Dimensionless-growth-constant}Dimensionless growth
constant vs. dimensionless wave number.} Curves are plotted for the
elastically anisotropic case, but the curves marked $0\degree$ are
the same as for the isotropic cases. In (a), $\beta=0$. In (b) $\beta=0.2$. }
\end{figure}

\begin{table}

\caption{\label{tab:CharAndDim}Characteristic wave-numbers, characteristic
times and associated dimensionless variables for the three cases addressed
in Eq.~\eqref{eq:fiso}.}

\begin{centering}\begin{tabular}{|c|c|c|c|c|}
\hline 
&
$k_{c}$&
$t_{c}$&
$\av$&
$\beta$\tabularnewline
\hline
\hline 
case a&
$\frac{2M(1+\nu)\epsilon_{m}^{2}}{\gamma}$&
$\frac{\gamma^{3}}{16\mathcal{D}\Omega M^{4}(1+\nu)^{4}\epsilon_{m}^{8}}$&
$\vec{k}/k_{c}$&
$\frac{\gamma W''(\bar{\h})}{4M^{2}(1+\nu)^{2}\epsilon_{m}^{4}}$\tabularnewline
\hline 
case b&
$\frac{2M(1+\nu)\epsilon_{m}^{2}}{F_{sw}^{(02)}}$&
$\frac{\left(F_{sw}^{(02)}\right)^{3}}{16\mathcal{D}\Omega M^{4}(1+\nu)^{4}\epsilon_{m}^{8}}$&
$\vec{k}/k_{c}$&
$\frac{F_{sw}^{(02)}F_{sw}^{(20)}}{4M^{2}(1+\nu)^{2}\epsilon_{m}^{4}}$\tabularnewline
\hline 
case c&
$a/b$&
$b^{3}/(\mathcal{D}\Omega a^{4})$&
$\vec{k}/k_{c}$&
$cb/a^{2}$\tabularnewline
\hline
\end{tabular}\par\end{centering}
\end{table}
The time dependent behavior of the film height fluctuations is facilitated
by using a characteristic wave number, characteristic time and related
dimensionless variables. For the {}``general'' case c of Eq.~\eqref{eq:fiso},
the characteristic wavenumber is $k_{c}=a/b$, and the characteristic
time is $t_{c}=1/(\mathcal{D}\Omega bk_{c}^{4})=b^{3}/(\mathcal{D}\Omega a^{4})$.
These characteristic dimensions can be used to define a dimensionless
wave vector, $\av=\vec{k}/k_{c}$ and a dimensionless wetting parameter
$\beta=c/(bk_{c}^{2})=cb/a^{2}$. One can also define a dimensionless
time, $\tau=t/t_{c}$. To obtain the corresponding characteristic
scales for cases a and b, one merely has to plug in the appropriate
substitutes for $a$, $b$ and $c$ and follow the pattern. For example,
for case a, make the substitution $a\rightarrow\Omega2M(2+\nu)\epsilon_{m}^{2}$,
etc. Table~\ref{tab:CharAndDim} summarizes these values for all
three cases. For all three cases, $f(\vec{k},\bar{\h})$ and the growth
constant $\sigma_{\vec{k}}$ reduce to the following forms:\begin{eqnarray}
f(\vec{k},\bar{\h}) & = & f(k_{c}\av,\bar{\h})=\Omega bk_{c}^{2}\left(-\alpha+\alpha^{2}+\beta\right)\label{eq:fdim}\\
\sigma_{\vec{k}} & = & \sigma_{k_{c}\av}=t_{c}^{-1}\alpha^{2}\left(\alpha-\alpha^{2}-\beta\right),\label{eq:sigdim}\end{eqnarray}
where $\alpha=\left\Vert \av\right\Vert =k/k_{c}$ is the dimensionless
wave number. These forms are plotted in Figs.~\ref{fig:Dimensionless-Diffusions-Potential}a
and~\ref{fig:Dimensionless-growth-constant}.%
 Fig.~\ref{fig:Dimensionless-Diffusions-Potential}a shows $f(\vec{k},\bar{\h})/\Omega bk_{c}^{2}$
vs. $\alpha$ for an isotropic or one dimensional surface. Figs.~\ref{fig:Dimensionless-growth-constant}
shows $t_{c}\sigma_{\vec{k}}$ vs. $\alpha$ for a 2D anisotropic
surface (Sec.~\ref{sub:2D-Anisotropic-case}). However, the curves
marked $0\degree$ are identical to the dispersion relation for a
1D or 2D isotropic surface (compare Eqs.~\eqref{eq:fiso} and~\eqref{eq:fanis}).

\subsubsection{\label{sub:Peaks.}Peaks}

\begin{figure}
\begin{centering}\includegraphics[width=3.375in,keepaspectratio]{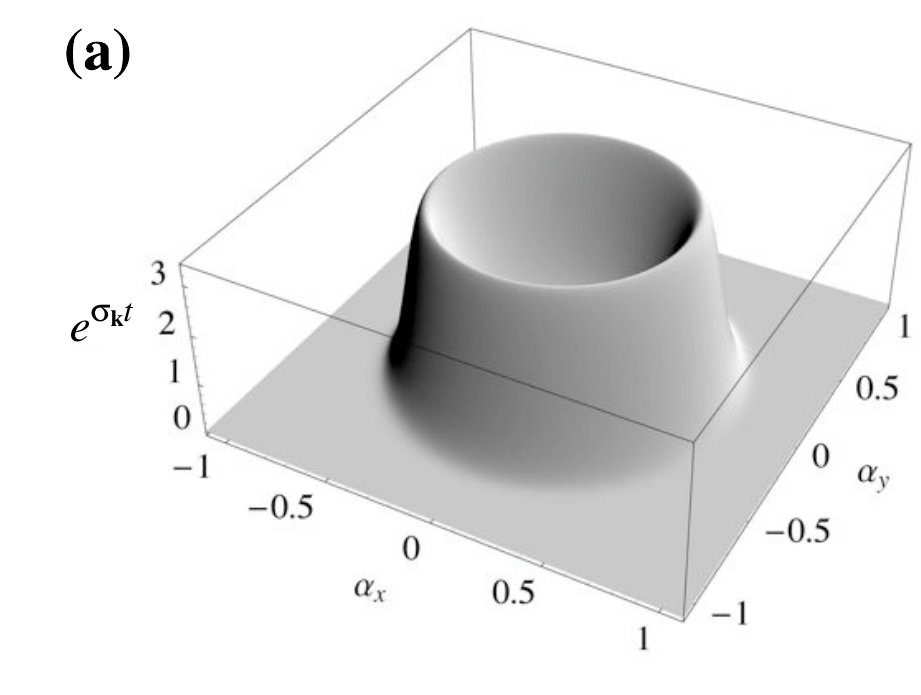}\par\end{centering}

\begin{centering}\includegraphics[width=3.375in,keepaspectratio]{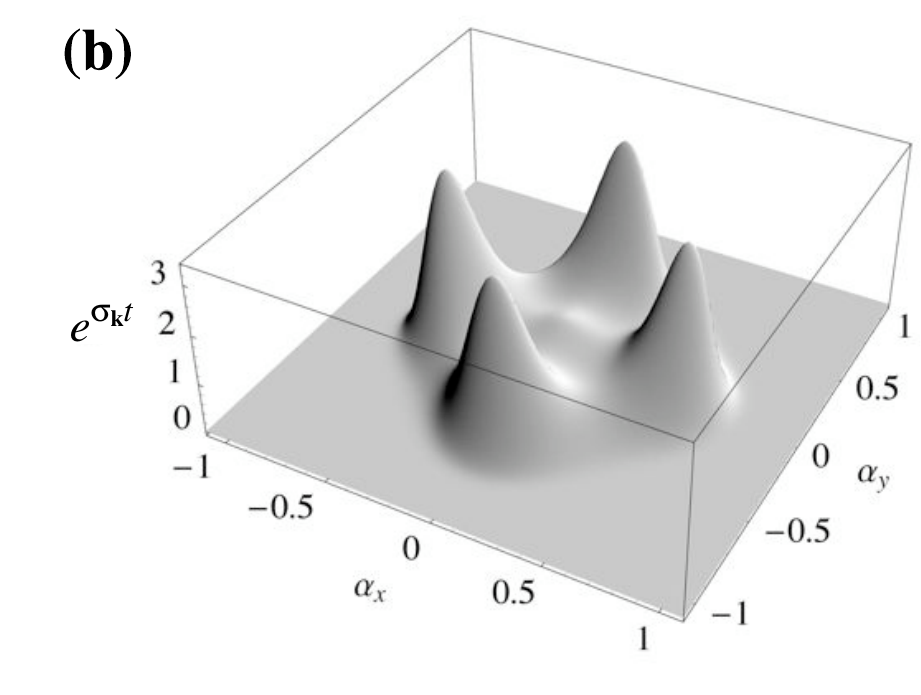}\par\end{centering}

\caption{\emph{\label{fig:Exponential-Envelope}Exponential Envelope $e^{\sigma_{\vec{k}}t}$
as function of $\av$} for \emph{$\beta=0.208$} and \emph{$t/t_{c}=100$.}
(a) 2D isotropic surface. (b) 2D anisotropic surface with \emph{$\epsilon=0.1236$}.}
\end{figure}

The peak growth rate and the corresponding wavenumber $k$ can be
found from Eq.~\eqref{eq:sigdim}. $\sigma_{\vec{k}}$ has a peak
at $k_{0}=k_{c}\alpha_{0}$ where\begin{equation}
\alpha_{0}=\frac{1}{8}\left(3+\sqrt{9-32\beta}\right).\label{eq:alpha0}\end{equation}
 Expanding $\sigma_{\vec{k}}$ about this peak to second order in
$k-k_{0}$, \[
\sigma_{\vec{k}}\approx\sigma_{0}-\frac{1}{2}\sigma_{2}(k-k_{0})^{2}\]
The two constants are \begin{equation}
\sigma_{0}=\frac{1}{4}t_{c}^{-1}\alpha_{0}^{2}\left(\alpha_{0}-2\beta\right),\label{eq:sigma0}\end{equation}
and \begin{equation}
\sigma_{2}=t_{c}^{-1}k_{c}^{-2}\left(3\alpha_{0}-4\beta\right).\label{eq:sigma2}\end{equation}
 Inserting this approximation for $\sigma_{\vec{k}}$ into Eq.~\eqref{eq:td},\begin{equation}
h_{\vec{k}}(t)=h_{\vec{k}}(0)e^{\sigma_{0}t}e^{-\frac{1}{2}\sigma_{2}t(k-k_{0})^{2}}.\label{eq:a0exp}\end{equation}
The individual initial surface fluctuation components grow with a
gaussian shaped envelope. An example of this envelope is plotted in
Fig.~\ref{fig:Exponential-Envelope}(a).%
 Notice that in two dimensions, the envelope forms a ring as the peak
is about the wave-number $k_{0}$ but not about any particular point
in the $\vec{k}$-plane.

\subsubsection{Stability and wetting potential\label{sub:Stability-and-wetting}}

Stranski-Krastanow growth is marked by a transition from stable two-dimensional
growth to unstable three-dimensional growth once a critical height
$\h_{c}$ is reached.~\cite{Bimberg99} Eqs.~\eqref{eq:alpha0},
\eqref{eq:sigma0} and~\eqref{eq:a0exp} are useful for analyzing
the transition from stable to unstable growth. In order for this transition
to occur, there must be some stabilizing term in the diffusion potential.
In the present model, this means that there must be some surface energy-like
term that varies strongly with film height. This condition equates
to stating that $W''(\bar{\h})$ or $F_{sw}^{20}$ or $c$ (Eq.~\eqref{eq:fiso})
must be rather large if $\bar{\h}<\h_{c}$. However, as $\bar{\h}$
increases, these terms are reduced. Finally, when $\bar{\h}>\h_{c}$,
this term is no longer capable of stabilizing the film against fluctuations
of all possible wavelengths. 

The critical value $\h_{c}$ can be found using the analysis from~\cite{Golovin:2003ms}.
By inspection of Eqs.~\eqref{eq:mulin}, \eqref{eq:fiso} and~\eqref{eq:gov_lin},
modes with $f>0$ increase the total free energy $\mathcal{F}$ as
they grow; thus, they are stable and decay with time. Modes with $f<0$
decrease the total free energy $\mathcal{F}$ as they grow; thus,
they are unstable and grow with time. This growth and decay rule is
easily verified by inspection of Eq.~\eqref{eq:gcgen}. Thus, stable
growth occurs when $f(\vec{k},\bar{\h})>0$ for \emph{all} values
of $\vec{k}$, and unstable growth occurs when $f(\vec{k},\bar{\h})<0$
for \emph{some} values of $\vec{k}$. Thus, the transition from stable
to unstable growth occurs when the minimum value of $f(\vec{k},\bar{\h})$
just becomes negative. Using the same dimensional analysis as in the
previous section and following the discussion of~\cite{Golovin:2003ms},
one finds that the minimum value, $f_{\text{min}}=\Omega bk_{c}^{2}\left(\beta-1/4\right)$,
occurs at $k_{\min}/k_{c}=\alpha_{\text{min}}=1/2$. $f_{\text{min}}$
first becomes negative, and the transition to unstable growth occurs
when the dimensionless wetting parameter (Table~\ref{tab:CharAndDim})
drops to a critical value, $\beta=1/4$ . $\beta>1/4$ stable 2D growth,
and $\beta<1/4$ unstable 3D growth. It is reasonable to suppose that
$W(\bar{\h})$, $W''(\bar{\h})$, and thus $\beta$ are positive monotonically
decreasing functions of $\bar{\h}$ so that the interface becomes
less important for large values of $\bar{\h}$. For example, in~\cite{Zhang:2001cr}
it is assumed that $W(\h)=B/\h$, where $B$ is constant. When $\beta\rightarrow0$,
corresponding to large $\bar{\h}$, the case discussed in~\cite{Spencer:1993vt}
is obtained. A similar analysis can be done for cases b and c once
one specifies how the terms $F_{sw}^{(20)}$ and $F_{sw}^{(02)}$
or $a$, $b$ and $c$ depend on $\bar{\h}$. 

Using a guessed form for a wetting potential, one can find the critical
film height $\h_{c}$ by setting $\beta=1/4$ . Applying this condition
to case a in Eq.~\eqref{eq:mu1}\[
W''(\h_{c})=\gamma k_{c}^{2}/4.\]
Using the wetting potential of ~\cite{Zhang:2001cr} as an example,
$W(\h)=B/\h$,\begin{equation}
\h_{c}=\sqrt[3]{8B/(\gamma k_{c}^{2})}=\sqrt[3]{8B\gamma/(2M(1+\nu)\epsilon_{m}^{2})^{2}}.\label{eq:HcPred}\end{equation}
Conversely, one can fit a wetting potential to an observed or reasonable
critical layer thickness from the same condition. Using the example
wetting potential from~\cite{Zhang:2001cr},\[
B=\frac{(2M(1+\nu)\epsilon_{m}^{2})^{2}\h_{c}^{3}}{8\gamma},\]
 as stated in~\cite{Zhang:2001cr}.%
\footnote{This result from~\cite{Zhang:2001cr} corresponds to the choice $F_{sw}(\h,\grad\h)=\sqrt{1+\left(\grad\h\right)^{2}}\gamma+W(\h)$.
However, the numerical model in~\cite{Zhang:2001cr} appears to use
$F_{sw}(\h,\grad\h)=\sqrt{1+\left(\grad\h\right)^{2}}\left[\gamma(\grad\h)+W(\h)\right]$.
This difference should lead to a slightly different critical film
height in their numerical model from the one that they predicted (Eq.~\eqref{eq:HcPred}). %
}

\subsection{2D Anisotropic case\label{sub:2D-Anisotropic-case}}

\newcommand{\sk}{\sigma_{\vec{k}}}

Crystal anisotropy leads to a dispersion relation $\sigma_{\vec{k}}$
that is both quantitatively and qualitatively different from the isotropic
case. Here the effect of elastic anisotropy is discussed in most detail.
Other sources of anisotropy are the surface and wetting energies.
For example, in~\cite{Zhang:2003tg} the surface energy density is
orientation dependent which introduces a possible anisotropy in the
dispersion relation. Possible sources of anisotropy are an anisotropic
elastic stiffness tensor, an orientation dependent surface energy
or wetting potential or anisotropic diffusion. As discussed below,
the form of anisotropy to linear order in the height fluctuation,
$h$, is somewhat restricted. Results are presented for 4-fold symmetric
surfaces, that is surfaces that have invariant dynamic evolution laws
when rotated by $90\degree$. Possible complications arising from
2-fold symmetric anisotropic terms (with  $180\degree$ rotational
symmetry) are also discussed. As for the isotropic case, first the
energetics are discussed, then the dynamics, and finally the expansion
about the peaks in the dispersion relation, $\sigma_{\vec{k}}$.

\subsubsection{Energetics}

The discussion of energetics will first treat the effects of elastic
anisotropy and then anisotropy resulting from surface or wetting like
terms. 

%
{}

\paragraph{Elastic anisotropy\label{par:Elastic-anisotropy}}

One would like to obtain a simple symbolic expression for the elastic
energy density at the free surface, $\omega_{\vec{k}}$, to first
order in $h_{\vec{k}}$ for the elastically anisotropic case. Similar
discussions can be found in~\cite{Obayashi:1998fk,Ozkan:1999gf}.
For the isotropic case, $\omega_{\vec{k}}=-2M(1+\nu)\epsilon_{m}^{2}h_{\vec{k}}$.
For the anisotropic case, \[
\omega_{\vec{k}}=-\mathcal{E}_{\theta_{\vec{k}}}kh_{\vec{k}}\]
where the prefactor $\mathcal{E}_{\theta_{\vec{k}}}$ is the decrease
in elastic energy at the surface per unit wave number ($k\rightarrow1$)
and unit amplitude ($h_{\vec{k}}\rightarrow1$) . It is not constant,
but instead depends on the $\theta_{\vec{k}}$, the angle that $\vec{k}$
makes with the $x-$direction. The case of a cube-symmetry elastic
stiffness tensor such as for Si is considered where one must specify
three elastic constants $c_{11}$, $c_{12}$ and $c_{44}$.~\cite{Vorbyev:1996fk}.
Growth on a (100) surface will produce an elastic energy prefactor
$\mathcal{E}_{\theta_{\vec{k}}}$ that is four-fold symmetric (symmetric
upon rotations by $90\degree$). A procedure similar to \cite{Obayashi:1998fk,Ozkan:1999gf}
based on a first order perturbation analysis is followed (Appendix~\ref{sec:Elastic-Anisotropy}).
A relatively simple interpolation formula~\cite{Friedman:fk} is
hypothesized and then verified numerically. 

\begin{figure}
\begin{centering}\includegraphics[width=3.375in,keepaspectratio]{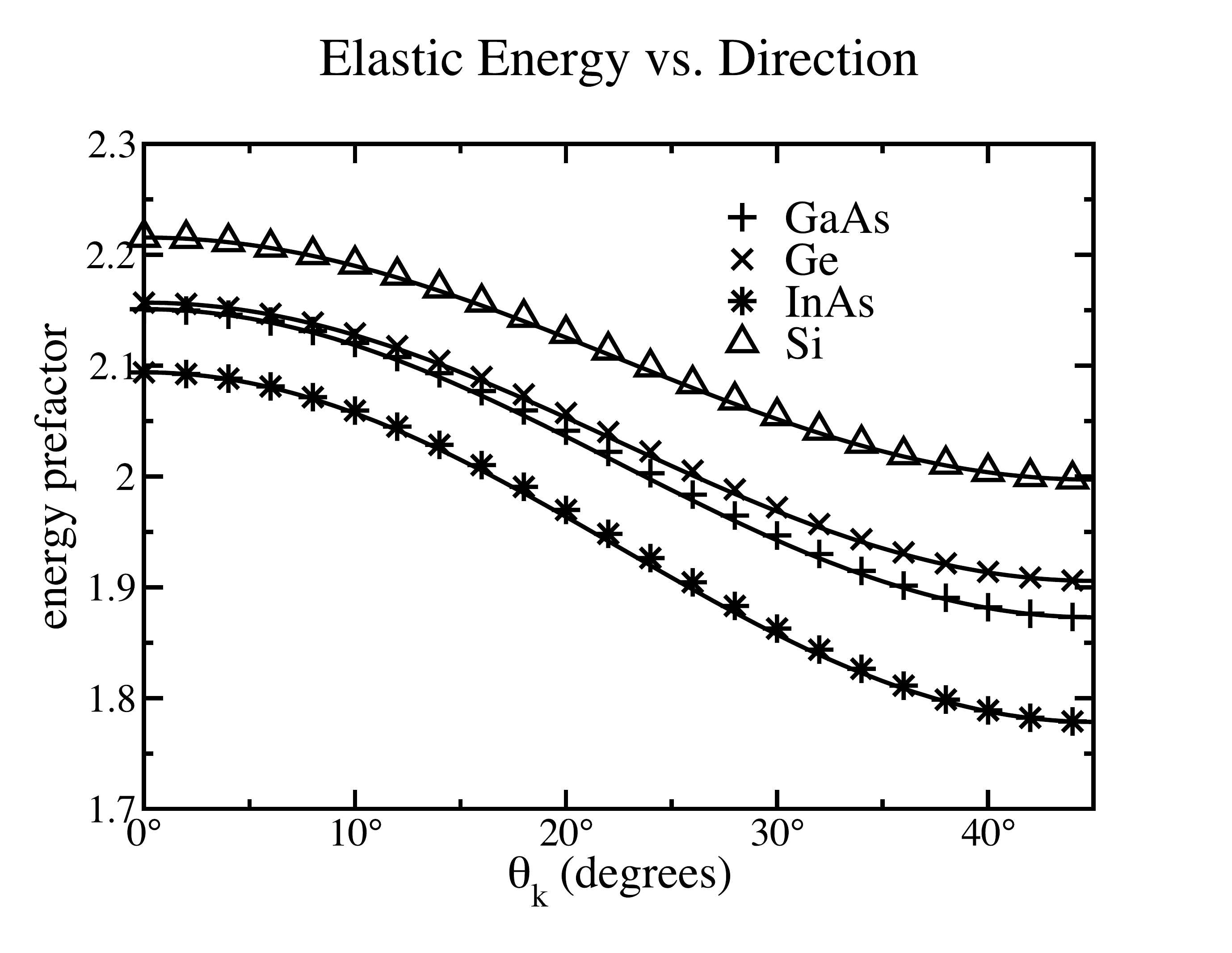}\par\end{centering}

\caption{\emph{\label{fig:omegavtheta}Plot of $\mathcal{E}_{\theta_{\vec{k}}}/(M\epsilon_{m}^{2})$
for various materials.} Symbols indicate values calculated using Appendix~\ref{sec:Elastic-Anisotropy}.
Solid lines are the interpolation (Eq.~\eqref{eq:einterp}) using
the values from Table~\ref{tab:Elastic-constants-and}.}
\end{figure}
\begin{table*}

\caption{Elastic constants~\cite{Vorbyev:1996fk} and calculated values (see
Appendix~\ref{sec:Elastic-Anisotropy}) for various materials of
interest at $T=300\mbox{K}$.\label{tab:Elastic-constants-and}}

\begin{centering}\begin{tabular}{|c|c|c|c|c|c|c|c|}
\hline 
&
$c_{11}$&
$c_{12}$&
$c_{44}$&
$M$&
$\frac{\mathcal{E}_{0\degree}}{M\epsilon_{0}^{2}}$&
$\frac{\mathcal{E}_{45\degree}}{M\epsilon_{0}^{2}}$&
$\epsilon_{A}$\tabularnewline
&
$10^{11}\frac{\text{erg}}{\text{cm}^{3}}$&
$10^{11}\frac{\text{erg}}{\text{cm}^{3}}$&
$10^{11}\frac{\text{erg}}{\text{cm}^{3}}$&
$10^{11}\frac{\text{erg}}{\text{cm}^{3}}$&
&
&
\tabularnewline
\hline
\hline 
Ge&
12.60&
4.40&
6.77&
13.93&
2.16&
1.906&
0.1176\tabularnewline
\hline 
Si&
16.60&
6.40&
7.96&
18.07&
2.22&
1.997&
0.1005\tabularnewline
\hline 
InAs&
8.34&
4.54&
3.95&
7.94&
2.70&
2.09&
0.226\tabularnewline
\hline 
GaAs&
11.90&
5.34&
5.96&
12.45&
2.15&
1.87&
0.1302\tabularnewline
\hline
\end{tabular}\par\end{centering}
\end{table*}
The interpolation procedure, suggested in~\cite{Friedman:fk} uses
the lowest possible order expansion in $\sin(\theta_{\vec{k}})$ and
$\cos(\theta_{\vec{k}})$ that has the appropriate four-fold symmetry
and then interpolates between $\theta_{\vec{k}}=0\degree$ and $\theta_{\vec{k}}=45\degree$.
Thus,\begin{equation}
\mathcal{E}_{\theta_{\vec{k}}}=\mathcal{E}_{0\degree}\left(1-\epsilon_{A}\sin^{2}\left(2\theta_{\vec{k}}\right)\right)\label{eq:einterp}\end{equation}
where $\epsilon_{A}=(\mathcal{E}_{0\degree}-\mathcal{E}_{45\degree})/\mathcal{E}_{0\degree}$
is an anisotropy factor. This lowest order form turns out to be a
very good fit to numerical calculations (Fig.~\ref{fig:omegavtheta}).
Table~\ref{tab:Elastic-constants-and} gives values of $\mathcal{E}_{0\degree}$
and $\epsilon_{A}$ for some systems of interest. In the elastically
isotropic case, $\mathcal{E}_{0\degree}=\mathcal{E}_{45\degree}=2M(1+\nu)$
so that $\epsilon_{A}=0$. 

There are two important differences from the elastically isotropic
case. The first is obvious, that $\mathcal{E}_{\theta_{\vec{k}}}$
depends on angular orientation, $\theta_{\vec{k}}$. The second is
that the peak value of $\omega_{\vec{k}}$ is not the same as that
for the elastically isotropic case because in general, $\mathcal{E}_{0\degree}\neq2M(1+\nu)$.
In~\cite{Friedman:fk}, where the purpose was simply to investigate
the mechanism by which elastic anisotropy effects order, this second
difference was neglected.

\paragraph{Surface and Wetting Energy Anisotropy\label{par:Surface-and-Wetting}}

The surface energy and wetting potential can be additional sources
of anisotropy if they depend on the surface orientation so that $\gamma\rightarrow\gamma(\grad\h)$
or $W(\h)\rightarrow W(\h,\grad\h)$ (for example,~\cite{Golovin:2004sa,Zhang:2003tg}).
Then, to first order in $h$ ,\[
\mu_{\text{surf.},\vec{k}}=\Omega\left(\gamma k^{2}+\vec{k}\cdot\tilde{\boldsymbol{\gamma}}''\cdot\vec{k}\right)h_{\vec{k}}\]
where $\tilde{\boldsymbol{\gamma}}''$ is the ($2\times2$) matrix
or Hessian matrix that results from taking the second derivatives
of $\gamma(\grad\h)$ with respect to the two components of $\grad\h$
(Appendix~\ref{sub:Linearizing-the-simple}). Similarly \[
\mu_{\text{wet},\vec{k}}=\Omega\left(W^{(20)}+\vec{k}\cdot\tilde{\mathbf{W}}^{(02)}\cdot\vec{k}\right)h_{\vec{k}}\]
where $W^{(20)}$ and $\tilde{\mathbf{W}}^{(02)}$ are the second
derivatives of $W(\h,\grad\h)$ with respect to $\h$ and $\grad\h$
(Appendix~\ref{sub:Linearizing-the-simple}). For both $\mu_{\text{surf.},\vec{k}}$
and $\mu_{\text{wet},\vec{k}}$, the first term is isotropic, and
the second term contains any possible anisotropy. 

The rank of the $\tilde{\boldsymbol{\gamma}}''$ and $\tilde{\mathbf{W}}^{(02)}$
matrices greatly restricts the possible forms of the additional anisotropy.
These $(2\times2)$ matrices must be either two-fold symmetric or
perfectly isotropic. Thus, if the surface energy and wetting potential
are four-fold symmetric as $\mathcal{E}_{\theta_{\vec{k}}}$ is, then
$\tilde{\boldsymbol{\gamma}}''\rightarrow\gamma''$, a scalar, and
$\tilde{\mathbf{W}}^{(02)}\rightarrow W^{(02)}$, a scalar, and neither
one contributes any additional anisotropy. They do, however, help
to stabilize or further destabilize the 2D surface as they add terms
proportional to $k^{2}$. The effect of these additional terms is
indistinguishable from the effect of varying the value of the surface
energy density, $\gamma$.~\cite{Golovin:2004sa,Gao:1999ve}

It should be noted that the (100) surface of a diamond or zinc-blend
structures allows for anisotropy that is only 2-fold symmetric (rotations
by $180\degree$). Thus, they could {}``break'' the four-fold symmetry
that occurs when one considers the elastic anisotropy alone. However,
this {}``broken'' symmetry is somewhat dubious because even the
diamond and zinc-blend structures have a screw symmetry (rotations
by $90\degree$ and translation in the {[}100] direction by half a
lattice vector). Thus, if for example, $W(\h,\grad\h)$ is anisotropic
with two-fold symmetry to linear order, there must be a fast oscillation
with changes in the film height $\h$. In Appendix~\ref{sec:Diffusion-Anisotropy},
a similar term related to anisotropic diffusion is discussed. There
does not appear to be any evidence for this two-fold symmetry in the
case of (100) surfaces of IV/IV systems such as Ge/Si, but in III-V/III-V
systems the four-fold symmetry of the (100) surface may indeed be
{}``broken'' in this way corresponding to either a surface energy
anisotropy or a diffusional anisotropy.~\cite{Liang:2006fk,Meixner:2003lz}.
Further analysis of such terms in any more detail would greatly complicate
the present discussion, so it is left for future work. Most of the
modeling literature avoids this complication by not including the
symmetry-breaking of the zinc-blend surface, for example~\cite{Obayashi:1998fk,Ozkan:1999gf,Zhang:2003tg}.

One can perform a similar analysis of the combined surface and wetting
potential, $F_{sw}(\h,\grad\h)$ (case b). To linear order the resulting
anisotropic diffusion potential is (Appendix~\ref{sub:Linearizing-the-general})
\[
\mu_{sw,\vec{k}}=\Omega\left(F_{sw}^{(20)}+\vec{k}\cdot\tilde{\mathbf{F}}_{sw}^{(02)}\cdot\vec{k}\right)h_{\vec{k}}.\]
Again, $\tilde{\mathbf{F}}_{sw}^{(02)}$ is a rank 2 tensor, and all
of the same symmetry considerations apply here as well.

Because the two-fold symmetry anisotropic terms are excluded from
the current discussion, and isotropic terms simply {}``renormalize''
the effective of surface energy, there will be no further consideration
of anisotropy resulting from the surface energy or wetting potential
in this discussion. Further calculations will proceed assuming that
the surface energy density, $\gamma$, nor the wetting potential,
$W(\h)$, depend on $\grad\h$ or similarly that $F_{sw}(\h,\grad\h)$
has a purely isotropic dependence on $\grad\h$. This assumption can
be made without affecting any of the qualitative results.

\paragraph{total diffusion potential\label{par:total-diffusion-potential}}

Having dispensed with the discussion of the various sources of anisotropy,
the total diffusion potential is stated for the case of 4-fold symmetric
elastic anisotropy and a completely isotropic surface energy and wetting
potential. $\mu_{\vec{k}}=f(\vec{k},\bar{\h})$ with\begin{equation}
f(\vec{k},\bar{\h})=\begin{cases}
\Omega\left[-\mathcal{E}_{0\degree}\left(1-\epsilon_{A}\sin^{2}(2\theta_{\vec{k}})\right)k+\gamma k^{2}+W''(\bar{\h})\right] & ;\text{ case a (Eq.\,\eqref{eq:mu1})}\\
\Omega\left[-\mathcal{E}_{0\degree}\left(1-\epsilon_{A}\sin^{2}(2\theta_{\vec{k}})\right)k+F_{sw}^{(02)}k^{2}+F_{sw}^{(20)}\right] & ;\text{ case b (Eq.\,\eqref{eq:mu2}})\\
-a\left(1-\epsilon_{A}\sin^{2}(2\theta_{\vec{k}})\right)+bk^{2}+c & ;\text{ case c (general)}\end{cases}.\label{eq:fanis}\end{equation}

\subsubsection{Dynamics}

\begin{table}

\caption{\label{tab:CharAndDim2}Characteristic wave-numbers, characteristic
times and associated dimensionless variables for the three cases addressed
in Eq.~\eqref{eq:fiso}}

\begin{centering}\begin{tabular}{|c|c|c|c|c|}
\hline 
&
$k_{c}$&
$t_{c}$&
$\av$&
$\beta$\tabularnewline
\hline
\hline 
case a&
$\mathcal{E}_{0\degree}/\gamma$&
$\gamma^{3}/(\mathcal{D}\Omega\mathcal{E}_{0\degree}^{4})$&
$\vec{k}/k_{c}$&
$\gamma W''(\bar{\h})/\mathcal{E}_{0\degree}^{2}$\tabularnewline
\hline 
case b&
$\ez/F_{sw}^{(02)}$&
$\left(F_{sw}^{(02)}\right)^{3}/(\mathcal{D}\Omega\ez^{4})$&
$\vec{k}/k_{c}$&
$F_{sw}^{(02)}F_{sw}^{(20)}/\ez^{2}$\tabularnewline
\hline 
case c&
$a/b$&
$b^{3}/(\mathcal{D}\Omega a^{4})$&
$\vec{k}/k_{c}$&
$cb/a^{2}$\tabularnewline
\hline
\end{tabular}\par\end{centering}
\end{table}
The dynamics is governed by surface diffusion, just as for the fully
isotropic case. It is assumed that the diffusivity is isotropic as
was done for the surface energy and the wetting energies; thus, all
anisotropy in the film evolution dynamics comes from elastic effects
alone. The possibility and effects of an anisotropic diffusion potential
is discussed in Appendix~\ref{sec:Diffusion-Anisotropy} (also see~\cite{Meixner:2003lz}).
The time dependence of the surface perturbations simply follows Eqs.~\eqref{eq:td}
and~\eqref{eq:gcgen}, but with Eq.~\eqref{eq:fanis} used for $f(\vec{k},\bar{\h})$.
%
{} As for the isotropic case, appropriate characteristic wave numbers
($k_{c}$) and time scales ($t_{c}$) can be found for each of the
three cases along with the associated dimensionless wave vector $\av$
and dimensionless wetting parameter $\beta$. These are listed in
Table~\ref{tab:CharAndDim2}. The dispersion relation, $\sigma_{\vec{k}}$
can be expressed in terms of these dimensionless variables ($\av$
and $\beta$), giving \begin{equation}
\sigma_{\vec{k}}=\sigma_{k_{c}\av}=t_{c}^{-1}\alpha^{2}\left[\alpha\left(1-\epsilon_{A}\sin^{2}(2\theta_{\vec{k}})\right)-\alpha^{2}-\beta\right].\label{eq:askd}\end{equation}
 The stability behavior is essentially the same as for the isotropic
case with a transition occurring at $\beta=1/4$ corresponding to
$\bar{\h}=\h_{c}$.

\subsubsection{Expansion about peaks\label{sub:Expansion-about-peaks}}

$\sigma_{\vec{k}}$ has $4$ peaks at $(\vec{k},\theta_{\vec{k}})=(k_{0},\pi[n-1]/2)$
with $k_{0}=k_{c}\alpha_{0}$ (Eq.~\eqref{eq:alpha0}) and $n=1\dots4$.
In vector form, there are four peaks at\[
\vec{k}_{n}=k_{0}\left(\cos(\pi(n-1)/2)\mathbf{i}+\sin(\pi(n-1)/2)\mathbf{j}\right).\]
Similar to the isotropic case, $\sigma_{\vec{k}}$ can be expanded
about individual peaks so that in the vicinity of peak $n$, $\sigma_{\vec{k}}\approx\sigma_{n}$
with\[
\sigma_{n}=\sigma_{0}-\frac{1}{2}\sigma_{\parallel}(k-k_{0})^{2}-\frac{1}{2}\sigma_{\perp}k_{0}^{2}(\theta_{\vec{k}}-n\pi/2)^{2},\]
where $\sigma_{0}$ is given by Eq.~\eqref{eq:sigma0}, $\sigma_{\parallel}=\sigma_{2}$
given by Eq.~\eqref{eq:sigma2}, and \[
\sigma_{\perp}=8\epsilon_{A}\alpha_{0}t_{c}^{-1}k_{c}^{-2}.\]
In terms of the vector components parallel and perpendicular to $\vec{k}_{n}$,
$k_{\parallel}$ and $k_{\perp}$ respectively,\[
\sigma_{n}=\sigma_{0}-\frac{1}{2}\sigma_{\parallel}(k_{\parallel}-k_{0})^{2}-\frac{1}{2}\sigma_{\perp}k_{\perp}^{2},\]
$k_{\parallel}=\cos[\pi(n-1)/2]k_{x}+\sin[\pi(n-1)/2]k_{y}$, and
$k_{\perp}=-\sin[\pi(n-1)/2]k_{x}+\cos[\pi(n-1)/2]k_{y}$ . The time
evolution of $h_{\vec{k}}$in the vicinity of one of the $\vec{k}_{n}$
is \[
h_{\vec{k}}(t)\approx h_{\vec{k}}(0)e^{t\left(\sigma_{0}-\frac{1}{2}\sigma_{2}(k_{\parallel}-k_{0})^{2}-\frac{1}{2}\sigma_{\perp}k_{\perp}^{2}\right)}.\]

\section{Correlation Functions}

\label{sec:Correlation-Functions}%
\begin{sidewaysfigure}

\noindent \begin{centering}\begin{tabular}{|c|c|c|}
\hline 
\includegraphics[width=2.25in,keepaspectratio]{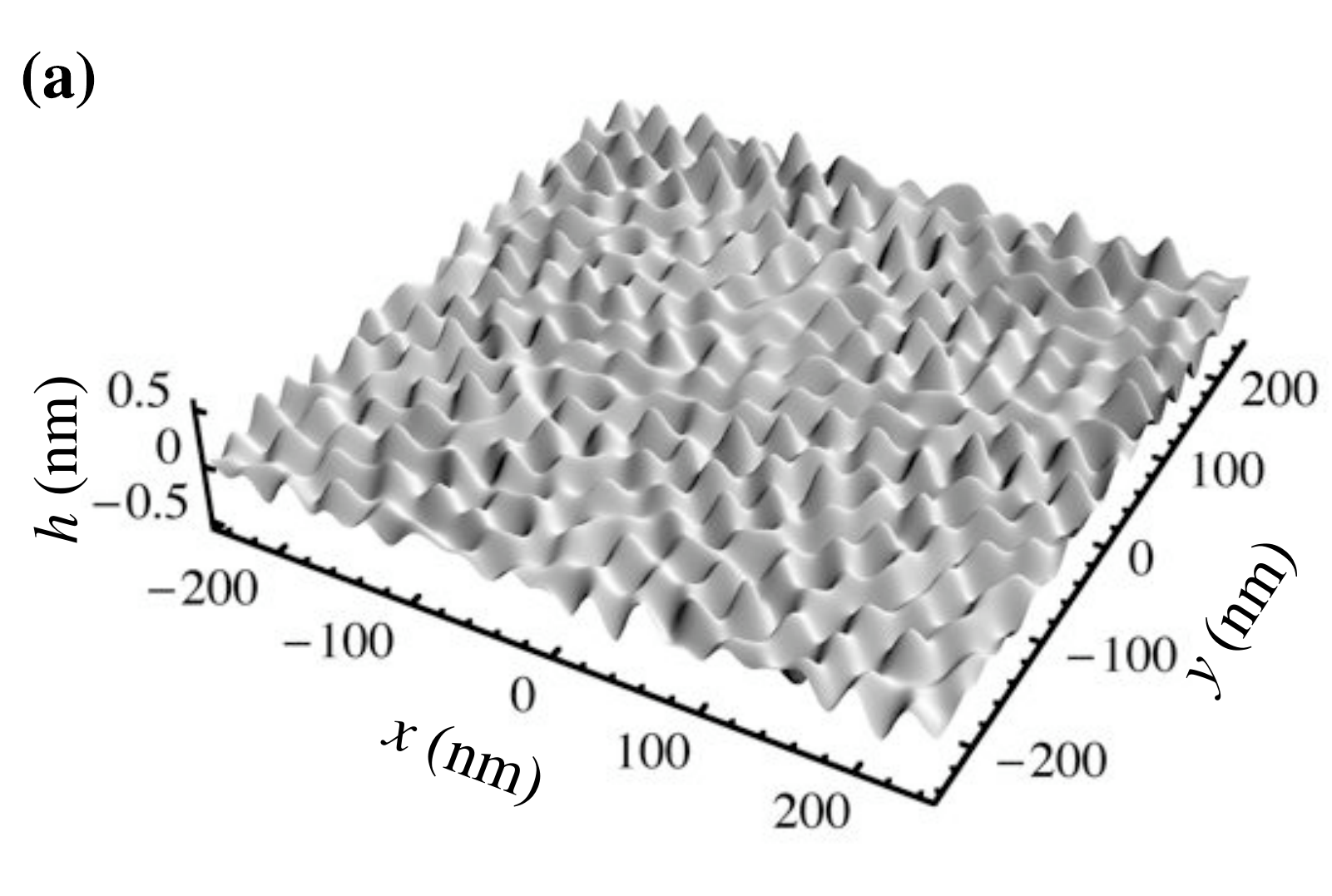}&
\includegraphics[width=2.24in,keepaspectratio]{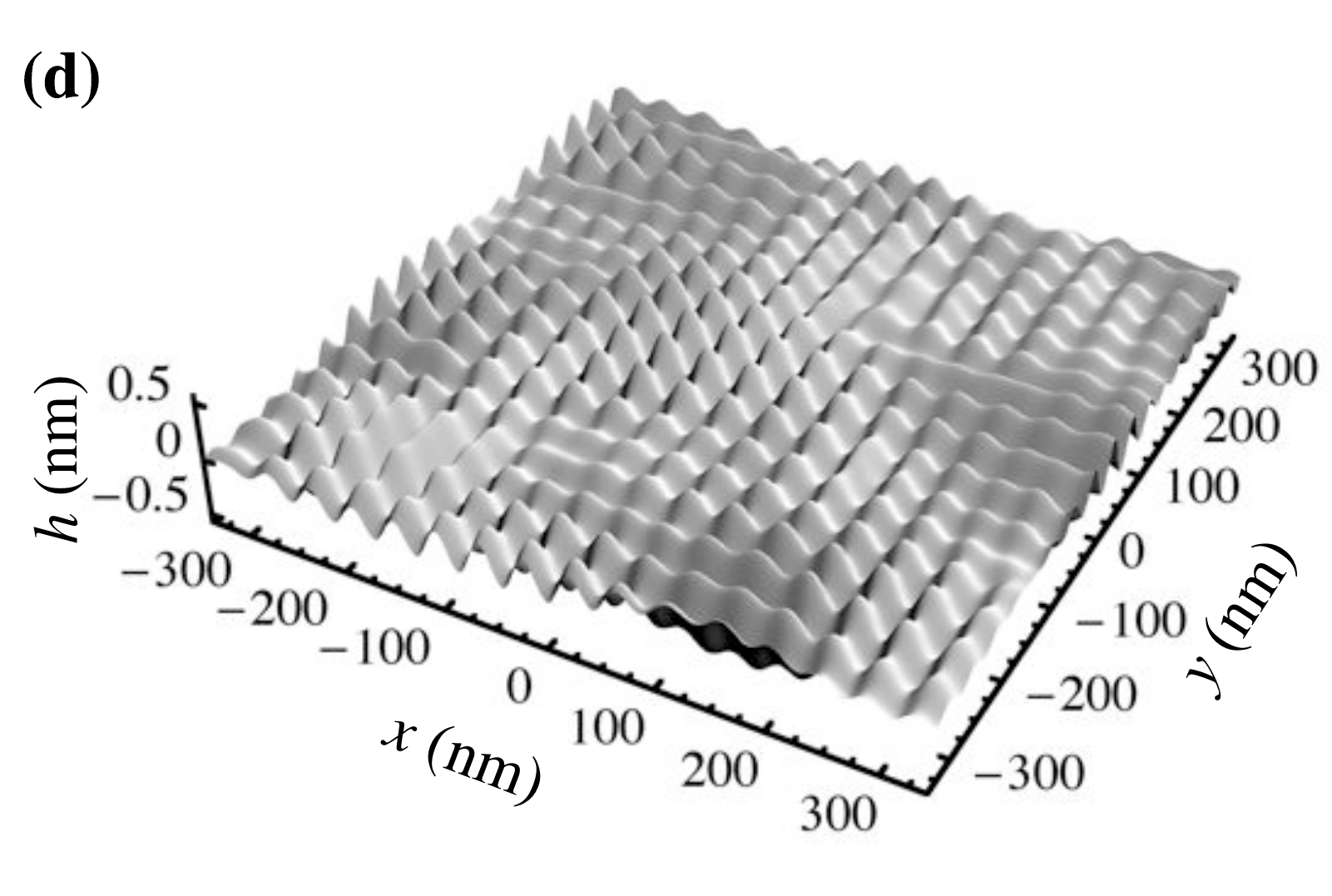}&
\includegraphics[width=2.25in,keepaspectratio]{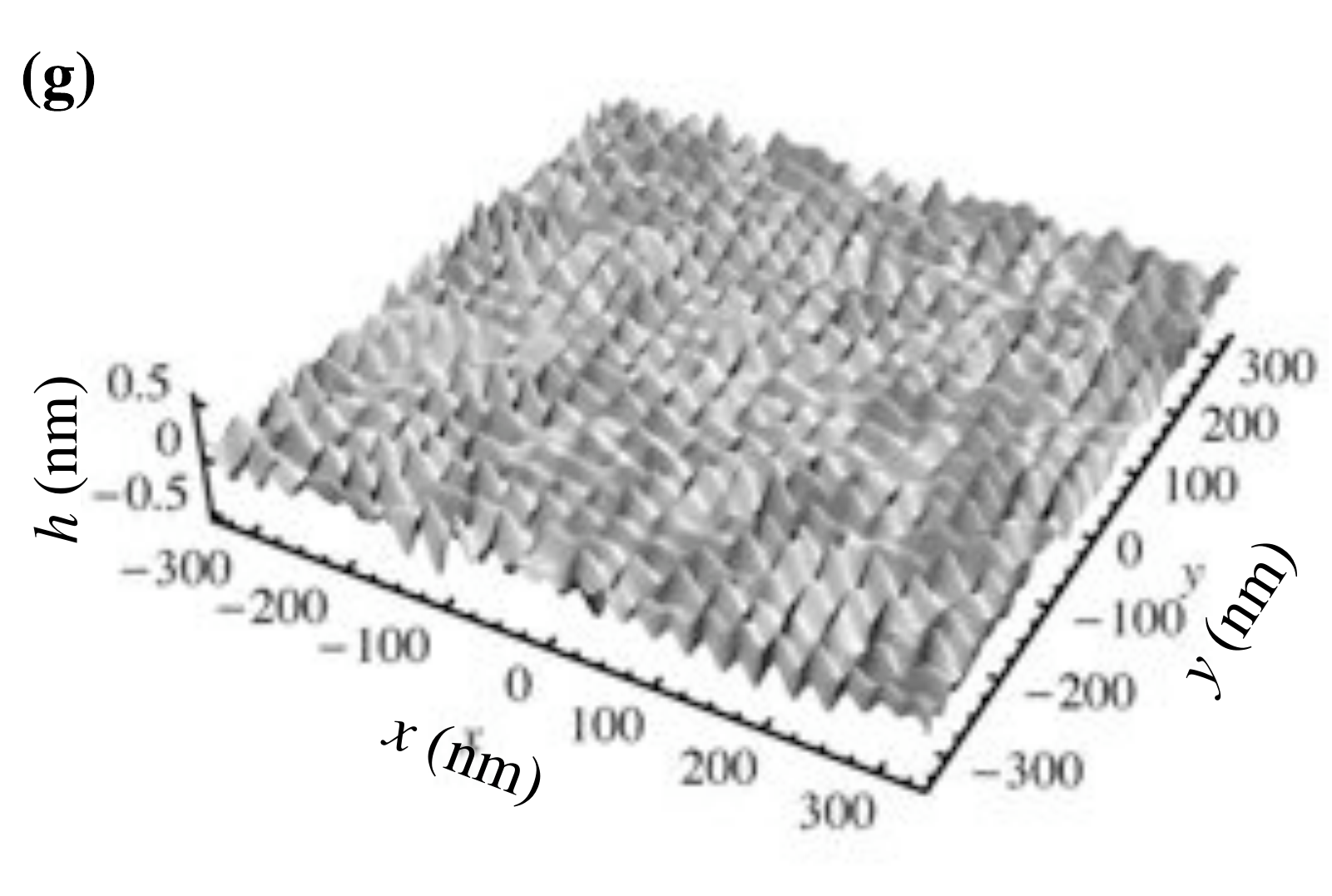}\tabularnewline
\hline 
\includegraphics[width=2.25in,keepaspectratio]{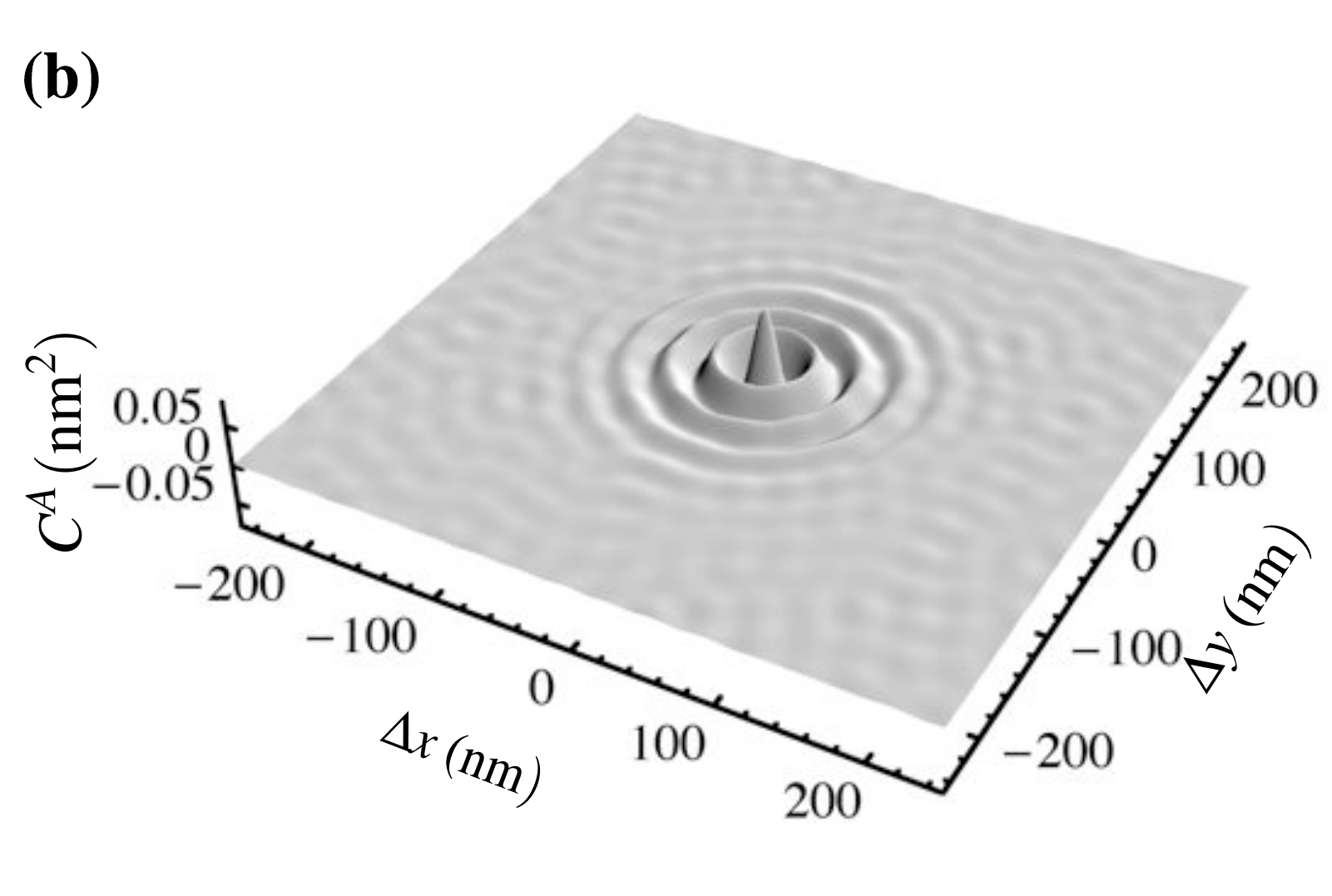}&
\includegraphics[width=2.25in,keepaspectratio]{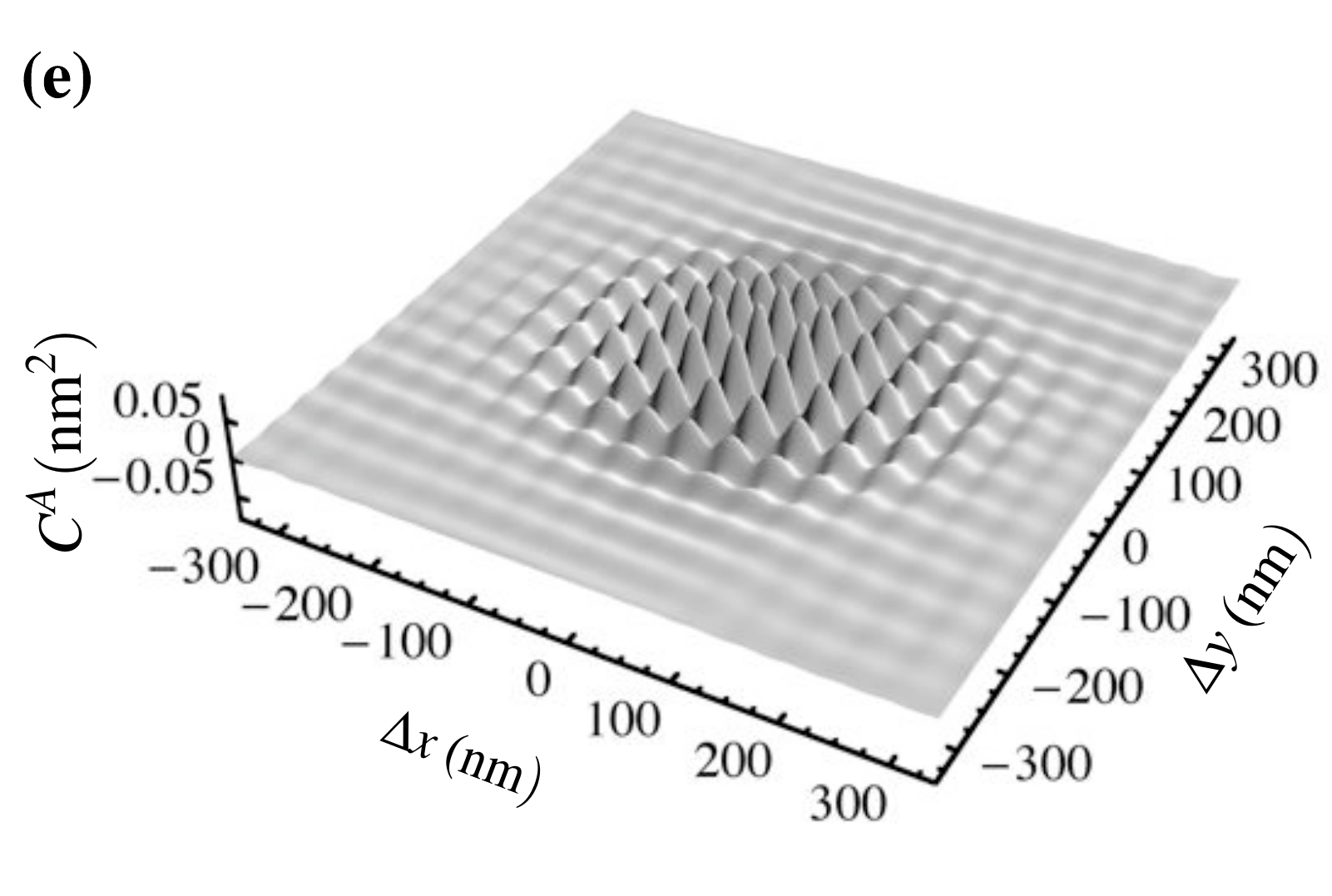}&
\includegraphics[width=2.25in,keepaspectratio]{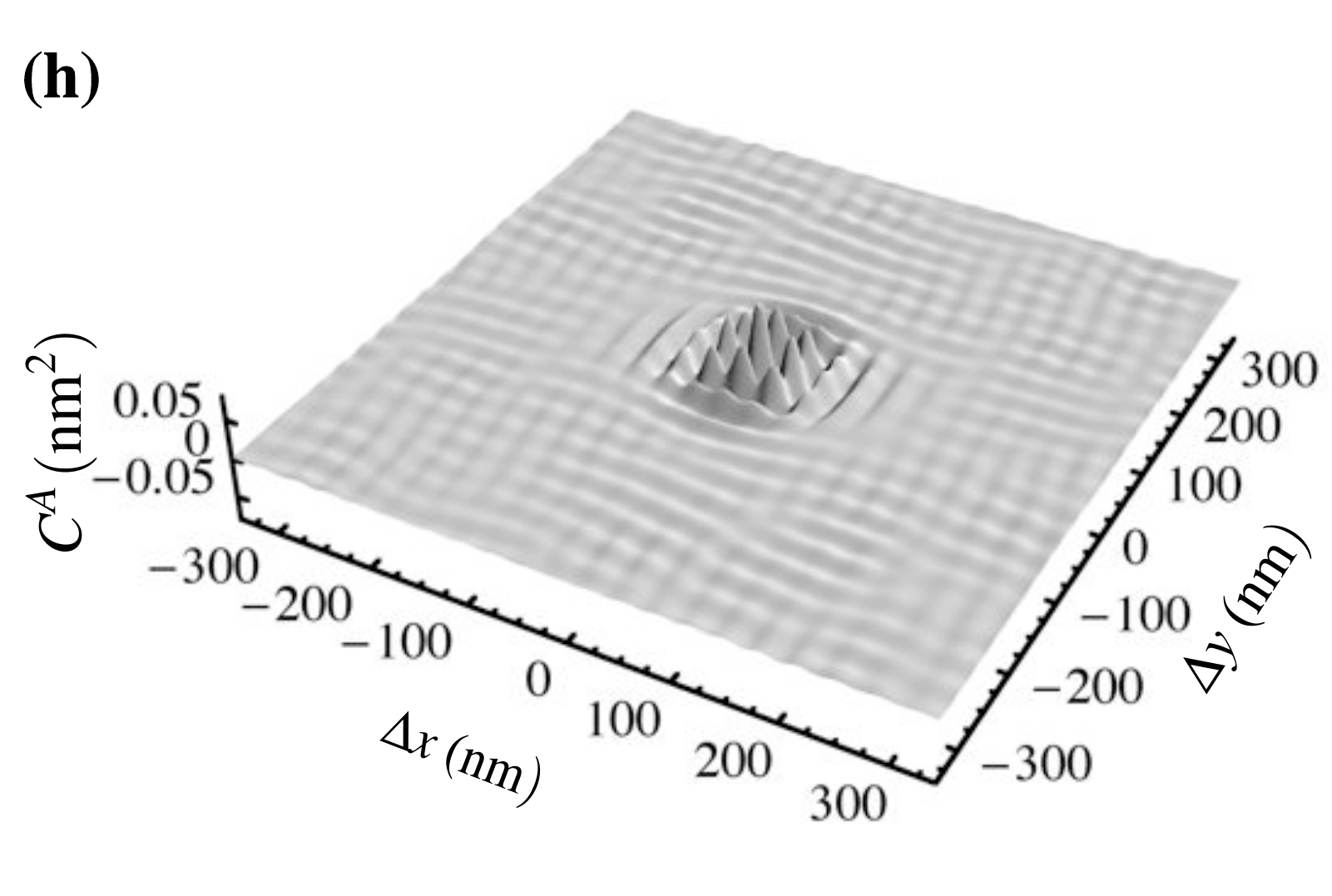}\tabularnewline
\hline 
\includegraphics[width=2.25in,keepaspectratio]{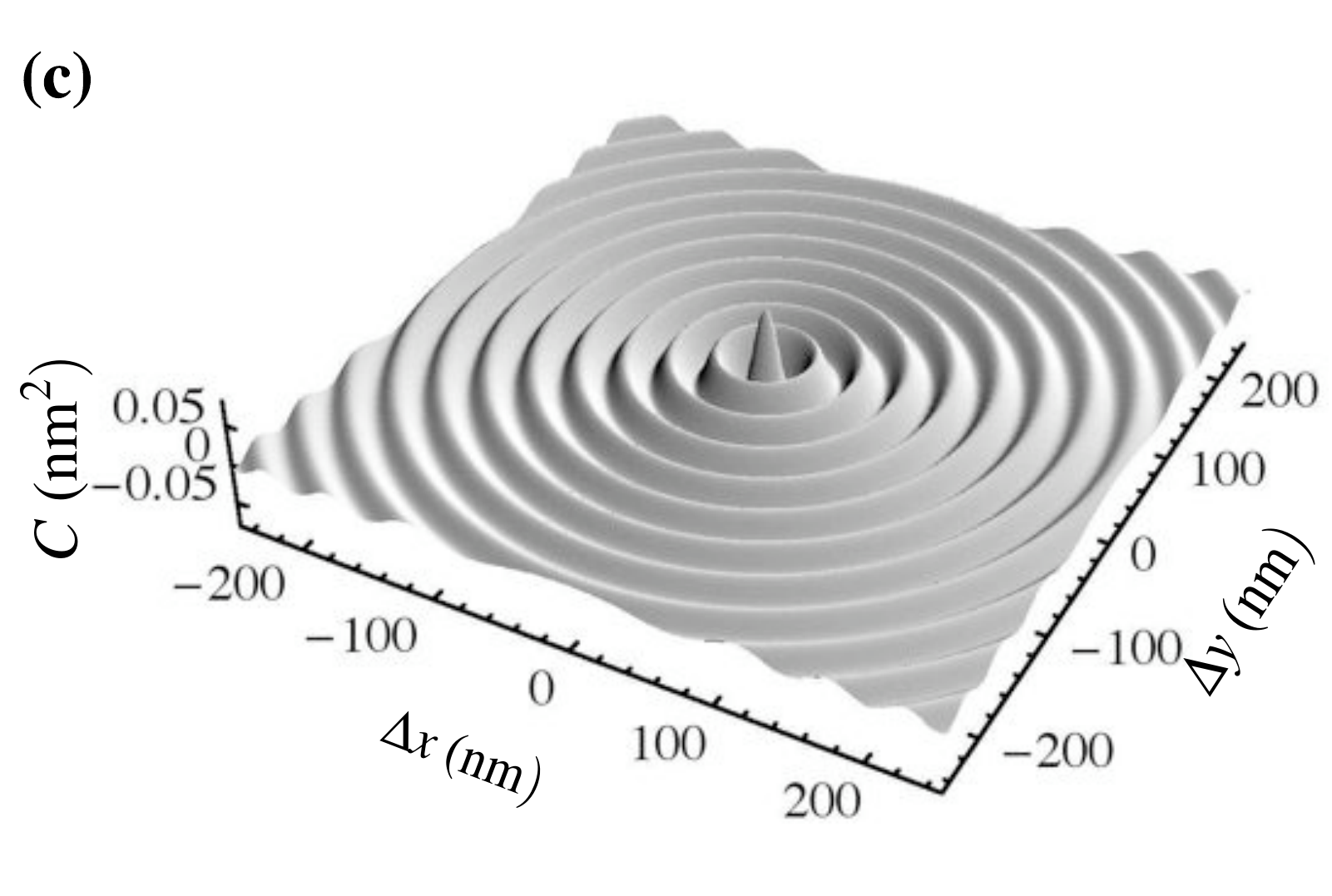}&
\includegraphics[width=2.25in,keepaspectratio]{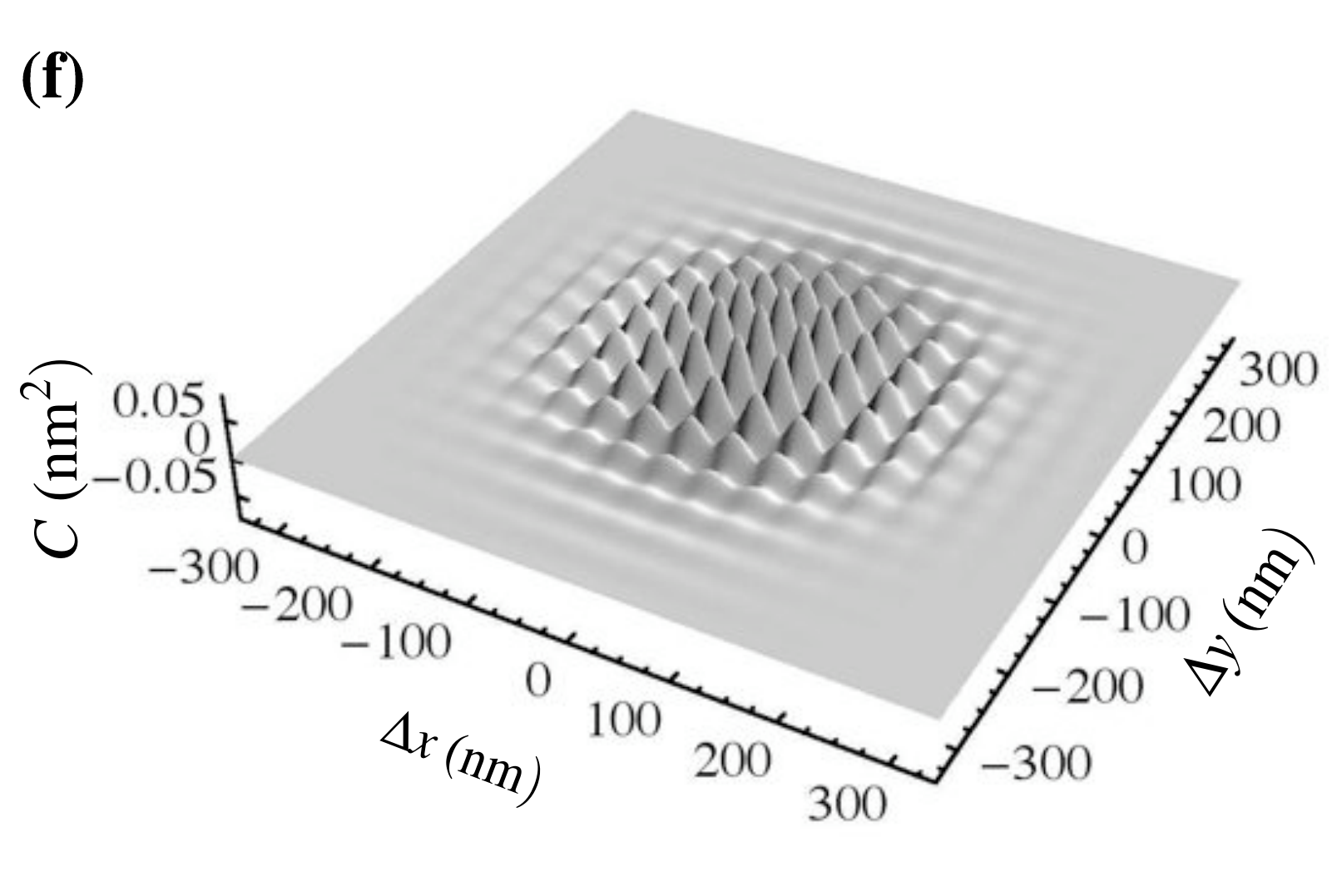}&
\includegraphics[width=2.25in,keepaspectratio]{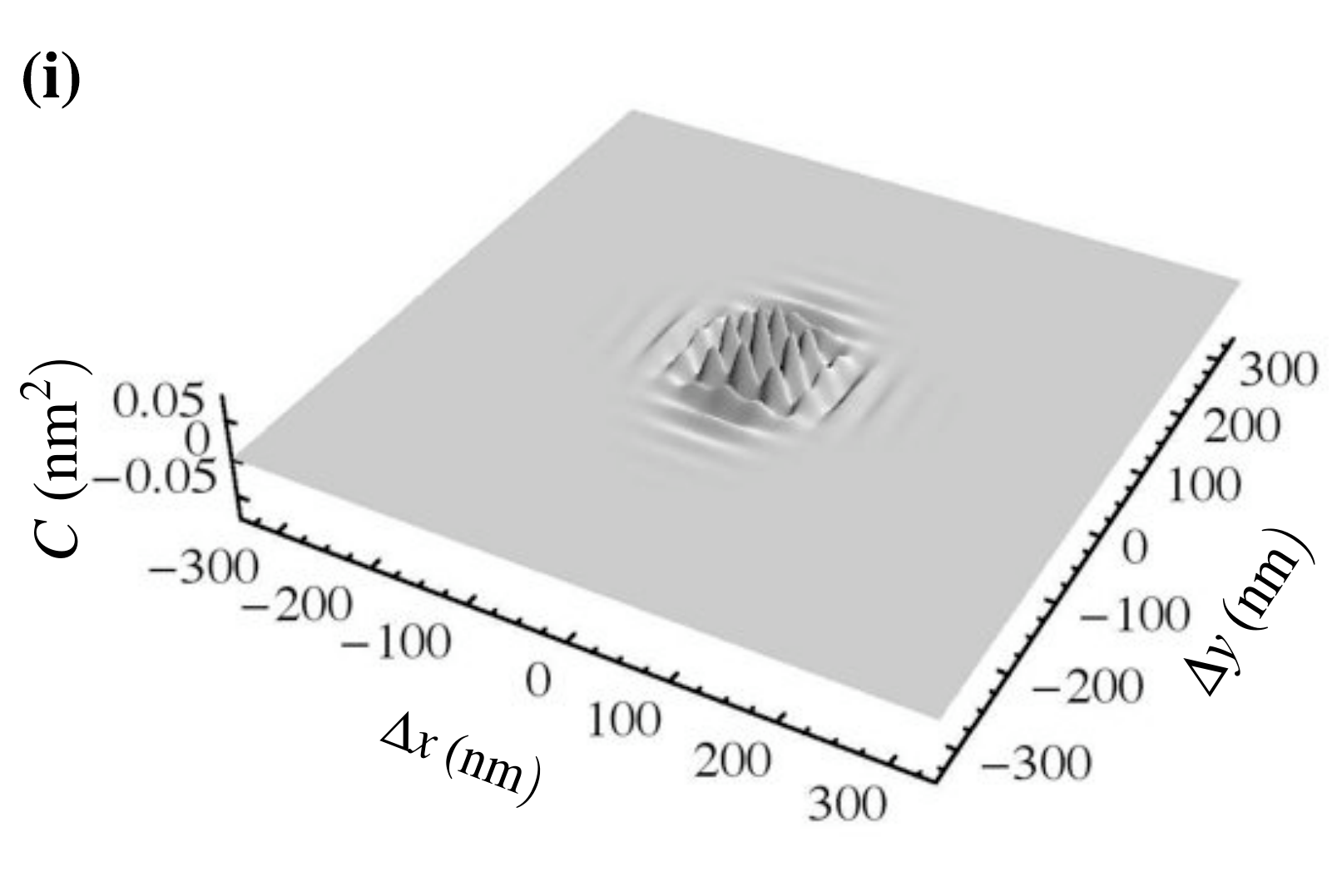}\tabularnewline
\hline
\end{tabular}\par\end{centering}

\caption{\emph{\label{fig:Film-heights-and}Film heights and real-space correlation
functions for Ge/Si} as discussed in Sec.~\ref{sec:Application-to-Materials}.
All units are in $\text{nm}$ or $\text{nm}^{2}$. (a-c) show respectively
example $h(\vec{x})$ , corresponding $C^{A}(\dx)$ and calculated
$C(\dx)$ for the 2D isotropic surface with $\beta=0.208$ and $t/t_{c}=306$.
Fig.~c uses Eq.~\eqref{eq:Cx2Di} because the corresponding formula
for $C(\dx)$ with finite $L_{\text{cor}}$ is not available. (d-f)
show the same for the 2D anisotropic surface with $\beta=0.208$ and
$t/t_{c}=430$. Eq.~\eqref{eq:Cxa1} is used for plot (f). (g-i)
show the same for $\beta=0$ and $t/t_{c}=40.3$. Eq.~\eqref{eq:Cxa1}
is used for plot (i). }
\end{sidewaysfigure}
Correlation functions and associated constants such as correlation
lengths can be very useful for characterizing order. In particular,
the autocorrelation function (Eq.~\eqref{eq:acdef1}) and its Fourier
transform (Eq.~\eqref{eq:ckaft}) also known as the spectrum function
can give a very good characterization of dot order (Figs.~\ref{fig:Ck}a
and~c and \ref{fig:Film-heights-and}b, e and~h). The autocorrelation
function is denoted $C^{A}(\dx)$ where $\dx$ is the difference vector
between two points in the $\vec{x}-$plane. The spectrum function
is a function of $\vec{k}$, and it is denoted $C_{\vec{k}}^{A}$.
The goal here is to be able to predict these two functions and to
describe them quantitatively in a manner that can be used to characterize
SAQD order with just a few numbers. The autocorrelation function is
the result of a spatial average over one experiment or one simulation
(numerical experiment). It is regular and repeatable because it is
closely tied to the correlation function and spectrum function that
results from an ensemble average (Eqs.~X and~X). These are denoted
as $C(\dx)$ and the spectrum $C_{\vec{k}}$ respectively. Note that
the ensemble averaged functions do not have a superscript {}``$A$.''
These ensemble average correlation functions are useful in the analysis
of stochastic ordinary and partial differential equations.~\cite{Zwanzig:2001zf,Gardiner:2004fk}.
From a strictly technical viewpoint, the spatial average and the ensemble
average are not exactly the same; however, they are closely enough
connected that it is reasonable to use one as a substitute for the
other (Sec.~\ref{sub:Correlation-Functions-and} and Appendix~\ref{sec:Correlation-Functions}). 

In the following, the analysis of SAQD order via autocorrelation and
correlation functions is discussed (Sec.~\ref{sub:Correlation-Functions-and}).
Then, the stochastic initial conditions are discussed (Sec.~\ref{sub:Stochastic-Initial-Conditions}).
Then, the prediction of the Fourier transforms of the correlation
functions is discussed (Sec.~\ref{sub:Reciprocal-Space-Corelation}).
The real-space correlation functions are presented (Sec.~\ref{sub:Real-Space-Correlation}).
Finally, there are some notes regarding generalizing the analysis
method to any dispersion relation that has peaks (Sec.~\ref{sub:Generalizability}),
for example, peaks related to broken four-fold symmetry or growth
on a miscut substrate.

\subsection{\label{sub:Correlation-Functions-and}Correlation Functions and SAQD
order}

Auto-correlation functions are well-suited for investigating SAQD
order. The autocorrelation function is defined as \begin{equation}
C^{A}(\dx)=\frac{1}{A}\int d^{2}\vec{x}'\, h(\dx+\vec{x'})h(\vec{x}')^{*}.\label{eq:acdef1}\end{equation}
Its Fourier transform sometimes called the spectrum~\cite{Gardiner:2004fk},
spectrum function or power spectrum is \begin{equation}
C_{\vec{k}}^{A}=\frac{1}{(2\pi)^{d}}\int d^{2}\dx\, e^{-i\vec{k}\cdot\dx}C(\dx)=\frac{(2\pi)^{d}}{A}\left|h_{\vec{k}}\right|^{2},\label{eq:ckaft}\end{equation}
where $A$ is the projected area of the film in the $x-y-$plane.
A periodic array of SAQDs leads to a periodic auto-correlation function.
A nearly periodic array leads to a range-limited periodic auto-correlation
function. The ensemble-mean of these autocorrelation functions can
be calculated, and it is a good predictor of a SAQD order.

\subsubsection{\label{sub:Periodic-array}Periodic array}

Consider a perfectly periodic height fluctuation corresponding to
a perfect lattice of SAQDs, \begin{equation}
h(\vec{x})=\frac{h_{0}}{N}\sum_{i=n}^{N}\exp\left[i\vec{k}_{n}\cdot\left(\vec{x}-\vec{x}_{O}\right)\right]\label{eq:idealh}\end{equation}
plus higher order harmonic, where the dots have a height proportional
to $h_{0}$, $N$ is the degree of symmetry, probably, 4-fold or 6-fold,
$\vec{x}_{O}$ is a random origin offset.\[
\vec{k}_{n}=k_{0}\left(\cos\left(\frac{2\pi(n-1)}{N}\right)\vec{i}+\sin\left(\frac{2\pi(n-1)}{N}\right)\vec{j}\right),\]
In a linear analysis, the higher order harmonics do not come into
play, so they are neglected here. In reciprocal space,\[
h_{\vec{k}}=\frac{h_{0}}{N}\sum_{n=1}^{N}e^{-i\vec{k}_{n}\cdot\vec{x}_{O}}\delta^{d}(\vec{k}-\vec{k}_{n})\]
plus higher order harmonic. The autocorrelation function is found
by plugging Eq.~\eqref{eq:idealh} into Eq.~\eqref{eq:acdef1} and
simplifying,

\begin{equation}
C^{A}(\dx)=\left(\frac{h_{0}}{N}\right)^{2}\sum_{n=1}^{N}\exp\left[i\vec{k}_{n}\cdot\dx\right]\label{eq:scor_ideal}\end{equation}
plus higher order harmonic. In finding Eq.~\eqref{eq:scor_ideal},
the relation \begin{equation}
\int d^{2}\vec{x}'\, e^{i\left(\vec{k}_{m}-\vec{k}_{n}\right)\cdot\vec{x}'}=A\delta_{\vec{k}_{m}\vec{k}_{n}}=(2\pi)^{d}\delta^{d}(\vec{k}_{m}-\vec{k}_{n})\label{eq:DD1}\end{equation}
has been used. $\delta_{\vec{k}\vec{k}'}$ is the Kronecker Delta,
and $\delta^{d}(\vec{k}-\vec{k}')$ is the Dirac Delta. Eq.~\eqref{eq:DD1}
will be helpful whenever it is necessary to take an areal average
or sum over Dirac Delta functions. In reciprocal space,\begin{eqnarray}
C_{\vec{k}}^{A} & = & \frac{(2\pi)^{2}}{A}\frac{h_{0}^{2}}{N^{2}}\sum_{m,n=1}^{N}\delta^{2}(\vec{k}-\vec{k}_{m})\delta^{2}(\vec{k}-\vec{k}_{n})\nonumber \\
 & = & \frac{h_{0}^{2}}{N^{2}}\sum_{i=1}^{N}\delta^{2}(\vec{k}-\vec{k}_{i})\label{eq:rcor_ideal}\end{eqnarray}
plus higher order harmonics, where $\delta^{d}(\vec{k}-\vec{k}_{n})=(A/(2\pi)^{d})\delta_{\vec{k}\vec{k}_{n}}$.%
\footnote{Eq.~\eqref{eq:DD1} has been used to help with summation%
}. Thus, the order of the SAQD lattice manifests itself as periodic
functions in real-space (Eq.~\eqref{eq:scor_ideal}) and sharp peaks
in reciprocal space (Eq.~\eqref{eq:rcor_ideal}).

\subsubsection{\label{sub:Nearly-Periodic-array}Nearly Periodic array}

A nearly periodic arrays shows deviation from perfect order. This
deviation is shows itself by broadening of the peaks of the spectrum
function, $C_{\vec{k}}^{A}$, and by range limited periodicity of
the real-space autocorrelation function, $C^{A}(\dx)$. These two
measure of disorder are naturally related.

%
{}

The disorder in lateral dot size $\Delta_{\text{size}}$ and spacing,
$\Delta_{\text{spacing}}$ are related to each other and to the broadening
of the peaks in $C_{\vec{k}}^{A}$ (Fig.~\ref{fig:Ck}.a and~c).
Prior to ripening, the size and spatial order should be related, as
the volume of a dot should be proportional to the amount of nearby
material. If the SAQDs have nearly uniform size and spacing (peak-to-peak
distance) $L_{0}$, the reciprocal space autocorrelation function
will be tightly clustered around the wavenumber characterizing the
dot spacing $k_{0}=2\pi/L_{0}$. There are a number of such peaks
depending on the system symmetry (Fig.~\ref{fig:Ck}.a and~c), but
consider just one. Since the order is not perfect, the peak will have
a finite width. Consequently, there will be a scatter in the dot size.
Since $L_{\text{0}}=2\pi/k_{0}$, the scatter in dot spacing ($\Delta_{\text{spacing}}$)
is related to the scatter in Fourier components ($\Delta_{k}$). Taking
the derivative of the spacing-wavenumber relation and rearranging,%
{}\[
\frac{\Delta_{\text{spacing}}}{L_{0}}\approx\frac{\Delta_{k}}{k_{0}}.\]
It is reasonable to expect that the fractional disorder in size ($\Delta_{\text{size }}/L_{\text{size}}$)
is given by a similar (if not exactly the same) number.

%
{}

Another way to view spatial order (periodicity) is not by dot-dot
distances, but the distance over which the dot array can be considered
periodic. This limited periodicity is evident in the film height autocorrelation
function (Eq.~\eqref{eq:acdef1} and Figs.~\ref{fig:Film-heights-and}.b,
e and~h). Consider two distant dots. Their position will be completely
uncorrelated, so it will be completely random as to whether one position
corresponds to a peak or a valley. Thus, for a large differences in
position the autocorrelation function tends to zero.\[
C^{A}(\dx_{\text{large}})=0\]
Similarly, the mean-square fluctuation of the film height can be large
so that \[
C^{A}(\dx=0)\gg0.\]
The distance over which the autocorrelation function, $C^{A}(\dx)$
decays to $0$ is the correlation length, $L_{\text{cor}}$. Thus,
$L_{\text{cor}}$ is a reasonable measure of spatial order. 

%
{}

The two measures of order $\Delta_{\text{spacing}}$ and $L_{\text{cor}}$
are intrinsically linked. The well known rule of Fourier transforms
states that the product of the real-space and reciprocal space widths
must be greater than or equal to unity. Thus, $\Delta_{k}L_{\text{cor}}\geq1$,
or $\Delta_{\text{spacing}}\ge2\pi L_{0}^{2}/L_{\text{cor}}$. Similarly,
one can expect that $\Delta_{\text{size}}\sim L_{\text{size}}^{2}/L_{\text{cor}}$.
Thus, assuming that dot size is governed by the amount of nearby material,
small dispersions in dot size are only possible if there is long correlation
length.

\subsubsection{\label{sub:Ensemble-Correlation-Functions}Ensemble Correlation Functions
/ ergodicity}

SAQDs are seeded by random fluctuations. Consequently, each experiment
or simulation must be treated as just one possible \emph{realization},
and the autocorrelation function will be different for each realization.
Thus, for analytic predictions, one must rely on ensemble averages.
In~\cite{Friedman:fk}, it was assumed that the ensemble average
correlation function was a good description of a SAQD order, an assumption
that was born out by numerical calculations. Now, this relation is
put on a more solid ground. In particular, it is found that the ensemble
correlation functions provide good estimates of the auto correlation
function and spectrum function produced by any particular realization.
First, it is shown that the mean value of the film-height fluctuation
is zero. Then the method to calculate the ensemble-averaged autocorrelation
function and spectrum function is presented. Additional mathematical
details are presented in Appendix~\ref{sec:Correlation-Functions}.

\paragraph{\label{par:Mean-fluctuation}Mean fluctuation}

It is fairly straightforward to show that the ensemble mean film-height
fluctuation is zero. The governing dynamics (Eq.~\eqref{eq:gov_lin})
is invariant upon the substitution $h(\vec{x},t)\rightarrow-h(\vec{x},t)$.
Thus, assuming that one does not bias the initial conditions the mean
fluctuations must be zero for all time,\[
\left\langle h(\vec{x},t)\right\rangle =\left\langle -h(\vec{x},t)\right\rangle =0\text{, and }\left\langle h_{k}(t)\right\rangle =0.\]
This is a common situation, and it is most appropriate to characterize
the film height fluctuations using the two-point correlation function
(or simply {}``the correlation function'').~\cite{Zwanzig:2001zf}

\paragraph{\label{par:Correlation-Function}Correlation Function}

The autocorrelation function can be estimated by its ensemble average.
Furthermore, this ensemble average is equivalent to the correlation
function that can be easily calculated analytically. These relations
are first discussed for the real-space correlation functions and then
their Fourier transforms. First, the statistical properties of the
autocorrelation function are discussed. Then the statistical properties
of the spectrum function. Finally, the method to The main results
are reported here, and details of derivations are reported in Appendix~\ref{app:Correlation-Functions}.

First it is noted that the autocorrelation function averaged over
all realizations is equal to the ensemble correlation function.

\begin{equation}
\left\langle C^{A}(\dx)\right\rangle =C(\dx)\text{, where }C(\dx)=\left\langle h(\dx)h(\vec{0})\right\rangle ,\label{eq:ergo-1}\end{equation}
where $\left\langle \dots\right\rangle $ indicate an ensemble average.
Eq.~\eqref{eq:ergo-1} assumes that the model of film-growth is translationally
invariant.%
\footnote{A quick survey of literature will find that, virtually all published
continuum models of SAQD formation are translationally invariant.%
} This relationship is fortunate, in that it allows one to predict
the {}``typical'' autocorrelation function using analytic tools
that apply only to ensemble averages.

Second, it is noted that as the area that is used to calculate the
autocorrelation function becomes large, the autocorrelation function
tends towards it mean value,\begin{equation}
C^{A}(\dx)\approx C(\dx)+O[A^{-1/2}],\label{eq:C-mean}\end{equation}
where $O[A^{-1/2}]$ indicates statistical fluctuations about the
mean value that become smaller and smaller as the area in an experiment
or the simulation area in a numerical experiment becomes larger. These
fluctuations  or noise die out as $A^{1/2}$. For example, the autocorrelation
functions in Figs.~\ref{fig:Film-heights-and}.e and~h are very
close to the ensemble average autocorrelation functions Figs.~\ref{fig:Film-heights-and}.f
and~h, but have random fluctuations that are most visible far from
the origin. This property, that averaging over a parameter such as
position is equivalent to averaging over all realizations, is known
as ergodicity. Individual realizations are tightly distributed about
a {}``typical'' behavior. This tight distribution lends credibility
to the notion that one can have representative experiments or simulations.
Unfortunately, the {}``demonstration'' of Eq.~\eqref{eq:C-mean}
in Appendix~\ref{app:Correlation-Functions} is not as general as
one might like. Rigorously, it applies when the Fourier components
of film height ($h_{\vec{k}}$) are independent and normally distributed;
however, it is reasonable to conjecture that a relationship like Eq.~\eqref{eq:C-mean}
holds whenever the statistical distribution of film heights is suitably
bounded as the boundedness of $C_{\vec{k}}^{A}$ plays an important
role in the derivations.

In reciprocal space, one finds that the ensemble-mean spectrum function
is \begin{equation}
\left\langle C_{\vec{k}}^{A}\right\rangle =C_{\vec{k}},\label{eq:cork}\end{equation}
where $C_{\vec{k}}$ is defined as the prefactor appearing in the
reciprocal-space two-point correlation function.\begin{equation}
C_{\vec{k}\vec{k}}=\left\langle h_{\vec{k}}h_{\vec{k}}^{*}\right\rangle =C_{\vec{k}}\delta^{d}(\vec{k}-\vec{k}')=C_{\vec{k}}\frac{A}{(2\pi)^{d}}\delta_{\vec{k}\vec{k}'},\label{eq:prefk}\end{equation}
where Eq.~\eqref{eq:DD1} has been used. This form for the two-point
correlation function in reciprocal space occurs if and only if the
system is translationally invariant. Eq.~\eqref{eq:cork} is valuable
because one can solve for $C_{\vec{k}}$ analytically in the linear
case or using various analytic approximations in the non-linear case.
Unlike the autocorrelation function, the spectrum function fluctuates
greatly about its mean. In fact, the fluctuations are about 100\%
(Appendix~\ref{sub:Variance-and-Convergence}). These large fluctuations
result in the commonly observed speckle pattern for the spectrum function
$C_{\vec{k}}^{A}$(Figs.~\ref{fig:Ck}.a and~c). Contrast this pattern
with ensemble-mean spectrum function $C_{\vec{k}}$ shown in Figs.~\ref{fig:Ck}.b
and~d. These speckles can be removed by a smoothing operation, and
a relation similar to Eq.~\eqref{eq:C-mean} results (Appendix~\ref{sub:Eq.-[eq:rssvar]}).
Finally, it should be noted that just as $C_{\vec{k}}^{A}$ is the
Fourier transform of $C^{A}(\dx)$, $C_{\vec{k}}$ is the Fourier
transform of $C(\dx)$ (Appendix~\ref{sub:Mean-Values}).

\subsection{Stochastic Initial Conditions\label{sub:Stochastic-Initial-Conditions}}

To model or simulate the formation of SAQDs, it is absolutely essential
to include some sort of stochastic effect. %
{}An initially flat film $h(\vec{x},0)=0$ is in unstable equilibrium.
Thus, to seed the formation of quantum dots, it is necessary to perturb
the flat surface. The simplest method to do this is to use stochastic
initial conditions with deterministic evolution. One can tenuously
suppose that white noise initial conditions do not {}``bias'' the
ultimate evolution of the film.~\cite{Cross:1993ti} Thus, the initial
conditions are taken from an ensemble with zero mean,\begin{equation}
\left\langle h(\vec{x},0)\right\rangle =0.\label{eq:h0mean}\end{equation}
 and a spatial correlation function, \begin{equation}
C(\vec{x},\vec{x}',0)=\left\langle h(\vec{x},0)h(\vec{x}',0)^{*}\right\rangle =\Delta^{2}\delta^{d}\left(\vec{x}-\vec{x}'\right),\label{eq:h0c1}\end{equation}
where the brackets $\left\langle \dots\right\rangle $ indicate an
ensemble average, $\Delta$ is the noise amplitude, and $\delta^{d}(x)$
is the $d-$dimensional Dirac Delta function. White noise conditions
have an infinite amplitude which is not physical. Thus, a minimum
modification can be made to {}``cut off'' the infinite fluctuations.\begin{equation}
C(\vec{x},\vec{x}',0)=\frac{\Delta^{2}}{(2\pi b_{0}^{2})^{d/2}}\exp\left(-\frac{\left(\vec{x}-\vec{x}'\right)^{2}}{2b_{0}^{2}}\right)\label{eq:h0c2}\end{equation}
In the limit $b_{0}\rightarrow0$, this correlation function reverts
to the white noise correlation functions.

In reciprocal space,\begin{eqnarray*}
C_{\vec{k}\vec{k}'}(0) & = & \left\langle h_{\vec{k}}(0)h_{\vec{k}'}^{*}(0)\right\rangle \\
 & = & (2\pi)^{-2d}\int d^{d}\vec{x}\int d^{d}\vec{x}'\, e^{\left(-i\vec{k}\cdot\vec{x}+i\vec{k}'\cdot\vec{x}'\right)}C(\vec{x},\vec{x}',0)\\
 & = & \frac{\Delta^{2}}{(2\pi)^{d}}e^{-\frac{1}{2}b_{0}^{2}k^{2}}\delta^{d}(\vec{k}-\vec{k}')\end{eqnarray*}
Letting $b_{0}\rightarrow0$, the white noise reciprocal space correlation
function is obtained. Thus, the initial spectrum function is\[
C_{\vec{k}}(0)=\frac{\Delta^{2}}{(2\pi)^{d}}e^{-\frac{1}{2}b_{0}^{2}k^{2}}.\]

The atomic-scale has a small and short-lived influence on the final
film morphology (Appendix~\ref{sec:Atomic-Scale-Cutoff}), but the
cutoff procedure is useful for choosing a reasonable value of $\Delta^{2}$.
It seems reasonable to choose $\Delta^{2}$ so that the initial r.m.s.
fluctuation $\sqrt{C(\vec{0},0)}=\left\langle h(\vec{0},0)h(\vec{0},0)^{*}\right\rangle ^{1/2}$is
one monolayer ($1\text{ ML}$). Also, choosing $b_{0}=1\mbox{\text{ ML}}$
as the atomic scale cutoff is \begin{equation}
\Delta^{2}=(2\pi)^{d/2}(1\text{ ML})^{2+d},\label{eq:namp}\end{equation}
where the natural unit $1\mbox{\text{ ML}}$ is, of course, material
dependent.

Using stochastic initial conditions, one can integrate individual
initial conditions to obtain representative samples and then average
over many realizations, the Monte Carlo approach, or one can calculate
analytically, the statistical measures of the ensemble. The ensemble
statistical measures are strongly related to the statistical measures
of order for an individual realization, so the second approach is
opted for here. Thus, the predicted SAQD order is ultimately stated
in terms of ensemble correlation functions.

\subsection{Reciprocal Space Correlation Functions}

\newcommand{\ckk}{C_{\vec{k}\vec{k}'}}

\label{sub:Reciprocal-Space-Corelation}The reciprocal space correlation
function, $C_{\vec{k}\vec{k}'}$, and spectrum function, $C_{\vec{k}}$,
are calculated for the 1D and 2D isotropic case and then for the 2D
anisotropic case. Generally $C_{\vec{k}}$ includes the length scales
introduced in Sec.~\ref{sub:Peaks.} as well as the atomic scale
cutoff $b_{0}$. 

\begin{eqnarray}
\ckk & = & \left\langle h_{\vec{k}}(t)h_{\vec{k}'}^{*}(t)\right\rangle =e^{(\sigma_{\vec{k}}+\sigma_{\vec{k}'})t}\left\langle h_{\vec{k}}(0)h_{\vec{k}'}(0)^{*}\right\rangle \nonumber \\
 & = & \frac{\Delta^{2}}{(2\pi)^{2}}e^{(\sigma_{\vec{k}}+\sigma_{\vec{k}'})t-\frac{1}{2}b_{0}^{2}k^{2}}\delta^{2}(\vec{k}-\vec{k}').\label{eq:Ckkgen}\end{eqnarray}
Without much error, $b_{0}$ can be neglected in the exponential (Appendix~\ref{sec:Atomic-Scale-Cutoff}).
Using Eq.~\eqref{eq:prefk}, the spectrum function is then identified
as \begin{equation}
C_{\vec{k}}=\frac{\Delta^{2}}{(2\pi)^{d}}e^{2\sigma_{\vec{k}}t}.\label{eq:Ck}\end{equation}
$C_{\vec{k}}$ is now calculated for each model: 1D isotropic, 2D
isotropic and 2D anisotropic.

\subsubsection{\label{sub:one-dimensional}one-dimensional}

\newcommand{\ihat}{\mathbf{i}}

The one dimensional surface is the simplest, so it is treated first.
The spectrum function is simply\[
C_{\vec{k}}=\frac{\Delta^{2}}{2\pi}e^{2\sigma_{0}t-\frac{1}{2}(2\sigma_{2}t)(k-k_{0})^{2}}.\]
$C_{\vec{k}}$ has a peak at $\vec{k}=\pm k_{0}\ihat$. One can easily
read off the correlation length as \begin{equation}
L_{\text{cor}}=\sqrt{2\sigma_{2}t}=k_{c}^{-1}\sqrt{2(3\alpha_{0}-4\beta)(t/t_{c})}.\label{eq:lcor}\end{equation}
so that \[
C_{\vec{k}}=\frac{\Delta^{2}}{2\pi}e^{2\sigma_{0}t-\frac{1}{2}L_{\text{cor}}^{2}(k-k_{0})^{2}}.\]
This approximation is valid when $k_{0}L_{\text{cor}}\gg1$. In terms
of $k_{x}$, \[
C_{\vec{k}}=\frac{\Delta^{2}}{2\pi}e^{2\sigma_{0}t}\left(e^{-\frac{1}{2}L_{\text{cor}}^{2}(k_{x}-k_{0})^{2}}+e^{-\frac{1}{2}L_{\text{cor}}^{2}(k_{x}+k_{0})^{2}}\right)\]

\subsubsection{\label{sub:2D-isotropic}2D isotropic}

\begin{figure*}
\noindent \begin{centering}\begin{tabular}{cc}
\includegraphics[width=2.5in,keepaspectratio]{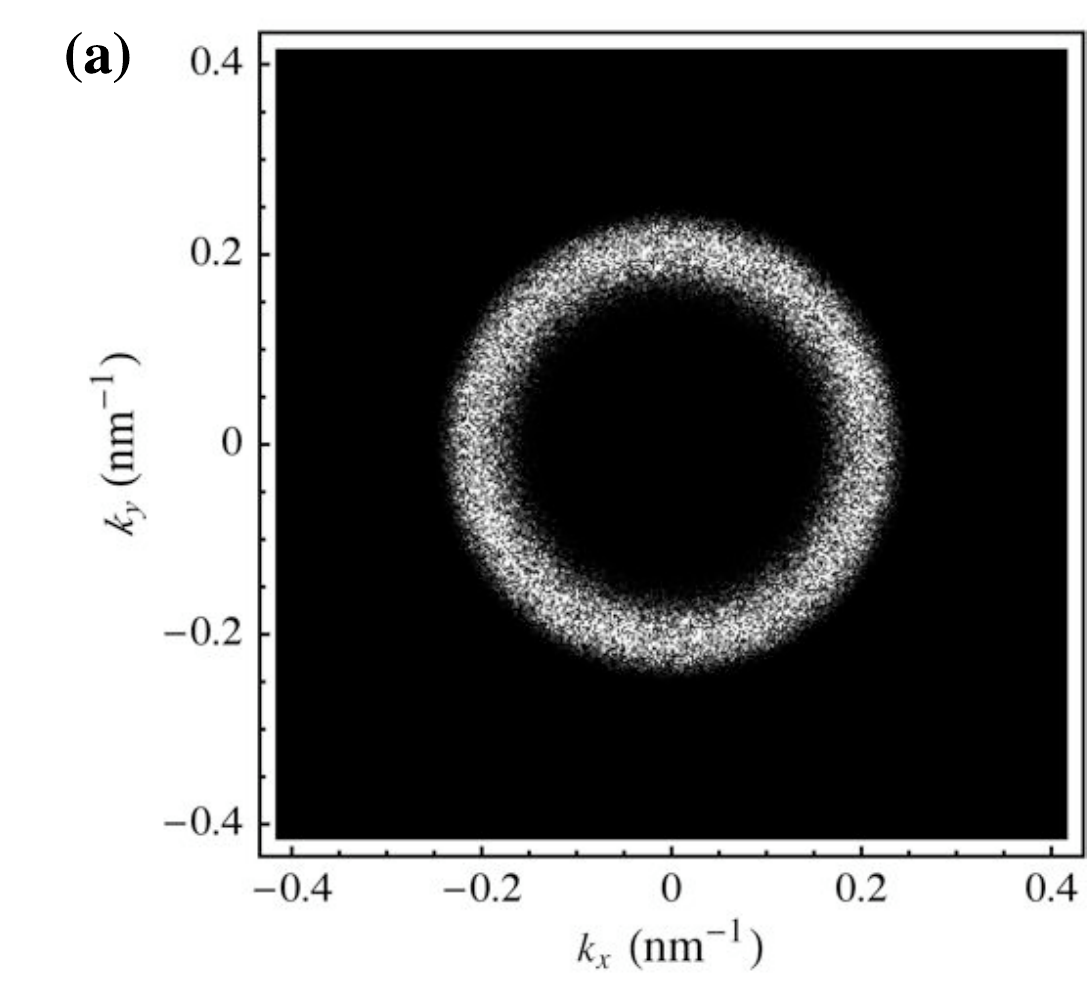}&
\includegraphics[width=2.5in,keepaspectratio]{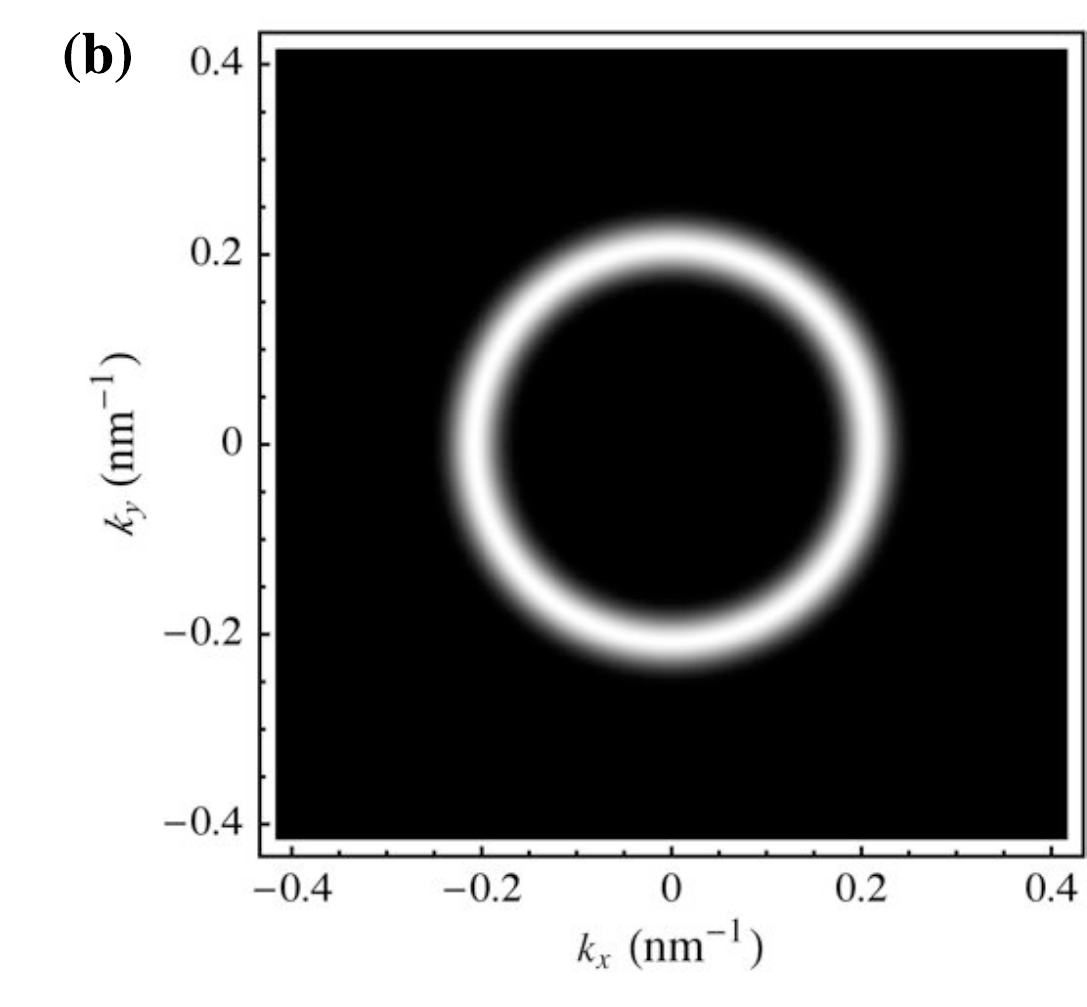}\tabularnewline
\includegraphics[width=2.5in,keepaspectratio]{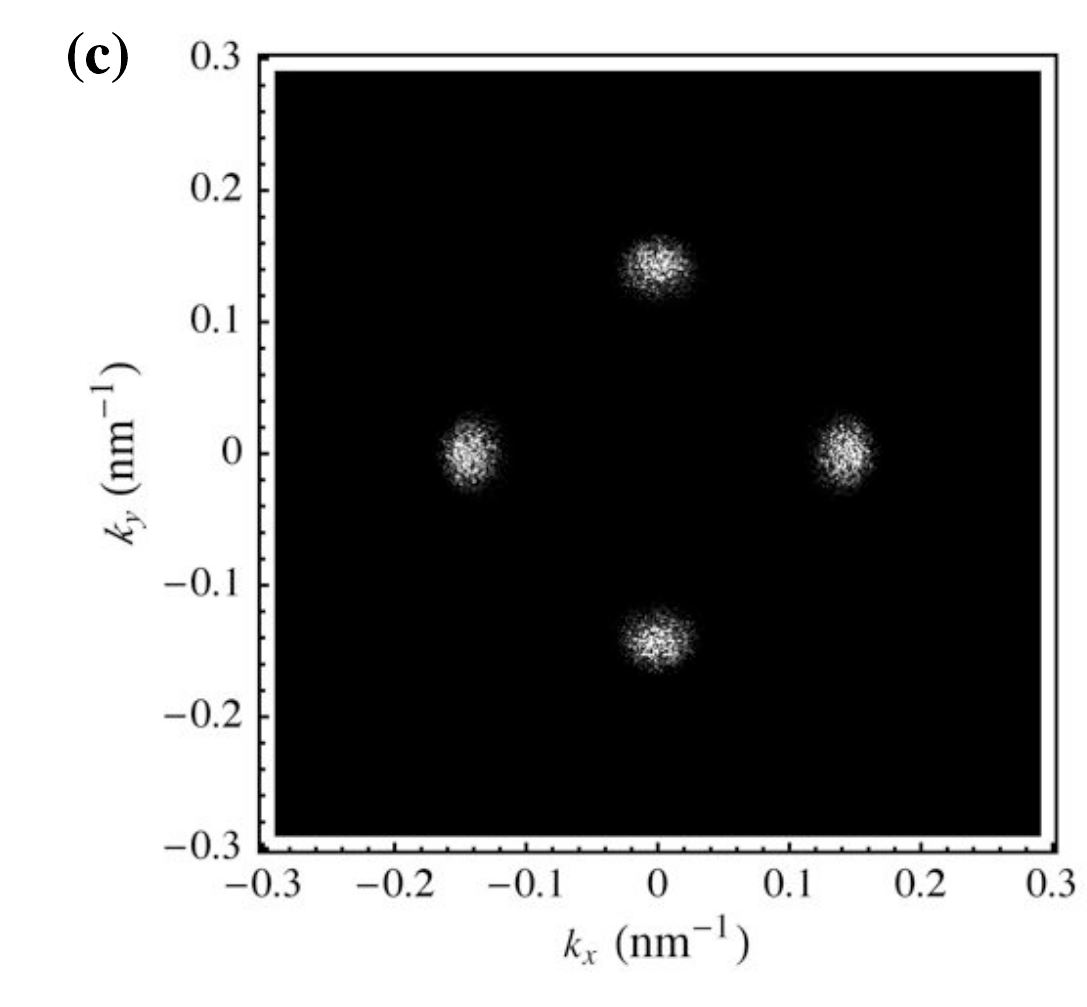}&
\includegraphics[width=2.5in,keepaspectratio]{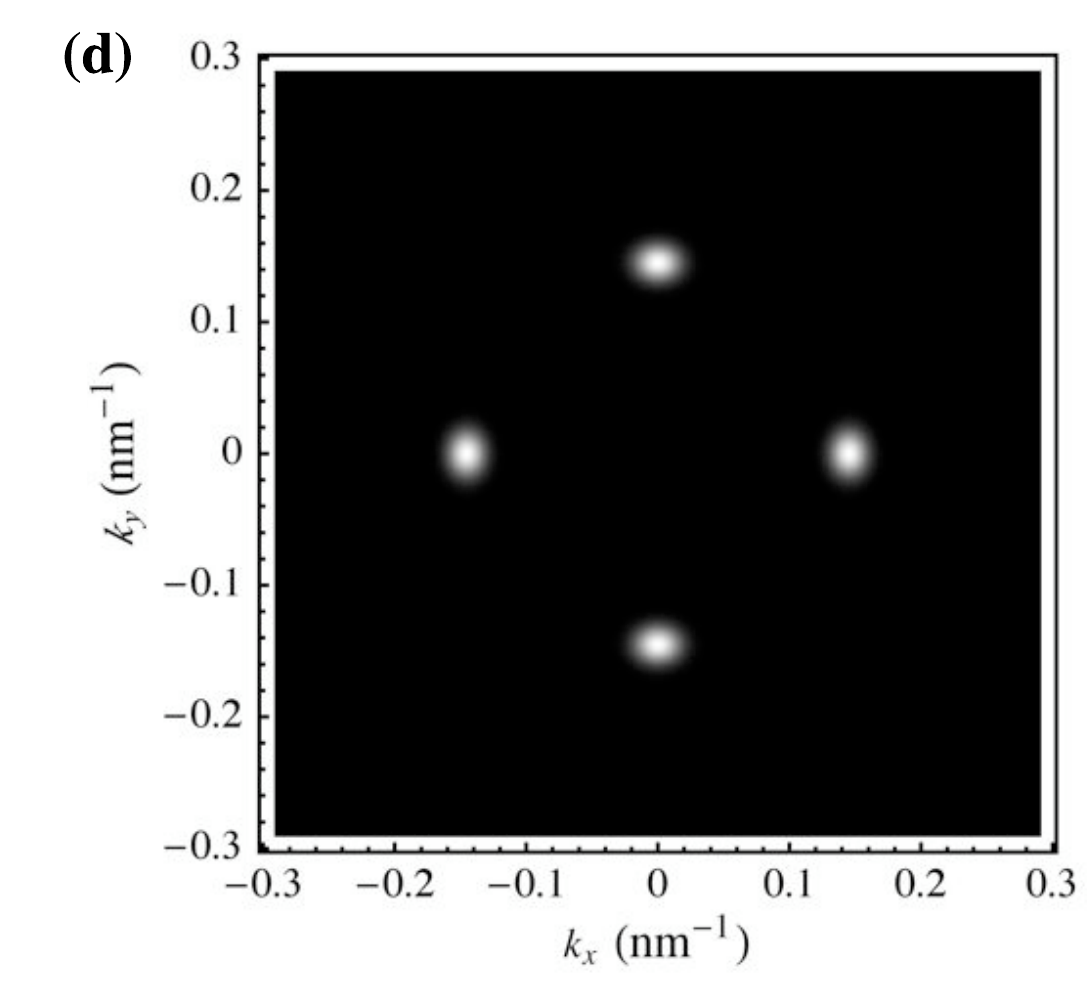}\tabularnewline
\end{tabular}\par\end{centering}

\caption{\emph{\label{fig:Ck}$C_{\vec{k}}^{A}$ and $C_{\vec{k}}$ for Ge/Si}
as discussed in Sec.~\ref{sec:Application-to-Materials}. (a,b) 2D
isotropic surface. Eq.~\eqref{eq:c2di} is used for $C_{\vec{k}}$.
(c,d) 2D anisotropic surface. Eq.~\eqref{eq:Ckanis} is used for
$C_{\vec{k}}$.}
\end{figure*}
The 2D isotropic case is very similar;\begin{equation}
C_{\vec{k}}=\frac{\Delta^{2}}{(2\pi)^{2}}e^{2\sigma_{0}t-\frac{1}{2}L_{\text{cor}}^{2}(k-k_{0})^{2}},\label{eq:c2di}\end{equation}
where $L_{\text{cor}}$ is the same as in Eq.~\eqref{eq:lcor}. It
has maximum that forms a ring in the $\vec{k}-$plane as graphed in
Fig.~\ref{fig:Ck}.b.

\subsubsection{\label{sub:anisotropic}anisotropic}

The anisotropic spectrum function is\begin{eqnarray}
C_{\vec{k}} & = & \frac{\Delta^{2}}{(2\pi)^{2}}e^{2\sigma_{0}t}\sum_{n=1}^{4}e^{-\frac{1}{2}L_{\parallel}^{2}\left(k_{\parallel}-k_{0}\right)^{2}-\frac{1}{2}L_{\perp}^{2}k_{\perp}^{2}},\label{eq:Ckanis}\end{eqnarray}
where\begin{eqnarray}
 &  & L_{\parallel}=\sqrt{2\sigma_{\parallel}t}=k_{c}^{-1}\sqrt{(6\alpha_{0}-8\beta)(t/t_{c})},\label{eq:lpar}\\
 &  & L_{\perp}=\sqrt{2\sigma_{\perp}t}=k_{c}^{-1}\sqrt{16\epsilon\alpha_{0}(t/t_{c})},\label{eq:lperp}\end{eqnarray}
$k_{\parallel}=\cos[\pi(n-1)/2]k_{x}+\sin[\pi(n-1)/2]k_{y}$, and
$k_{\perp}=-\sin[\pi(n-1)/2]k_{x}+\cos[\pi(n-1)/2]k_{y}$ and it is
graphed in Fig.~\ref{fig:Ck}.d. This approximation is valid when
$k_{0}L_{\parallel}\gg1$ and $k_{0}L_{\perp}\gg1$.

Loosely speaking, one can argue that the isotropic case is similar
to letting $\epsilon_{A}\rightarrow0$ in Eq.~\eqref{eq:lperp} so
that the perpendicular correlation length is always 0 regardless of
time. A more conservative approach would be to argue that $L_{\perp}\approx2\pi/k_{0}$
for the isotropic model via inspection of Figs.~\ref{fig:Ck}(a)
and (b). Even still, the more conservative result guarantees that
the perpendicular correlation length will always be the same as the
dot spacing; thus, it will always limit SAQD order to the first nearest
neighbor at best.

\subsection{Real Space Correlation Functions}

\label{sub:Real-Space-Correlation}The real space correlation functions
$C(\dx)$ are now calculated for the 1D and 2D isotropic cases and
the 2D elastically anisotropic case.

\subsubsection{one-dimensional}

\begin{figure}
\noindent \begin{centering}\includegraphics[width=3.375in,keepaspectratio]{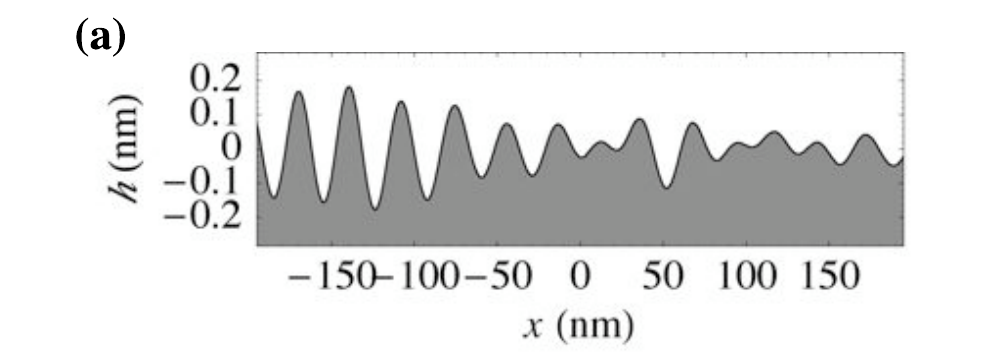}\par\end{centering}

\noindent \begin{centering}\includegraphics[width=3.375in,keepaspectratio]{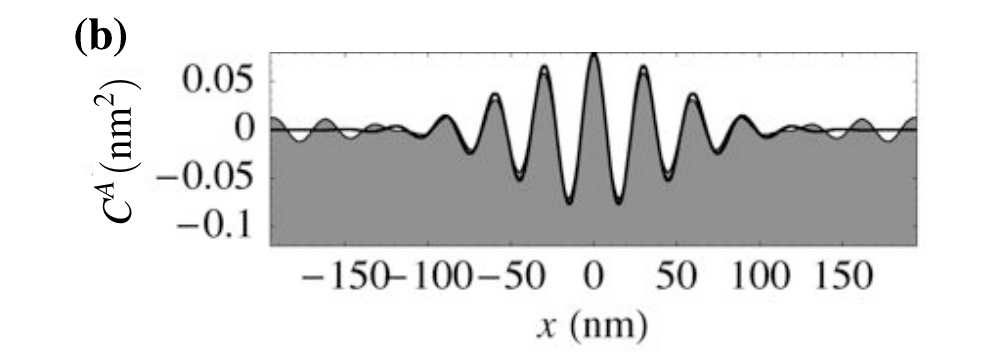}\par\end{centering}

\caption{\label{fig:1D-real-space}1D isotropic surface in real space for
Ge/Si as discussed in Sec.~\ref{sec:Application-to-Materials}. (a)
Example of $h(\vec{x})$ plotted over a length of $8L_{\text{cor}}$.
(b) corresponding reals space correlation functions plotted for range
$\pm4L_{\text{cor}}$. Filled plot is an example of $C^{A}(\dx)$.
Solid line is$C(\dx)$ (Eq.~\eqref{eq:Cx1D}). }
\end{figure}
In one dimension, \begin{eqnarray}
C(\dx) & = & \int_{-\infty}^{\infty}dk_{x}\, e^{ik_{x}\dx}C_{\vec{k}}\nonumber \\
 & = & \frac{\Delta^{2}}{\sqrt{2\pi}}\frac{1}{L_{\text{cor}}}e^{2\sigma_{0}t-\frac{1}{2}\Delta x^{2}/L_{\text{cor}}^{2}}2\cos\left(k_{0}x\right).\label{eq:Cx1D}\end{eqnarray}
Thus, $C(\dx)$ has a damped periodicity indicating that it is imperfectly
periodic (Fig.~\ref{fig:1D-real-space}).

\subsubsection{2D isotropic}

In two dimensions with elastic isotropy,\begin{eqnarray*}
C(\dx) & = & \int d^{2}\vec{k}\, e^{i\vec{k}\cdot\dx}C_{\vec{k}}\\
 & = & \frac{\Delta^{2}}{(2\pi)^{2}}e^{2\sigma_{0}t}\int_{0}^{2\pi}d\theta_{\vec{k}}\int_{0}^{\infty}dk\, ke^{i(k\Delta x\cos(\theta_{\vec{k}}-\theta_{\dx})}e^{-\frac{1}{2}L_{\text{cor}}^{2}(k-k_{0})^{2}}.\end{eqnarray*}
Performing the angular integration first,\[
C(\dx)=\frac{\Delta^{2}}{2\pi}e^{2\sigma_{0}t}\int_{0}^{\infty}dk\, kJ_{0}(k\Delta x)e^{-\frac{1}{2}L_{\text{cor}}^{2}(k-k_{0})^{2}},\]
where $J_{0}$ is the zeroth Bessel function. In general, this integral
is best performed numerically; however, it can be solved in two important
cases: $\dx\rightarrow\vec{0}$ and $L_{\text{cor}}\rightarrow\infty$
(corresponding to long times). In the first case, \begin{eqnarray*}
C(\dx) & = & \frac{\Delta^{2}}{2\pi}e^{2\sigma_{0}t}\int_{0}^{\infty}dk\, ke^{-\frac{1}{2}L_{\text{cor}}^{2}(k-k_{0})^{2}}.\end{eqnarray*}
Under the same conditions that Eq.~\eqref{eq:c2di} is valid ($k_{0}L_{\text{cor}}\gg1$),
the lower limit of the integral can be approximated as $-\infty$
so that \begin{equation}
C(\dx=\vec{0})=\frac{\Delta^{2}k_{0}}{\sqrt{2\pi}L_{\text{cor}}}e^{2\sigma_{0}t}.\label{eq:var2Di}\end{equation}
This function gives the mean square surface height fluctuation. In
the second case where $L_{\text{cor}}\rightarrow\infty$, $e^{-\frac{1}{2}L_{\text{cor}}^{2}(k-k_{0})^{2}}\rightarrow(2\pi)^{1/2}L_{\text{cor}}^{-1}\delta(k-k_{0})$,
so that\begin{eqnarray}
C(\dx) & = & \frac{\Delta^{2}k_{0}}{\sqrt{2\pi}L_{\text{cor}}}e^{2\sigma_{0}t}J_{0}\left(k_{0}\Delta x\right).\label{eq:Cx2Di}\end{eqnarray}
This correlation function is the most ordered case for a 2D isotropic
surface. It is graphed in Fig.~\ref{fig:Film-heights-and}c. %
{}

\subsubsection{anisotropic}

To find the real-space correlation function for the elastically anisotropic
case, it is best to find the contribution from each peak and then
sum so that\begin{equation}
C(\dx)=\frac{\Delta^{2}}{(2\pi)^{2}}e^{2\sigma_{0}t}\sum_{n=1}^{4}C^{n}(\dx)\label{eq:canis1}\end{equation}
where\[
C^{n}(\dx)=\int d^{2}\vec{k}\, e^{i\vec{k}\cdot\vec{x}}e^{-\frac{1}{2}L_{\parallel}^{2}\left(k_{\parallel}-k_{0}\right)^{2}-\frac{1}{2}L_{\perp}^{2}k_{\perp}^{2}}.\]
$\dx$ can be decomposed into the directions parallel and perpendicular
to $\vec{k}_{n}$, so that $\Delta x_{\parallel}=\cos(\pi(n-1)/2)\Delta x+\sin(\pi(n-1)/2)\Delta y$
and $\Delta x_{\perp}=-\sin(\pi(n-1)/2)\Delta x+\cos(\pi(n-1)/2)\Delta y$.
Thus,

\begin{eqnarray*}
C^{n}(\dx) & = & \left(\int dk_{\parallel}\, e^{ik_{\parallel}\Delta x_{\parallel}-\frac{1}{2}L_{\parallel}^{2}\left(k_{\parallel}-k_{0}\right)^{2}}\right)\left(\int dk_{\perp}\, e^{ik_{\perp}\Delta x_{\perp}-\frac{1}{2}L_{\perp}^{2}k_{\perp}^{2}}\right)\\
 & = & \frac{2\pi}{L_{\parallel}L_{\perp}}e^{-\frac{1}{2}\left(x_{\parallel}^{2}/L_{\parallel}^{2}+x_{\perp}^{2}/L_{\perp}^{2}\right)}e^{ik_{0}x_{\parallel}}.\end{eqnarray*}
Plugging into Eq.~\eqref{eq:canis1},\begin{equation}
C(\dx)=\frac{\Delta^{2}}{\pi L_{\parallel}L_{\perp}}e^{2\sigma_{0}t}\left[e^{-\frac{1}{2}\left(x^{2}/L_{\parallel}^{2}+y^{2}/L_{\perp}^{2}\right)}\cos(k_{0}x)+e^{-\frac{1}{2}\left(x^{2}/L_{\perp}^{2}+y^{2}/L_{\parallel}^{2}\right)}\cos(k_{0}y)\right].\label{eq:Cxa1}\end{equation}

\subsection{Generalizability}

\label{sub:Generalizability}The dynamics and analysis used here were
for a specific model, but the general procedure for analyzing the
order resulting from a linearized model should hold for any model
with well-separated peaks in the dispersion relation, $\sigma_{\vec{k}}$.
The procedure to follow is: 

\begin{enumerate}
\item Generate the dispersion relation, $\sigma_{\vec{k}}$ as some function
of $\vec{k}$. 
\item Find the peaks in the dispersion relation, $\vec{k}_{n}$, ($n=1\dots N$) 
\item Expand about the peaks to generate the peak values, $\sigma_{n}$,
and local Hessian matrix, \[
\left(\tilde{H}_{n}\right)_{ij}=\left.\frac{\partial^{2}}{\partial k_{i}\partial k_{j}}\sigma_{\vec{k}}\right|_{\vec{k}=\vec{k}_{n}}.\]
The spectrum function is then approximately\begin{equation}
C_{\vec{k}}(t)\approx\frac{\Delta^{2}}{(2\pi)^{2}}\sum_{n=1}^{N}e^{2\sigma_{n}t}\exp\left[t\,\left(\vec{k}-\vec{k}_{n}\right)\cdot\tilde{H}_{n}\cdot\left(\vec{k}-\vec{k}_{n}\right)\right].\label{eq:spectrum}\end{equation}

\item Find the Eigenvalues of the local Hessian matrix, $\left(H_{n}\right)_{I}$
and $\left(H_{n}\right)_{II}$ . They should be negative, if there
is a peak at $\vec{k}_{n}$ 
\item Use the eigenvalues to determine the correlation lengths, $\left(L_{n}\right)_{I}=\sqrt{2\left|\left(H_{n}\right)_{I}\right|t}$
and $\left(L_{n}\right)_{II}=\sqrt{2\left|\left(H_{n}\right)_{II}\right|t}$.
The real-space correlation function is\begin{equation}
C(\dx,t)\approx\frac{\Delta^{2}}{(2\pi)}\sum_{n=1}^{N}\frac{1}{4t\sqrt{\left(H_{n}\right)_{I}\left(H_{n}\right)_{II}}}e^{2\sigma_{n}t}\exp\left(\frac{\vec{x}\cdot\tilde{H}_{n}^{-1}\cdot\vec{x}}{4t}\right)e^{i\vec{k}_{n}\cdot\vec{x}}.\label{eq:correlation}\end{equation}
The {}``goodness'' of these approximate forms requires that $\left(L_{n}\right)_{I}^{-1}$
and $\left(L_{n}\right)_{II}^{-1}$ be much less than the spacing
between peaks in the correlation function so that the gaussians do
not overlap greatly. A reasonable test for this no-overlap condition
is $\left\Vert \vec{k}_{n}\right\Vert \left(L_{n}\right)_{I}\ll1$
and $\left\Vert \vec{k}_{n}\right\Vert \left(L_{n}\right)_{II}\ll1$,
assuming that the peaks are not large in number or very closely spaced. 
\end{enumerate}

\section{Order Predictions}

\label{sec:Application-to-Materials}The real-space correlation function
formulas (Eqs.~\eqref{eq:Cx1D}, \eqref{eq:var2Di}, and~\eqref{eq:Cxa1})
and correlation length formulas (Eqs.~\eqref{eq:lcor}, \eqref{eq:lpar}
and~\eqref{eq:lperp}) can now be used to estimate the order of SAQDs.
Ge on Si is chosen for this example because this system has received
the most attention from theoretical work~\cite[ and others]{Spencer:1993ve,Zhang:2003tg,Gao:1999ve,Friedman:2006bc,Wang:2004dd,Beck:2004yq,Tersoff:1994fk,Obayashi:1998fk,Ozkan:1999gf},
and it is the simplest since it involves the diffusion of a single
species. The procedure described below tries to predict the amount
of order when an initial atomic-scale fluctuation becomes {}``large''.
{}``Large'' is taken to be greater than atomic-scale. Beyond this
point, one would expect non-linear terms to become important. An example
is presented for Ge on Si at $600\text{K}$ to compare and contrast
the 2D anisotropic results with the 1D isotropic and 2D isotropic
results. The predictions are also compared with a linear numerical
calculation on a discrete reciprocal-space grid to test the approximations
made and to illustrate the relation between the surface profile ($h(\vec{x})$),
the example autocorrelation functions ($C^{A}(\vec{x})$ and $C_{\vec{k}}^{A}$)
and the ensemble correlation functions ($C(\dx)$ and $C_{\vec{k}}$).
Figs.~\ref{fig:Ck}, \ref{fig:1D-real-space} and~\ref{fig:Film-heights-and}
show these results. Finally, the relation between average film height
and order is investigated.

\subsection{Ge at 600K}

\label{sub:Ge-at-600K}The formulations for the three discussed cases
are implemented for Ge/Si at $600\text{K}$. The correlation lengths
are estimated for the end of the linear regime where fluctuations
become large (greater than atomic scale). First, appropriate physical
constants are used to give the corresponding correlation length and
correlation functions vs. time. These include an initial average film
height $\bar{\h}$ and a white noise amplitude $\Delta$ (Eq.~\eqref{eq:namp}).
These initial conditions approximate a film at the beginning of an
anneal that immediately follows a rapid deposition. The time $t_{\text{large}}$
is found by solving for the time where the mean-square fluctuations
are atomic scale, $\left\langle h(\vec{x},t)^{2}\right\rangle =C(\dx=\vec{0})=1\text{ ML}^{2}$.
At this point, the correlation lengths are calculated. 

Physical constants for the 2D anisotropic calculation are taken as
follows. The elastic constants for Ge at $600\text{ K}$ are $c_{11}=1.199\times10^{12}$,
$c_{12}=4.01\times10^{11}$(from $c_{S}=3.991$), $c_{44}=6.73$.~\cite{Vorbyev:1996fk}
Using $a_{\text{Ge}}=0.5658\text{nm}$ and $a_{\text{Si}}=0.5431\text{nm}$,
it is found that $\epsilon_{m}=0.0418$. Using the procedure from
(Appendix~\ref{sec:Elastic-Anisotropy}), $M=1.332\times10^{12}\text{dyn/cm}^{2}$.
$\mathcal{E}_{0\degree}=4.96\times10^{9}\text{erg/cm}^{3}$, and $\mathcal{E}_{45\degree}=4.35\times10^{9}\text{erg/cm}^{3}$
, giving $\epsilon_{A}=0.1236$. The atomic volume is $\Omega=2.27\times10^{-23}\text{ cm}^{3}$.
The estimated surface energy density is $\gamma=1927\text{ erg/cm}^{2}$.
The wetting potential is estimated by picking a plausible critical
surface height, $\h_{c}\approx4\text{ ML}=1.132\text{ nm}$ and setting
$W(\h)=\mathcal{E}_{0\degree}^{2}\h_{c}^{3}/(8\gamma\h)=2.315\times10^{-6}/\h\text{ erg/cm}^{2}$.
The resulting characteristic wave number is $k_{c}=0.257\text{ nm}^{-1}$.
The initial film height is taken to be $\bar{\h}=\h_{c}+0.25\text{ ML}=1.203\text{ nm}$
and then allowed to evolve naturally. Thus, $\beta=0.208$, $\alpha_{0}=0.5658$,
$k_{0}=0.1456\text{ nm}^{-1}$, $\sigma_{0}=0.1192/t_{c}$, $\sigma_{\parallel}=0.864/(k_{c}^{2}t_{c})$,
$\sigma_{\perp}=0.559/(k_{c}^{2}t_{c})$, $L_{\parallel}=0.744k_{0}^{-1}(t/t_{c})^{1/2}$,
and $L_{\perp}=0.599k_{0}^{-1}(t/t_{c})^{1/2}$. The unspecified diffusivity
has been absorbed into the characteristic time $t_{c}$. From Eq.~\eqref{eq:namp},
$\Delta^{2}=0.0403\text{ nm}^{4}$, and Eq.~\eqref{eq:Cxa1} gives
\[
C(\vec{0})=\left(1.223\times10^{-3}t_{c}/t\right)e^{0.02385t/t_{c}}\text{ nm}^{2}.\]
The initial infinitely rough surface undergoes a smoothing described
by the $t_{c}/t$ factor. Then the surface roughens due to the exponential.
The initial divergent roughness is an artifact of the non-physical
white noise with the atomic scale cutoff $b_{0}$ neglected (Appendix~\ref{sec:Atomic-Scale-Cutoff}).
The time for the fluctuations to become {}``large'' again are found
by setting

\begin{equation}
C(\vec{0})=h_{\text{large}}^{2}\label{eq:hlarge}\end{equation}
where $h_{\text{large}}=1\text{ ML}=0.283\text{ nm}$. The solutions
are $t_{1}=0.01527t_{c}$ or $t_{2}=430t_{c}$. The first solution
is discarded since it is due to the non-physical white noise. At $t_{\text{large}}=t_{2}$,
$L_{\parallel}=105.8\text{ nm}$, and $L_{\perp}=85.2\text{ nm}$.
Taking $L_{\perp}$ as more limiting, the correlation spans about
$n=k_{0}L_{\perp}/\pi=3.95$ islands across. The corresponding reciprocal
space (Eq.~\eqref{eq:Ckanis}) and real-space correlation function
(Eq.~\eqref{eq:Cxa1}) are shown in Figs.~\ref{fig:Ck}.d and~\ref{fig:Film-heights-and}.f
respectively.

A corresponding numerical experiment is performed. A periodic surface
of size $l=96(2\pi/k_{0})$ is used. Random initial conditions consistent
with Eq.~\eqref{eq:namp} are used for $k-$space points on a square
grid bounded by $k_{x},\, k_{y}=\pm2k_{0}$. The relation between
discrete and continuous Fourier components is used, $(h_{\vec{k}})_{\text{discrete}}=[(2\pi)^{d}/A]h_{\vec{k}}$.
Eqs.~\eqref{eq:td} and~\eqref{eq:gcgen} are used without any additional
approximation to find $h_{\vec{k}}$ at time $t=t_{\text{large}}$.
The resulting $C_{\vec{k}}^{A}$, a portion of the height profile
$h(\vec{x})$ and $C^{A}(\dx)$ are plotted in Figs.~\ref{fig:Ck}(c),
\ref{fig:Film-heights-and}(d) and~\ref{fig:Film-heights-and}(f)
respectively.

Similar calculations can be performed for the one-dimensional and
two-dimensional elastically isotropic cases. Isotropic values used
previously~\cite{Spencer:1993ve,Friedman:fk} are about $E=1.361\times10^{12}\text{ dyn/cm}^{2}$
and $\nu=0.198$ giving $M=E/(1-\nu)=1.697\times10^{12}\text{ dyn/cm}^{2}$
and $\mathcal{E}=2M(1+\nu)=7.10\times10^{9}\text{ erg/cm}^{3}$. Using
the same critical surface height, $\h_{c}=4\text{ ML}$, $W(\h)=4.74\times10^{-6}/\h\text{ erg/cm}^{2}$.
The resulting characteristic wave number is $k_{c}=0.368\text{ nm}^{-1}$.
If the film is grown to $\bar{\h}=\h_{c}+0.25\text{ ML}=1.203\text{ nm}$
and then allowed to evolve naturally, $\beta=0.208$; thus, $\alpha_{0}=0.5658$,
$k_{0}=0.208\text{ nm}^{-1}$, $\sigma_{0}=0.1192/t_{c}$, $\sigma_{2}=0.864/(k_{c}^{2}t_{c})$,
and $L_{\text{cor}}=0.744k_{0}^{-1}(t/t_{c})^{1/2}$. In one dimension,
Eq.~\eqref{eq:Cx1D} is used to find the mean square height fluctuation.
Using Eq.~\eqref{eq:namp} with $d=1$, $\Delta^{2}=0.0568\text{ nm}^{3}$,
and\[
C(\vec{0},t)=0.01271(t/t_{c})^{-1/2}e^{0.0238t/t_{c}}.\]
Setting $C(\vec{0},t)=(1\text{ ML})^{2}=0.0801\text{ nm}^{2}$, $t_{1}=0.0252t_{c}$,
and $t_{2}=186.9t_{c}$. At $t_{2}$, $L_{\text{cor}}=48.8\text{ nm}$,
and $n=k_{0}L_{\text{cor}}/\pi=3.24$, so about 3 dots in a row should
be well correlated. The corresponding numerical calculation of size
$l=96(2\pi/k_{0})$ is performed. A portion of $h(\vec{x})$, $C^{A}(\dx)$
and $C(\dx)$ are shown in Fig.~\ref{fig:1D-real-space}. In two
dimensions, Eq.~\eqref{eq:var2Di} is used to find $\left\langle h(\vec{x},t)^{2}\right\rangle $,\[
C(\vec{0},t)=9.40\times10^{-4}(t/t_{c})^{-1/2}e^{0.0238t/t_{c}}.\]
Setting $C(\vec{0},t)=0.0801\text{ nm}^{2}$, $t_{1}=1.376\times10^{-4}t_{c}$,
and $t_{2}=306t_{c}$. At $t_{2}$, $L_{\text{cor}}=62.4\text{ nm}$,
and $n=k_{0}L_{\text{cor}}/\pi=4.14$, and correlation is expected
to extend about 4 dots. However, it should be noted that this correlation
is not lattice-like. Corresponding numerical results and ensemble
correlation functions are shown in Figs.~\ref{fig:Ck} and~\ref{fig:Film-heights-and}.a-c.

\subsection{General case of $\beta$}

\label{sub:General-case-of}%
\begin{figure}
\noindent \begin{centering}\includegraphics[width=3.375in]{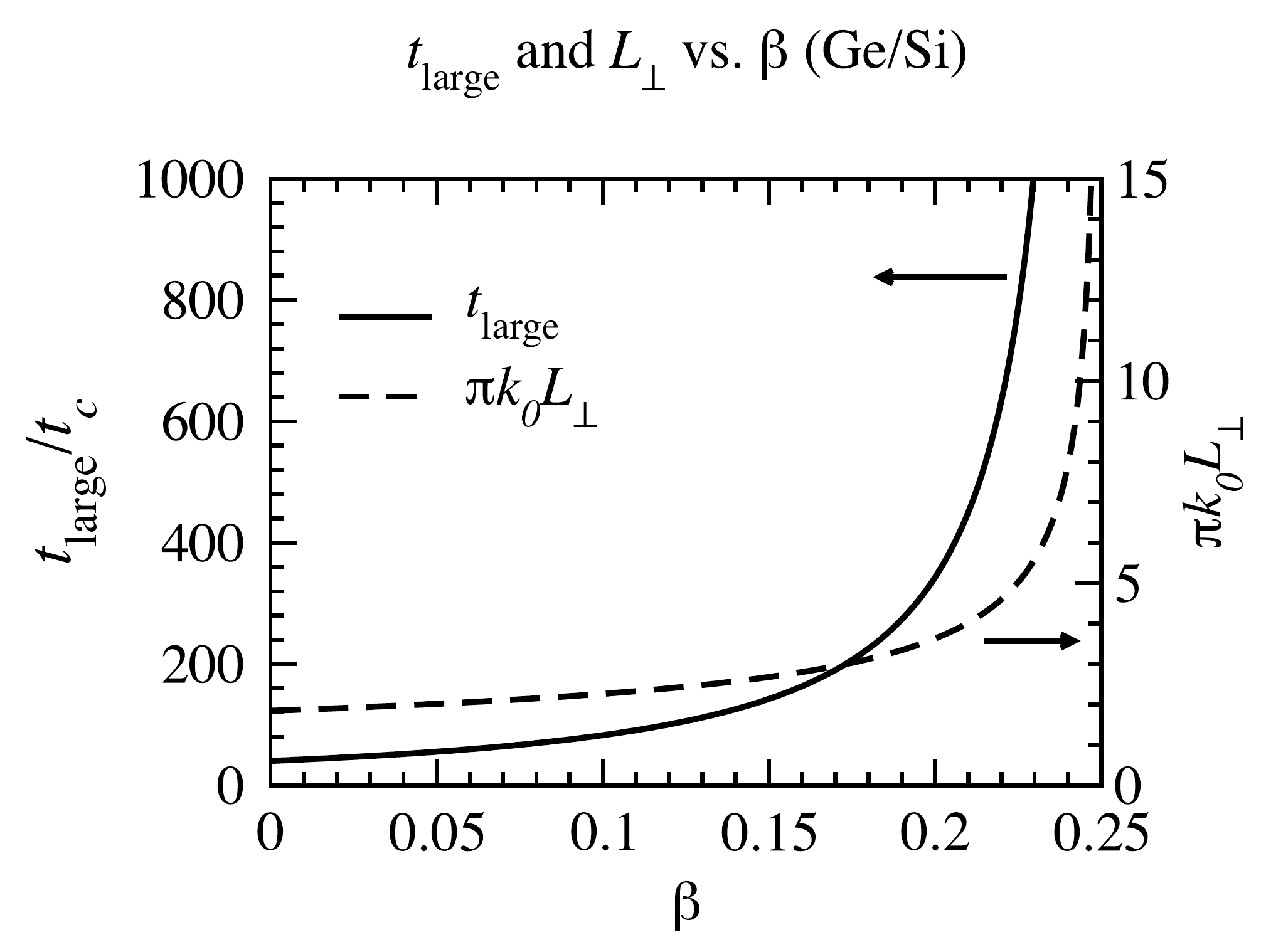}\par\end{centering}

\caption{\label{fig:beta-figure}\emph{$t_{\text{large}}$ and $L_{\perp}$
vs. $\beta$ for Si/Ge} using the 2D anisotropic model as described
in Sec.~\ref{sec:Application-to-Materials}. Units are normalize
to characteristic time $t_{c}$ and predicted number of correlated
dots ($n=k_{0}L_{\perp}/\pi$). }
\end{figure}
In~\cite{Friedman:fk} it was suggested that allowing the film to
evolve with $\beta$ close to the stability threshold could enhance
the SAQD correlation. It is interesting to note what happens for different
values of $\beta$. Similar analytic and numerical calculations are
performed for the large film-height limit, $\beta=0$, for the 2D
anisotropic Ge/Si surface. For $\beta=0$, $t_{\text{large}}=40.3t_{c}$,
$L_{\perp}=30.0\text{ nm}$, and $n=k_{0}L_{\perp}/\pi=1.84$, so
one to two dots in a row are expected to be well correlated. $h(\vec{x})$
and real-space correlation functions are shown in Figs.~\ref{fig:Film-heights-and}g-i.
The range of order is significantly less than for the case $\beta=0.208$
(Sec.~\ref{sub:Ge-at-600K}). For Si/Ge at $600\text{K}$, the 2D
anisotropic predictions for $t_{\text{large}}$ and $L_{\perp}$ are
shown in Fig.~\ref{fig:beta-figure}. In general, the closer $\beta$
is to the critical value $0.25$, the longer the correlation length.
One can manipulate equation~\eqref{eq:hlarge} to find that $t_{\text{large}}/t_{c}$
varies approximately but not exactly as $(\beta-1/4)^{-1}\times\ln[h_{\text{large}}^{2}\sqrt{\epsilon_{A}}/(\Delta^{2}k_{c}^{2})]$.
Consequently, $L_{\perp}\sim(\beta-1/4)^{-1/2}$. Furthermore, the
appearance of $h_{\text{large}}$ and $\Delta^{2}$ inside the logarithm
shows that the final order estimates are not overly sensitive to the
guesses for $\Delta^{2}$ and $h_{\text{large}}^{2}$. The divergence
of $L_{\perp}$ with $\beta-1/4$ is initially encouraging, but it
is clear that for the parameters used for Ge/Si, subatomic control
of the film height is needed to yield significantly enhanced long
range correlations. Also as one approaches this threshold, one can
probably expect thermal activation to nucleate subcritical SAQDs whose
effect on supercritically formed SAQDs is uncertain. There should
be some interesting phenomena at the the $\h\rightarrow\h_{c}$.

\section{Discussion/Conclusions }

\label{sec:Conclusions}The order of epitaxial self-assembled quantum
dots during initial stages of growth has been studied using a common
model of surface diffusion with stochastic initial conditions. It
has been shown that correlation functions of small surface height
fluctuations can be predicted analytically using corresponding ensemble
average correlation functions. These correlation functions are characterized
by correlation lengths that can be predicted by analytic formulas
given certain reasonable assumptions about the diffusion potential
and the height and lateral scale of initial atomic scale random fluctuations.
Thus, the linear model of film surface height evolution via surface
diffusion has enabled analytic predictions of epitaxial SAQD order
that are valid for small film height fluctuations. To what extent
the initial degree of order persists into later stages of growth remains
to be studied, but the order of initial stages should certainly have
a strong influence on final outcomes. Furthermore, the linear analysis
should provide insight into the less tractable non-linear behavior.
These predictions of SAQD order have been used to investigate the
role of crystal anisotropy and initial film height.

Crystal anisotropy has been shown to play an important role in enhancing
SAQD order as observed in previous numerical simulations continuum
and atomistic numerical simulations.~\cite{Holy:1999th,Liu:2003kx,Liu:2003ig,Springholz:2001nx}
If a four-fold symmetry is assumed for the governing dynamics, the
effect of crystal anisotropy to linear order is felt through elastic
anisotropy alone. It is shown that elastic anisotropy is required
to produce a lattice-like structure of SAQDs. The enhanced spatial
order should in turn lead to enhanced size order, a consequence that
must be confirmed with non-linear studies, but appears to be true
based on the present available literature. 

The role of initial film height has been shown to greatly influence
order. Growth near the critical film height for dot formation can
enhance order. This order enhancement comes from increasing the duration
of the linear small-fluctuation stage of growth. In fact, the predicted
correlation lengths diverge when the initial film height approaches
the critical film height from above. Achieving large correlation lengths
in this manor is of course practically limited by ability to control
film heights to subatomic accuracy. Additionally, one should be careful
when interpreting the continuum model in such a context, as the effect
of atomic discreteness might be greater at the transition film height.
Finally, it is likely that additional randomizing effects of thermal
activation will effectively cut off this divergence when the critical
film height is approached from below during deposition. 

Finally, the presented method may be useful as a first step in the
analysis of methods to enhance SAQD order. It is reasonable to suppose
that under some circumstances initial growth stages will be very important
while for others they will not. For example, prior work on vertical
stacking appears to confirm the presented ordering mechanism.~\cite{Liu:2003ig}.
Vertical stacking not only achieves vertical correlation of dots,
but each layer is more ordered horizontally than the one below. Additionally,
a {}``growth window'' was found, whereby to achieve enhanced order,
the evolution of each layer be terminated before ripening begins.
The reported simulation~\cite{Liu:2003ig} supports the following
scenario for SAQD order development. Order is enhanced during the
small fluctuation stage as described here. Once the fluctuations are
sufficiently large, the seeded dots evolve towards their equilibrium
shapes. Finally, the dots begin to ripen and order diminishes. Order
is transfered via strain to the next layer so that the next layer
gets a head start on its initial ordering. Thus, the multiple layers
of dots effectively draws out the linear growth stage. It may be possible
to modify the present model to predict the correlation length of each
SAQD layer.

\appendix

\section{Diffusion Potential}

\label{sec:Diffusion-Potential}The diffusion potential is calculated
in terms of the film height $\h$ that is a function of the in plane
coordinates $\vec{x}=x\mathbf{i}+y\mathbf{j}$. The elastic and surface
energy portions of the diffusion potential can be found in~\cite{Freund:2003ih}
\[
\mu_{\text{elast}}(\vec{x})=\Omega\omega(\vec{x})\text{, and }\mu_{surf}=-\Omega\gamma\kappa(\vec{x}),\]
where $\Omega$ is the atomic volume, $\omega(\vec{x})$ is the elastic
energy density at the film surface, $\gamma$ is the surface energy
density, and $\kappa$ is the total surface curvature. However, other
calculations need to be included:

\begin{enumerate}
\item $\mu_{\text{wet}}$ for the two wetting potential cases, Eq.~\eqref{eq:mu1}
and~\eqref{eq:mu2},
\item and $\mu_{\text{surf}}$ and $\mu_{\text{wet}}$ when the surface
energy density $\gamma$ and wetting energy density $W$ also depend
on surface orientation.
\end{enumerate}
Before these case are addressed, a general form for the diffusion
potential is justified.

\subsection{General Form $\mu=\Omega\delta\mathcal{F}/\delta\h(\vec{x})$}

\label{sub:General-form-of}The diffusion potential, $\mu(\vec{x})$,
is the change in free energy, $\mathcal{F}$, when a particle is added
at a position, $\vec{x}$. Note that $\mu(\vec{x})$ and $\mathcal{F}$
are relative energies. They can be used to compare the binding energy
of one site on the surface in comparison with another site, but should
not be interpreted as an absolute binding energy or total formation
energy of the surface. If a particle has a volume $\Omega$, then
the diffusion potential at $\vec{x}$ is related to the variation
of free energy with volume, \begin{equation}
\delta\mathcal{F}=\Omega^{-1}\int d^{d}\vec{x}\,\mu(\vec{x})\delta V(\vec{x}),\label{eq:DeltaF}\end{equation}
where $\delta V(\vec{x})$ is the volume variation at $\vec{x}$.
Calculating $\delta V(\vec{x})$, $V=\int d^{d}\vec{x}\,\h(\vec{x}).$Therefore,
$\delta V(\vec{x})=\delta\h(\vec{x}).$ Substituting into $\delta\mathcal{F}$
(Eq.~\eqref{eq:DeltaF}), $\delta\mathcal{F}=\Omega^{-1}\int d^{d}\vec{x}\,\mu(\vec{x})\delta\h(\vec{x})$
or $\mu(\vec{x})=\Omega\delta\mathcal{F}/\delta\h(\vec{x}).$

\subsection{Simple Model}

\label{sub:Simple-Model}Starting from Eq.~\eqref{eq:simplest},
$\mu(\vec{x})$ is found by taking the variational derivative,\[
\mu_{\text{elast.}}(\vec{x})=\Omega\frac{\delta}{\delta\h(\vec{x})}\int_{\text{volume}}d^{d}\vec{x}dz\,\omega[\h](\vec{x},z)=\Omega\omega\left(\vec{x}\right)\]
where the {}``$[\h]$'' indicates that the elastic energy, $\omega$,
is a nonlocal functional of the film height $\h$, and $\omega(\vec{x})=\omega[\h]\left(\vec{x},\h(\vec{x})\right)$,
the elastic energy density evaluated above lateral position $\vec{x}$
at the free surface ($z=\h(\vec{x})$). See~\cite{Freund:2003ih}
for details of the derivation. The surface energy diffusion potential
is \begin{eqnarray*}
\mu_{\text{surf.}}(\vec{x,}t) & = & \Omega\frac{\delta}{\delta\h(\vec{x})}\int d^{d}\vec{x}\,\left[1+(\grad\h(\vec{x}))^{2}\right]^{1/2}\gamma\\
 & = & -\Omega\grad\cdot\left[1+(\grad\h(\vec{x}))^{2}\right]^{1/2}\gamma=-\Omega\gamma\kappa(\vec{x}).\end{eqnarray*}
The wetting energy diffusion potential is \begin{eqnarray*}
\mu_{\text{wet}}(\vec{x}) & = & \Omega\frac{\delta}{\delta\h(\vec{x})}\int d^{d}\vec{x}\, W(\h(\vec{x}))\\
 & = & \Omega W'(\h(\vec{x}))\end{eqnarray*}
Putting these three terms together, one obtains Eq.~\eqref{eq:mu1}

\subsection{\label{sub:General-Model}General Model}

Consider the general form for the combined surface energy and wetting
potential,\[
\mathcal{F}_{sw}=\int d^{d}\vec{x}\, F_{sw}(\h(\vec{x}),\grad\h(\vec{x}))\]
 as in Eq.~\eqref{eq:swcombo} so that the free energy is an integral
over the $\vec{x}-$plane of an energy density that depends on $\h(\vec{x})$
and $\grad\h(\vec{x})$ locally. The corresponding diffusion potential
is \[
\mu(\vec{x})=\Omega\frac{\delta\f_{sw}}{\delta\h(\vec{x})}=\Omega\left[F_{sw}^{(10)}(\h(\vec{x}),\grad\h(\vec{x}))-\grad\cdot\vec{F}_{sw}^{(10)}(\h(\vec{x}),\grad\h(\vec{x}))\right]\]

\section{Linearized Diffusion Potential and Anisotropy}

\label{sec:Linearized-Diffusion-Potential}The linearized diffusion
potential $\mu_{\text{lin, }\vec{k}}$ is found by finding $\mu(\vec{x})$
to first order in height fluctuations ($h$), to get $\mu_{\text{lin}}(\vec{x})$
and then taking the Fourier transform to get $\mu_{\text{lin},\vec{k}}$.
The linearization of the simple isotropic diffusion potential corresponding
to Eqs.~\eqref{eq:simplest} and~\eqref{eq:mu1} was discussed in
Sec.~\ref{par:simple-form}. Here, the more general diffusion potential
corresponding to Eqs~\eqref{eq:swcombo} and~\eqref{eq:mu2} is
linearized and then applied to the anisotropic simple model and the
anisotropic general model. Only the surface and wetting parts of the
diffusion potential are discussed in this appendix. See ref.~\cite{Freund:2003ih},
Sec.~\ref{par:Elastic-anisotropy} and Appendix~\ref{sec:Elastic-Anisotropy}
for discussion of $\mu_{\text{elast.}}$.

\subsection{Linearizing the simple model}

\label{sub:Linearizing-the-simple}Consider a wetting potential and
diffusion potential that both depend on the film height gradient $\grad\h$,
$\gamma\rightarrow\gamma(\grad\h$) and $W(\h)\rightarrow W(\h,\grad\h)$.
Starting from Eq.~\eqref{eq:fsw-simp} and expanding to second order
in the film height fluctuation using $\h(\vec{x})=\bar{\h}+h(\vec{x})$
(Eq.~\eqref{eq:heights}), \begin{eqnarray*}
\left[1+\left(\grad\h\right)^{2}\right]^{-1/2}\gamma(\grad\h) & = & \left(1-\frac{1}{2}\left(\grad h\right)^{2}+\dots\right)\left(\gamma+\vec{\boldsymbol{\gamma}}'\cdot\grad h+\tilde{\boldsymbol{\gamma}}'':\grad h\grad h+\dots\right)\\
 & = & \gamma+\vec{\boldsymbol{\gamma}}'\cdot\grad h-\frac{1}{2}\gamma\left(\grad h\right)^{2}+\tilde{\boldsymbol{\gamma}}'':\grad h\grad h+O[h^{3}]\end{eqnarray*}
where $\gamma$ is $\gamma(\vec{0})$, and the primes indicate the
derivatives with respect to the surface height gradient.\[
\boldsymbol{\gamma}'=\left.\partial_{\grad\h}\gamma(\grad\h)\right|_{\grad\h=\vec{0}}\text{, and }\tilde{\boldsymbol{\gamma}}''=\left.\partial_{\grad\h}\partial_{\grad\h}\gamma(\grad\h)\right|_{\grad\h=\vec{0}}.\]
 Taking the derivative with respect to $\grad h$ results in a tensor
of rank equal to the order of the derivative because $\grad h$ is
a vector (tank 1 tensor). Taking the variational derivative, $\mu_{\text{surf.}}(\vec{x})=\Omega\delta\f_{\text{surf.}}/\delta h(\vec{x})$,\[
\mu_{\text{surf., lin}}(\vec{x})=\Omega\left[\gamma\nabla^{2}h(\vec{x})-\tilde{\boldsymbol{\gamma}}'':\grad\grad h(\vec{x})\right]\]
The term with $\boldsymbol{\gamma}'$ vanishes because it is the divergence
of a constant ($\grad\cdot\boldsymbol{\gamma}'$). Taking the inverse
Fourier transform,\begin{equation}
\mu_{\text{surf., lin},\vec{k}}=\Omega\left(-\gamma k^{2}+\vec{k}\cdot\tilde{\boldsymbol{\gamma}}''\cdot\vec{k}\right)h_{\vec{k}}.\label{eq:mu_s_l}\end{equation}
The first term is isotropic. The second term is parameterized by a
rank 2 symmetric tensor. 

Going through the same process, one finds essentially the same result
for an orientation dependent wetting energy. The step details are
so close to the details for linearizing the more general form, $F_{sw}(\h,\grad\h)$,
they are deferred to  (Appendix~\ref{sub:Linearizing-the-general}).
One finds that \begin{equation}
\mu_{\text{wet,lin},\vec{k}}=\Omega\left(W^{(20)}+\vec{k}\cdot\tilde{\mathbf{W}}^{(02)}\cdot\vec{k}\right).\label{eq:mu_w_l}\end{equation}
where $W^{(mn)}=\left.\partial_{\h}^{m}\partial_{\grad\h}^{n}W(\h,\grad\h)\right|_{\h=\bar{\h},\grad\h=\vec{0}}$
is the $m^{\text{th}}$ and $n^{\text{th}}$ derivative of the wetting
energy density with respect to $\h$ and $\grad\h$ evaluated for
a perfectly flat film of height $\bar{\h}$. $W^{(mn)}$ is a tensor
of rank $n$.

\subsubsection{isotropic case}

\label{sub:isotropic-case}In the isotropic case, $\tilde{\boldsymbol{\gamma}}''\rightarrow\gamma''\tilde{\mathbf{I}}$,
where $\tilde{\mathbf{I}}$ is the identity operator, and $\gamma''$
is a scalar. Similarly, $\tilde{\mathbf{W}}^{(02)}\rightarrow W^{(02)}\tilde{\mathbf{I}}$.
One thus gets for the combined surface and wetting parts of the diffusion
potential,\[
\mu_{sw,\text{lin},\vec{k}}=\Omega\left[\left(-\gamma+\gamma''+W^{(02)}\right)k^{2}+W^{(20)}\right]h_{\vec{k}}.\]
Thus, in the isotropic case, the linear order effect of introducing
a surface orientation to either the surface energy or the wetting
potential is simply to change the apparent surface energy density
by $\gamma\rightarrow\gamma-\gamma''-W^{(02)}$.

\subsubsection{anisotropic case}

\label{sub:anisotropic-case}The surface and wetting parts of the
diffusion potential (Eqs.~\eqref{eq:mu_s_l} and~\eqref{eq:mu_w_l})
can admit only a limited anisotropy. They both contain rank 2 symmetric
tensors, $\tilde{\boldsymbol{\gamma}}''$ and $\tilde{\mathbf{W}}^{(02)}$
in the $\vec{x}-$plane. For a two-dimensional surface, this means
that they can either have two-fold-symmetric (rotations by $180\degree$)
anisotropy or none at all. Thus, for the case considered in Sec.~\ref{par:Surface-and-Wetting},
four-fold-symmetric anisotropy , the surface and wetting parts of
the diffusion potential must be completely isotropic. As discussed
in Sec.~\ref{par:Surface-and-Wetting}, the (100) surface of zinc-blend
structures, such as the mentioned Ge, Si, InAs and GaAs present a
rather complicated situation. For simplicity, it is assumed here that
the surface and wetting energies are at least four-fold symmetric.
Consequently, they are completely isotropic.

Finally, it should be noted that if $F_{sw}$ depends on higher order
derivatives, then the discussion is greatly complicated and a larger
class of anisotropic terms is admissible. For example, when $F_{sw}\rightarrow F_{sw}(\h,\grad\h,\grad\grad\h,\grad\grad\grad\h,\dots)$
is expanded about $\h(\vec{x})=\bar{\h}$ to quadratic order in $h$,
it would contains tensors of rank 6 and maybe even higher.

\subsection{Linearizing the general model}

\label{sub:Linearizing-the-general}The elastic part of the linearized
diffusion potential was discussed in Sec.~\ref{par:Elastic-anisotropy}
and Appendix~\ref{sec:Elastic-Anisotropy} . Eq.~\eqref{eq:mu_w_l}
can be found by using all of the following steps with the substitution
$F_{sw}\rightarrow W$. The surface-wetting part of the diffusion
potential $\mu(\vec{x})$ is found by expanding $F_{sw}$ to second
order in the film-height fluctuation, $h$, and then taking the variational
derivative. Expanding $F_{sw}$ about $h=0$ and $\grad h=\vec{0}$,\begin{eqnarray*}
F_{sw}(\bar{\h}+h,\grad h) & = & F_{sw}^{(00)}+F_{sw}^{(10)}h+\vec{F}_{sw}^{(01)}\cdot\grad h+{h\vec{F}}_{sw}^{(11)}\cdot\grad h\dots\\
 &  & \dots+\frac{1}{2}F_{sw}^{(20)}h^{2}+\frac{1}{2}\tilde{\mathbf{F}}_{sw}^{(02)}:\grad h\grad h+O[h^{3}].\end{eqnarray*}
Note that in this expansion, all the $F_{sw}^{(mn)}$ terms are constant
with respect to $h$ and depend implicitly on the average film height,
$\bar{\h}$. The first index indicates the $m^{\text{th}}$ derivative
with respect to $h$. The second index indicates the $n^{\text{th}}$
derivative with respect to $\grad h$. The derivatives are evaluated
for a perfectly flat surface of height $\bar{\h}$. Thus, \[
F_{sw}^{\left(mn\right)}=\left.\partial_{\h}^{m}\partial_{\grad\h}^{n}F_{sw}\left(\h,\grad\h\right)\right|_{\h=\bar{\h},\,\grad\h=\vec{0}}.\]
Since $\grad h$ is a vector in the $\vec{x}-$plane, $F_{sw}^{(mn)}$
is a tensor of rank $n$. Taking the variational derivative of $\f_{sw}=\int d^{d}\vec{x}\, F_{sw}(\h,\grad\h)$
and keeping terms to order $h^{1}$,\[
\frac{\delta\f}{\delta h(\vec{x})}=F_{sw}^{(10)}-\grad\cdot\vec{F}_{sw}^{(01)}+F_{sw}^{(20)}h-\grad\cdot\left(\tilde{\mathbf{F}}_{sw}^{(02)}\cdot\grad h\right).\]
Note that the $F_{sw}^{(00)}$ term vanishes because it is constant,
and the $\vec{F}_{sw}^{(11)}$ term vanishes upon simplification.
Additionally, the $F_{sw}^{(10)}$ can be neglected if one enforces
the condition that the film-height fluctuations do not add or subtract
material from the surface, namely that $\int d^{d}\vec{x}\,\delta h(\vec{x},t)=0$.
Alternatively, one can discard it in anticipation of taking the gradient
of the diffusion potential, since it is a constant. The term $\grad\cdot\vec{F}_{sw}^{(01)}=0$
for the same reasons, or because $F_{sw}^{(01)}$ is a constant. Multiplying
through by the atomic volume, \begin{equation}
\mu_{\text{lin}}(\vec{x})=\Omega\left[F_{sw}^{(20)}h-\tilde{\mathbf{F}}_{sw}^{(02)}:\grad\grad h\right].\label{eq:mulingen}\end{equation}

\subsubsection{isotropic case}

\label{sub:isotropic-case}In the isotropic case, $\tilde{\mathbf{F}}_{sw}^{(02)}$
must be proportional to the identity so that $\tilde{\mathbf{F}}_{sw}^{(02)}=F_{sw}^{(02)}\tilde{\mathbf{I}}$;
thus,\[
\mu_{sw,\text{lin}}(\vec{x})=\Omega\left[F_{sw}^{(20)}h(\vec{x})-F_{sw}^{(02)}\nabla^{2}h(\vec{x})\right].\]
Taking the inverse Fourier transform of this equation,\[
\mu_{sw,\text{lin},\vec{k}}=\Omega\left[F_{sw}^{(20)}+F_{sw}^{(02)}k^{2}\right]h_{\vec{k}}.\]
This gives case b in Eq.~\eqref{eq:fiso}.

\subsubsection{anisotropic case}

\label{sub:anisotropic-case}If the surface is anisotropic, then $\tilde{\mathbf{F}}_{sw}^{(02)}$
in Eq.~\eqref{eq:mulingen} is a rank 2 symmetric tensor in the $\vec{x}-$plane.
Thus, it can have two distinct eigenvalues, and automatically has
2-fold rotational symmetry (rotations by $180\degree$). If any other
symmetry is assumed such as 4-fold symmetry (rotations by $90\degree$),
then $\tilde{\mathbf{F}}_{sw}^{(02)}$ must be fully isotropic. Taking
the inverse Fourier transform,

\[
\mu_{sw,\text{lin},\vec{k}}=\Omega\left[F_{sw}^{(20)}+\vec{k}\cdot\tilde{\mathbf{F}}_{sw}^{(02)}\cdot\vec{k}\right]h_{\vec{k}}.\]
In Eq.~\eqref{eq:fanis}, case b, it is assumed that there is four-fold
symmetry, resulting in a surface-wetting part of the diffusion potential
that is completely isotropic.

\section{Elastic Anisotropy}

\label{sec:Elastic-Anisotropy}In principal, the anisotropic elastic
energy $\omega_{\vec{k}}$ is found in the same fashion as the isotropic
elastic energy.~\cite{Freund:2003ih} The flat film, initially in
a state of biaxial stress, is perturbed by a small periodic surface
fluctuation of amplitude $h_{0}$. An appropriate elastic field is
added to satisfy the perturbed traction-free boundary condition at
the free surface. Finally, the elastic energy is evaluated at the
free surface to first order in $h_{0}$. The coefficient $h_{0}$
is the sought after $\omega_{\vec{k}}$. The equations themselves
are cumbersome and best solved using a numeric implementation, so
an abstract procedure for calculating $\omega_{\vec{k}}$ is outlined
here. $\omega_{\vec{k}}$ is found for $k=1$ but arbitrary $\theta_{\vec{k}}$.

Let the surface have a height variation\[
h(\vec{x})=h_{0}e^{ikx}.\]
To first order in $h_{0}$, the surface normal is \[
\mathbf{n}(\vec{x})=-ikh_{0}e^{ikx}\mathbf{i}+\mathbf{k}.\]
The elastic energy needs to be calculated to first order in $h_{0}$.
To find the elastic energy, it is necessary to find the perturbing
elastic field to first order in $h_{0}$. 

The initial unperturbed stress state is\[
\tilde{\sigma}_{m}=\left[\begin{array}{ccc}
\sigma_{m} & 0 & 0\\
0 & \sigma_{m} & 0\\
0 & 0 & 0\end{array}\right],\]
where $\sigma_{m}=\left(c_{11}+c_{12}-2c_{12}^{2}/c_{11}\right)\epsilon_{m}$.
Note that this stress state is isotropic in the $x-y$-plane and thus
independent of rotations about the vertical axis. Under this stress
state, a flat surface is traction-free. With the height perturbation,
the traction is\begin{equation}
t_{j}=\left(\mathbf{n}\cdot\tilde{\sigma}_{m}\right)_{j}=-ikh_{0}M\epsilon_{m}\delta_{j1}e^{ikx}.\label{eq:torig}\end{equation}

Next to find the perturbing elastic fields. These are not isotropic
in the $x-y-$plane, and it is necessary to take into account the
angle. First, the $3\times3\times3\times3$ elastic stiffness tensor
$c_{ijkl}$ is constructed for the cube orientation from the compact
$9\times9$ matrix $c_{ij}$. The tensor representation aids in rotation.
The stiffness tensor is then passively rotated in the $x-y-$plane
by an angel $\theta_{\vec{k}}$,\[
c_{ijkl}(\theta_{\vec{k}})=\sum_{m,n,p,q=1}^{3}R(\theta_{\vec{k}})_{im}R(\theta_{\vec{k}})_{jn}R(\theta_{\vec{k}})_{kp}R(\theta_{\vec{k}})_{lq}c_{mnpq}\]
where\[
R(\theta_{\vec{k}})=\left[\begin{array}{ccc}
\cos(\theta_{\vec{k}}) & \sin(\theta_{\vec{k}}) & 0\\
-\sin(\theta_{\vec{k}}) & \cos(\theta_{\vec{k}}) & 0\\
0 & 0 & 1\end{array}\right].\]
This passive rotation of $c_{ijkl}$ is equivalent to actively rotating
the wave vector $\vec{k}=k\mathbf{i}$ by $\theta_{\vec{k}}$. 

\newcommand{\fxz}{e^{k(ix+\kappa z)}}

The appropriate form for the perturbing displacement field is found.
Assume a displacement of the form\[
u_{i}(x,y,z)=U_{i}\fxz,\]
where $\kappa$ can have a complex value. The elastic equilibrium
equations are \[
\sum_{i,k,l=1}^{3}\frac{\partial}{\partial x_{i}}c_{ijkl}(\theta_{\vec{k}})\frac{\partial}{\partial x_{k}}u_{l}=0;\, j=1\dots3.\]

\begin{equation}
\left(\sum_{l=1}^{3}C_{jl}(\theta_{\vec{k}},\kappa)U_{l}\right)k^{2}\fxz=0\label{eq:equil}\end{equation}
where \[
C_{jl}(\theta_{\vec{k}},\kappa)=\sum_{i,k=1}^{3}c_{ijkl}(\theta_{\vec{k}})(i\delta_{i1}+\delta_{i3}\kappa)(i\delta_{k1}+\delta_{k3}\kappa).\]
Factoring out $k^{2}\fxz$, the part in parenthesis must be identically
zero. 

To obtain a non-trivial solution, the determinant of $C_{jl}(\theta_{\vec{k}},\kappa)$
to zero. Six complex values of $\kappa$ are found. The values of
$\kappa$ with $\text{Re}[\kappa]<0$ are discarded since the corresponding
displacements blow up as $z\rightarrow-\infty$. Each of the remaining
values $\kappa=\kappa^{p}$ with $p=1\dots3$ is substituted back
into $C_{jl}(\theta_{\vec{k}},\kappa)$, and Eq.~\eqref{eq:equil}
is solved to find the corresponding eigenvectors, $U_{l}^{p}$. The
total displacement is thus\[
u_{l}(x,y,z)=i\epsilon_{m}h_{0}\sum_{p=1}^{3}A_{p}U_{l}^{p}e^{k(ix+\kappa^{p}z)},\]
where it is assumed that the perturbing elastic displacement field
is proportional to $h_{0}$and $\sigma_{m}$, and the factor of $i$
is put in for convenience. The coefficients $A_{p}$ can be found
from the traction-free boundary condition at the free surface. The
traction formula is\begin{eqnarray}
t_{j} & = & \sum_{i,k,l=1}^{3}n_{i}c_{ijkl}(\theta_{\vec{k}})\frac{\partial}{\partial x_{k}}u_{l}(x,y,z)=ik\epsilon_{m}h_{0}\sum_{i,k,l,p=1}^{3}n_{i}c_{ijkl}(\theta_{\vec{k}})A_{p}U_{l}^{p}(i\delta_{k1}+\kappa^{p}\delta_{k3})e^{k(ix+\kappa^{p}z)}\label{eq:trac1}\end{eqnarray}
The traction is already proportional to $h_{0}$. Thus, all terms
in the sum must be kept to zeroth order in $h_{0}$ so that\[
h(\vec{x})=0\text{, and }\mathbf{n}(\vec{x})=\mathbf{k}.\]
Thus, plugging $z=0$ to Eq.~\eqref{eq:trac1},\begin{equation}
t_{j}=ik\epsilon_{m}h_{0}\sum_{p=1}^{3}\sum_{l=1}^{3}\left(ic_{3j1l}(\theta_{\vec{k}})+\kappa^{p}c_{3j3l}(\theta_{\vec{k}})\right)A_{p}U_{l}^{p}e^{ikx}.\label{eq:tpert}\end{equation}

Since the total traction (Eqs.~\eqref{eq:torig} and~\eqref{eq:tpert})
must be zero, the coefficients $A_{p}$ are found from\[
K_{jp}A_{p}=R_{j},\]
where\[
K_{jp}=\sum_{l=1}^{3}\left(ic_{3j1l}(\theta_{\vec{k}})+\kappa^{p}c_{3j3l}(\theta_{\vec{k}})\right)U_{l}^{p},\]
and\[
R_{j}=M\delta_{j1}\]
for $j=1\dots3$. It is worth noting that only for the symmetry directions,
$\theta_{\vec{k}}=0\degree$ and $\theta_{\vec{k}}=45\degree$ is
the strain purely plane-strain as it is for the elastically isotropic
case.

The elastic energy at the film surface is found to order $O(h_{0})$.
If the stress and strain are expanded to first order in $h_{0}$,
$\tilde{\sigma}=\tilde{\sigma}_{0}+\tilde{\sigma}_{1}$, and $\tilde{\epsilon}=\tilde{\epsilon}_{0}+\tilde{\epsilon}_{1}$,
then \[
U=\frac{1}{2}\tilde{\epsilon}:\tilde{c}:\tilde{\epsilon}=\frac{1}{2}\tilde{\sigma}_{0}:\epsilon_{0}+\tilde{\sigma}_{0}:\tilde{\epsilon}_{1}+O(h_{0}^{2}).\]
Thus,\[
U=U_{0}+M\epsilon_{m}\left((\epsilon_{1})_{11}+(\epsilon_{1})_{22}\right)\]
\[
(\epsilon_{1})_{11}=\frac{\partial u_{1}}{\partial x}=-\epsilon_{m}kh_{0}\sum_{p=1}^{3}A_{p}U_{1}^{p}.\]
$(\epsilon_{1})_{22}=\partial u_{2}/\partial y=0$. Thus,\[
U=U_{0}-\mathcal{E}_{\theta_{\vec{k}}}kh_{0}e^{ikx}\]
where \[
\mathcal{E}_{\theta_{\vec{k}}}=M\epsilon_{m}^{2}\sum_{p=1}^{3}A_{p}U_{1}^{p}\]
where $A_{p}$and $U_{1}^{p}$ are implicitly functions of $\theta_{\vec{k}}$.
This procedure has been used to find the values of $\mathcal{E}_{0\degree}$
and $\mathcal{E}_{45\degree}$ for Table.~\ref{tab:Elastic-constants-and}
and Sec.~\ref{sec:Application-to-Materials}.

\section{Diffusional Anisotropy}

\label{sec:Diffusion-Anisotropy}In general, the surface diffusivity
can depend on the film height $\h(\vec{x})$ and the surface orientation
$\grad\h(\vec{x})$ so that the surface current is\[
\vec{J}_{S}(\vec{x})=\tilde{\mathbf{D}}(\h(\vec{x}),\grad H(\vec{x}))\cdot\grad s\mu(\vec{x})\]
where $\grad s$ is the surface gradient, and $\tilde{\mathbf{D}}$
is a rank 2 tensor in the two-dimensional space tangent to the film
surface at $\vec{x}$. Linearizing the surface current about a flat
surface,\[
\vec{J}_{S}(\vec{x})=\tilde{\mathbf{D}}(\bar{\h})\cdot\grad\mu_{\text{lin}}(\vec{x})\]
where the diffusivity must be evaluated for $h=0$ and $\grad h=0$,
since $\mu_{\text{lin}}(\vec{x})$ is already proportional to $h(\vec{x})$.
The linearized diffusivity is a symmetric rank 2 tensor in the $\vec{x}-$plane.
Thus, it is similar to $\tilde{\mathbf{F}}_{sw}$ discussed in Appendix~\ref{sub:anisotropic-case}.
It is automatically either two-fold symmetry (rotations by $180\degree$)
or it is completely isotropic. In Eq.~\eqref{eq:fanis}, four-fold
symmetry of the surface is assumed. Thus, the diffusivity must be
completely isotropic; $\tilde{\mathbf{D}}\rightarrow\mathcal{D}$,
a scalar. Section~\ref{par:Surface-and-Wetting} and Appendix~\ref{sub:anisotropic-case}
contain discussions of the symmetry properties of the various rank
2 tensors that appear in the linear evolution equations. A limited
case of diffusional anisotropy has been modeled via kinetic Monte
Carlo technique.~\cite{Meixner:2003lz}

\section{Correlation Functions}

\label{app:Correlation-Functions}

\subsection{Mean Values}

\label{sub:Mean-Values}Equations~\eqref{eq:ergo-1} and~\eqref{eq:cork}
are central to the presented analysis. Here, they are derived. The
two-point correlation functions for a stochastic system are introduced.
Then, the average of the autocorrelation function is taken and expressed
in terms of the two-point correlation functions. Finally, this average
is simplified using the translational invariance of the system (governing
equations and ensemble of initial conditions).

The two-point real-space space correlation function is

\[
C(\vec{x},\vec{x}')=\left\langle h(\vec{x})h(\vec{x}')^{*}\right\rangle ,\]
and the reciprocal space correlation function is \[
C_{\vec{k}\vec{k}'}=\left\langle h_{\vec{k}}h_{\vec{k}'}^{*}\right\rangle .\]
These are related by the double Fourier transform,\begin{eqnarray}
 & C_{\vec{k}\vec{k}'}=\frac{1}{(2\pi)^{2d}}\int d^{d}\vec{x}d^{d}\vec{x}'\, e^{-i\vec{k}\cdot\vec{x}+i\vec{k}'\cdot\vec{x}'}C(\vec{x},\vec{x}');\label{eq:CF1}\\
 & C(\vec{x},\vec{x}')=\int d^{d}\vec{k}d^{d}\vec{k}'\, e^{i\vec{k}\cdot\vec{x}-i\vec{k}'\cdot\vec{x}'}C_{\vec{k}\vec{k}'}.\label{eq:CF2}\end{eqnarray}

These ensemble correlation functions can be used to give the ensemble-mean
autocorrelation function and spectrum function. In real space,

\begin{eqnarray}
\left\langle C^{A}(\dx)\right\rangle  & = & \frac{1}{A}\int d^{2}\vec{x}'\,\left\langle h(\dx+\vec{x}')h(\vec{x}')\right\rangle \nonumber \\
 & = & \frac{1}{A}\int d^{2}\vec{x}'\, C(\dx+\vec{x}',\vec{x}').\label{eq:CAav}\end{eqnarray}

\begin{equation}
\left\langle C_{\vec{k}}^{A}\right\rangle =\frac{(2\pi)^{d}}{A}\left\langle h_{\vec{k}}h_{\vec{k}}^{*}\right\rangle =\frac{(2\pi)^{d}}{A}C_{\vec{k}\vec{k}}.\label{eq:rcor7}\end{equation}

Fortunately, the translational invariance of the system simplifies
these relations. Inspecting the governing equations and invoking the
translational invariance of the stochastic initial conditions, the
resulting ensemble and its statistical measures must also be translationally
invariant. Thus under the translation by $\vec{x}'$,\begin{equation}
C(\dx+\vec{x}',\vec{x}')=C(\dx,\vec{0})=C(\dx),\label{eq:TIr}\end{equation}
so that the independent variable is reduced to just the difference
vector $\dx=\vec{x}-\vec{x}'$. This relation can be used to simplify
both the real and reciprocal space relations.

The real space relation simplifies as follows.Inserting Eq.~\eqref{eq:TIr}
into Eq.~\eqref{eq:CAav},\begin{equation}
\left\langle C^{A}(\dx)\right\rangle =\frac{1}{A}\int d^{2}\vec{x}'\, C(\dx,\vec{0})=C(\dx).\label{eq:Csimp}\end{equation}
The reciprocal space relation (Eq.~\eqref{eq:CF1}) simplifies to
\begin{equation}
C_{\vec{k}\vec{k}'}=C_{\vec{k}}\delta^{2}(\vec{k}-\vec{k}')=C_{\vec{k}}\frac{A}{(2\pi)^{d}}\delta_{\vec{k}\vec{k}'},\label{eq:RSCsimp}\end{equation}
where \[
C_{\vec{k}}=\frac{1}{(2\pi)^{d}}\int d^{2}\dx\, e^{-i\vec{k}\cdot\dx}C(\dx).\]
One can see immediately from Eq.~\eqref{eq:Csimp} that $C_{\vec{k}}$
is the Fourier transform of $\left\langle C^{A}(\dx)\right\rangle =C(\dx)$,
or one can plug Eq.~\eqref{eq:RSCsimp} into Eq.~\eqref{eq:rcor7},
to get $\left\langle C_{\vec{k}}^{A}\right\rangle =C_{\vec{k}}.$

\subsection{Variance and Convergence}

\label{sub:Variance-and-Convergence}The ergodic hypothesis is that
an average with respect to a parameter such as position or time tends
towards an ensemble average. In this case, \begin{eqnarray}
 &  & C_{\vec{k}}^{A}\approx\left\langle C_{\vec{k}}^{A}\right\rangle =C_{\vec{k}},\label{eq:ergodic}\\
\text{and } &  & C^{A}(\dx)\approx\left\langle C^{A}(\dx)\right\rangle =C(\dx).\nonumber \end{eqnarray}
when the surface area is very large. The ensemble average is a good
substitute if the variance about the average vanishes as the substrate
area $A$ becomes large. It is found that in reciprocal space,\begin{equation}
\text{Var}(C_{\vec{k}}^{A})=\left\langle \left(C_{\vec{k}}^{A}\right)^{2}\right\rangle -\left\langle C_{\vec{k}}^{A}\right\rangle ^{2}=C_{\vec{k}}^{2}.\label{eq:rsvar}\end{equation}
Thus, the ergodic hypothesis does not hold for $C_{\vec{k}}^{A}$.
In practice, $C_{\vec{k}}^{A}$ is a speckled version of $C_{\vec{k}}$
(Fig.~\ref{fig:Ck}) However, if one smooths $C_{\vec{k}}^{A}$ by
averaging over a small patch in reciprocal space of size $k_{\text{smooth}}=1/\Delta_{s}$,
so that \begin{equation}
C_{\vec{k}}^{A}(\Delta_{s})=\left(\frac{\Delta_{s}^{2}}{2\pi}\right)^{d/2}\int d^{d}\vec{k}'\, e^{-\frac{1}{2}\Delta_{s}^{2}(\vec{k}'-\vec{k})^{2}}C_{\vec{k}'}^{A},\label{eq:rss}\end{equation}
then $\text{Var}\left(C_{\vec{k}}^{A}(\Delta_{s})\right)$ diminishes
as $1/A$. For sufficiently large $\Delta_{s}$,\begin{equation}
\left\langle C_{\vec{k}}^{A}(\Delta_{s})\right\rangle \approx C_{\vec{k}},\label{eq:dsl}\end{equation}
and \begin{equation}
\text{Var}\left(C_{\vec{k}}^{A}(\Delta_{s})\right)\approx\frac{\pi^{d/2}\Delta_{s}^{d}}{A}C_{\vec{k}}^{2}.\label{eq:rssvar}\end{equation}
Thus, the ergodic hypothesis (Eq.~\eqref{eq:ergodic}) only holds
for a smoothed version of $C_{\vec{k}}^{A}$.

In real space, \begin{eqnarray}
\text{Var}\left(C^{A}(\dx)\right) & = & \left\langle \left(C^{A}(\dx)\right)^{2}\right\rangle -\left\langle C^{A}(\dx)\right\rangle ^{2}\nonumber \\
 & = & \frac{(2\pi)^{d}}{A}\int d^{d}\vec{k}\,\left(e^{2i\vec{k}\cdot\dx}C_{\vec{k}}^{2}+C_{\vec{k}}^{2}\right),\label{eq:cvar}\end{eqnarray}
where the integral is bounded (finite) provided that either $t>0$
or the atomic scale cutoff $b_{0}>0$. Thus, the ergodic hypothesis
holds for the real space autocorrelation function.

\subsubsection{Eq.~\eqref{eq:rsvar}}

\label{sub:Eq.-[eq:rsvar]}First, $\left\langle C_{\vec{k}}^{A}C_{\vec{k}'}^{A}\right\rangle $
is calculated.

\[
\left\langle C_{\vec{k}}^{A}C_{\vec{k}'}^{A}\right\rangle =\left(\frac{(2\pi)^{d}}{A}\right)^{2}\left\langle h_{\vec{k}}h_{\vec{k}}^{*}h_{\vec{k}'}h_{\vec{k}'}^{*}\right\rangle .\]
Assume that he distribution of $h_{\vec{k}}$ is gaussian. Also, assume
that $h(\vec{x})$ is real so that $h_{\vec{k}}h_{-\vec{k}}=\left|h_{\vec{k}}\right|^{2}$.
Then, \begin{eqnarray*}
\left\langle h_{\vec{k}_{1}}h_{\vec{k}_{2}}^{*}h_{\vec{k}_{3}}h_{\vec{k}_{4}}^{*}\right\rangle  & = & C_{\vec{k}_{1}}C_{\vec{k}_{2}}\delta^{d}(\vec{k}_{1}-\vec{k}_{4})\delta^{d}(\vec{k}_{2}-\vec{k}_{3})\dots\\
 & \dots & +C_{\vec{k}_{1}}C_{\vec{k}_{2}}\delta^{d}(\vec{k}_{1}+\vec{k}_{3})\delta^{d}(\vec{k}_{2}+\vec{k}_{4})\dots\\
 & \dots & +C_{\vec{k}_{1}}C_{\vec{k}_{3}}\delta^{d}(\vec{k}_{1}-\vec{k}_{2})\delta^{d}(\vec{k}_{3}-\vec{k}_{4}).\end{eqnarray*}
Thus,\begin{eqnarray}
\left\langle C_{\vec{k}}^{A}C_{\vec{k}'}^{A}\right\rangle  & = & \left(\frac{(2\pi)^{d}}{A}\right)^{2}\left(C_{\vec{k}}^{2}\left[\delta^{d}(\vec{k}-\vec{k}')\right]^{2}\dots\right.\nonumber \\
 &  & \dots\left.+C_{\vec{k}}^{2}\left[\delta^{d}(\vec{k}+\vec{k}')\right]^{2}+C_{\vec{k}}C_{\vec{k}'}\left[\delta^{d}(\vec{0})\right]^{2}\right).\label{eq:CAk1}\\
 & = & C_{\vec{k}}^{2}\left(\delta_{\vec{k}\vec{k}'}+\delta_{\vec{k}(-\vec{k}')}\right)+C_{\vec{k}}C_{\vec{k}'},\label{eq:CAk2}\end{eqnarray}
where Eq.~\eqref{eq:DD1} has been used liberally. Setting $\vec{k}=\vec{k}'$,
results in Eq.~\eqref{eq:rsvar}.

\subsubsection{Eq.~\eqref{eq:rssvar}}

\label{sub:Eq.-[eq:rssvar]}Now consider $C_{\vec{k}}^{A}$ smoothed
over a length $\Delta_{s}$ (Eq.~\eqref{eq:rss}). The mean value
is\begin{eqnarray*}
\left\langle C_{\vec{k}}^{A}(\Delta_{s})\right\rangle  & = & \left(\frac{\Delta_{s}^{2}}{2\pi}\right)^{d/2}\int d^{d}\vec{k}'\, e^{-\frac{1}{2}\Delta_{s}^{2}(\vec{k}'-\vec{k})^{2}}\left\langle C_{\vec{k}'}^{A}\right\rangle .\\
 & = & \left(\frac{\Delta_{s}^{2}}{2\pi}\right)^{d/2}\int d^{d}\vec{k}'\, e^{-\frac{1}{2}\Delta_{s}^{2}(\vec{k}'-\vec{k})^{2}}C_{\vec{k}'}.\end{eqnarray*}
For sufficiently small $k_{\text{smooth}}$, (sufficiently large $\Delta_{s}$),
Eq.~\eqref{eq:dsl} results. 

The variance of $C_{\vec{k}}^{A}(\Delta_{s})$ is now calculated.
First, it is necessary to calculate $\left\langle \left[C_{\vec{k}}^{A}(\Delta_{s})\right]^{2}\right\rangle $.\begin{eqnarray*}
\left\langle \left[C_{\vec{k}}^{A}(\Delta_{s})\right]^{2}\right\rangle  & = & \left(\frac{\Delta_{s}^{2}}{2\pi}\right)^{d}\int d^{d}\vec{k}'\, e^{-\frac{1}{2}\Delta_{s}^{2}(\vec{k}'-\vec{k})^{2}}\dots\\
 & \dots & \times\int d^{d}\vec{k}''\, e^{-\frac{1}{2}\Delta_{s}^{2}(\vec{k}''-\vec{k})^{2}}\left\langle C_{\vec{k}'}^{A}C_{\vec{k}''}^{A}\right\rangle .\end{eqnarray*}
Using Eq.~\eqref{eq:CAk1} and Eq.~\eqref{eq:DD1} as needed,\begin{eqnarray*}
\\\left\langle \left[C_{\vec{k}}^{A}(\Delta_{s})\right]^{2}\right\rangle  & = & \left(\frac{\Delta_{s}^{2}}{2\pi}\right)^{d}\int d^{d}\vec{k}'d^{d}\vec{k}''\, e^{-\frac{1}{2}\Delta_{s}^{2}(\vec{k}'-\vec{k})^{2}}e^{-\frac{1}{2}\Delta_{s}^{2}(\vec{k}''-\vec{k})^{2}}\left(\frac{(2\pi)^{d}}{A}\right)^{2}\dots\\
 &  & \dots\times\left\{ C_{\vec{k}}^{2}\left[\delta^{d}(\vec{k}'-\vec{k}'')\right]^{2}+C_{\vec{k}}^{2}\left[\delta^{d}(\vec{k}'+\vec{k}'')\right]^{2}+C_{\vec{k}}C_{\vec{k}'}\left[\delta^{d}(\vec{0})\right]^{2}\right\} \\
 & = & \frac{\Delta_{s}^{2d}}{A}\int d^{d}\vec{k}'\,\left\{ e^{-\Delta_{s}^{2}(\vec{k}'-\vec{k})^{2}}C_{\vec{k}'}^{2}+e^{-\frac{1}{2}\Delta_{s}^{2}\left[(\vec{k}'-\vec{k})^{2}+(\vec{k}'+\vec{k})^{2}\right]}C_{\vec{k}'}^{2}\right\} \dots\\
 &  & \dots+\left[\left(\frac{\Delta_{s}^{2}}{2\pi}\right)^{d/2}\int d^{d}\vec{k}'\, e^{-\frac{1}{2}\Delta_{s}^{2}(\vec{k}'-\vec{k})^{2}}C_{\vec{k}'}\right]^{2}\end{eqnarray*}
The first integral is bounded (finite) because $C_{\vec{k}}$ is bounded.
Let its finite value be denoted $I$. The second integral is simply
$\left\langle C_{\vec{k}}^{A}(\Delta_{s})\right\rangle .$Thus,

\[
\text{Var}(C_{\vec{k}}^{A}(\Delta_{s}))=\frac{\Delta_{s}^{2d}I}{A},\]
a finite value that decreases as $A^{-1}$ as required for the ergodic
hypothesis to hold. For sufficiently small $k_{\text{smooth }}$ (large
$\Delta_{s}$), $I\approx(\pi/\Delta_{s}^{2})^{d/2}C_{\vec{k}}^{2}$,
and Eq.~\eqref{eq:rssvar} results. It should also be noted that
the large $\Delta_{s}$ required for this approximation also creates
a more stringent requirement that $A$ be large.

\subsubsection{Eq.~\eqref{eq:cvar}}

\label{sub:Eq.-[eq:cvar]}Now, consider the real space auto-correlation
function. First, $\left\langle C^{A}(\dx)C^{A}(\dx)\right\rangle $
is needed. \[
\left\langle C^{A}(\dx)C^{A}(\dx)\right\rangle =\int d^{d}\vec{k}d^{d}\vec{k}'\, e^{i(\vec{k}+\vec{k}')\cdot\dx}\left\langle C_{\vec{k}}^{A}C_{\vec{k}'}^{A}\right\rangle \]
Proceeding in a fashion similar to the previous section (making use
of Eqs.~\eqref{eq:CAk1} and~\eqref{eq:DD1} as needed) ,\begin{eqnarray*}
\\\left\langle C^{A}(\dx)C^{A}(\dx)\right\rangle  & = & \frac{(2\pi)^{2d}}{A^{2}}\int d^{d}\vec{k}d^{d}\vec{k}'\, e^{i(\vec{k}+\vec{k}')\cdot\dx}\left(C_{\vec{k}}^{2}\left[\delta^{d}(\vec{k}-\vec{k}')\right]^{2}\dots\right.\\
 &  & \left.\dots+C_{\vec{k}}^{2}\left[\delta^{d}(\vec{k}+\vec{k}')\right]^{2}+C_{\vec{k}}C_{\vec{k}'}\left[\delta^{d}(\vec{0})\right]^{2}\right)\\
 & = & \frac{(2\pi)^{d}}{A}\int d^{d}\vec{k}\,\left(e^{2i\vec{k}\cdot\dx}C_{\vec{k}}^{2}+C_{\vec{k}}^{2}\right)\dots\\
 &  & \dots+\left(\int d^{d}\vec{k}\, e^{i\vec{k}\cdot\dx}C_{\vec{k}}\right)\left(\int d^{d}\vec{k}'\, e^{i\vec{k}'\cdot\dx}C_{\vec{k}}\right)\\
 & = & \frac{(2\pi)^{d}}{A}\int d^{d}\vec{k}\,\left(e^{2i\vec{k}\cdot\dx}C_{\vec{k}}^{2}+C_{\vec{k}}^{2}\right)+\left\langle C^{A}(\dx)\right\rangle ^{2}\end{eqnarray*}
Thus, Eq.~\eqref{eq:cvar} results. For the variance to be vanishing,
the integral in Eq.~\eqref{eq:cvar} must be bounded (finite). If
time, $t>0$, the exponential in Eq.~\eqref{eq:rcgen} guarantees
that the integral is bounded. For time $t=0$, the integral is only
bounded if the atomic scale cutoff $b_{0}>0$.

\section{Atomic Scale Cutoff}

\label{sec:Atomic-Scale-Cutoff}Starting from Eq.~\eqref{eq:Ckkgen},
\begin{equation}
C_{\vec{k}}=\frac{\Delta^{2}}{(2\pi)^{d}}e^{2\sigma_{\vec{k}}t-\frac{1}{2}b_{0}^{2}k^{2}}.\label{eq:rcgen}\end{equation}
The effect of the small scale cutoff is both small and short-lived,
as it only works to suppress fluctuations with large wavenumbers.
The most important fluctuations have wavenumbers between $0$ and
$2k_{c}$. Thus, the typical size of the cutoff term is about $b_{0}^{2}k_{c}^{2}.$
If a typical dot size or spacing size $10\text{ nm}$, and a typical
atomic scale is $10^{-1}\text{ nm}$, a typical value for this term
is about $10^{-3}-10^{-2}$. To calculate the effect of the cutoff,
it can absorbed into the time-dependent part with the substitution\[
b\rightarrow b\left(1+\frac{b_{0}^{2}}{4b\mathcal{D}t}\right)\]
so that its effect lasts only as long as a perturbation with atomic
scale curvature ($\kappa=b_{0}$). Thus, Eq.~\eqref{eq:Ck} is a
good approximation.

\section*{Acknowledgement}

Thanks to L. Fang and C. Kumar for useful comments during the writing
of this article.

\bibliographystyle{spiejour}
\bibliography{/Users/friedman/Documents/Bibliographies/General}

\end{document}